# Nuclear Security Applications of Antineutrino Detectors: Current Capabilities and Future Prospects

A. Bernstein[a*], G. Baldwin[b], B. Boyer[c], M. Goodman[d], J. Learned[e], J. Lund[b], D. Reyna[b], R. Svoboda[a,f]

a) Lawrence Livermore National Laboratory , b) Sandia National Laboratories c) Los Alamos National Laboratory, d) Argonne National Laboratory,
e) University of Hawaii, f) University of California, Davis
* corresponding author bernstein3@llnl.gov

## 1 Executive Summary

Antineutrinos are electrically neutral, nearly massless fundamental particles produced in large numbers in the cores of nuclear reactors and in nuclear explosions. In the half century since their discovery, major advances in the understanding of their properties, and in detector technology, have opened the door to a new discipline – Applied Antineutrino Physics. Because antineutrinos are inextricably linked to the process of nuclear fission, many applications of interest are in nuclear nonproliferation.

The current state of the art in antineutrino detection is such that it is now possible to monitor the operational status, power levels, and fissile content of nuclear reactors in real time with simple detectors at distances of a few tens of meters. This has already been demonstrated at civil power reactors in Russia and the United States, with detectors designed specifically for reactor monitoring and safeguards[1,2]. This existing near-field monitoring capability may be useful in the context of the International Atomic Energy Agency's (IAEA) Safeguards Regime[3], and other cooperative monitoring regimes, such as the proposed Fissile Material Cutoff Treaty[4]

Though not part of any existing treaty, today's technology would allow cooperative monitoring, discovery or exclusion of small (few MegaWatt thermal, MWt) reactors at standoff distances up to 10 kilometers. In principle, discovery and exclusion is also possible at longer ranges, as is standoff nuclear explosion detection at the kiloton level. However, the required detector masses are 10-100 times greater than the state of the art, and achieving these long range detection goals would require significant research and development on several fronts. Many elements of the necessary R&D program are already being pursued in the fundamental physics community, in the form of very large neutrino detection experiments.

Antineutrino detectors are likely not useful for detection or monitoring of quiescent, non-critical fissile materials, regardless of the amount of material or the size of the detector, because emission rates from these materials are vastly lower than from critical systems.

This white paper presents a comprehensive survey of applied antineutrino physics relevant for nonproliferation, summarizes recent advances in the field, describes the overlap of this nascent discipline with other ongoing fundamental and applied antineutrino research, and charts a course for research and development for future applications. It is intended as a resource for policymakers, researchers, and the wider nuclear nonproliferation community.

The conclusions and recommendations of this white paper are:

1) Practical mear-field (<100 m) monitoring of pressurized water reactors with antineutrino detectors has been demonstrated[1,2], and offers a promising complement to existing reactor monitoring methods for IAEA and other safeguards regimes. We recommend further investigation of near-field antineutrino monitoring capabilities for providing reactor operational status, thermal power and fissile content of reactors for safeguards. In particular, further R&D is appropriate in determining sensitivity levels at non-PWR reactors, in direct measurement and simulation of the evolution of antineutrino rates and spectra at various reactors, and in detectors with improved deployability characteristics. We further recommend close cooperation between the antineutrino physics community, the IAEA and relevant government agencies worldwide to ensure that development is well matched to safeguards needs.

2) Mid-field (1-10 kilometer) monitoring of the operational status, and presence or absence of 10 MWt reactors, and placing upper limits on plutonium production in such reactors, is possible with existing technology, assuming deeply buried (1 kilometer overburden) detectors and in parts of the world with few commercial reactor backgrounds. We recommend development of 1000-10,000 ton scale detectors with dual physics and nonproliferation aims. We further recommend R&D focus on reducing costs through, among other options, reduction in overburden while maintaining suitable signal to background levels, and improvements in collection of the light generated by antineutrino interactions in water and scintillator detectors. Since they will normally occur within the borders of a country, mid-field monitoring regimes are likely to be cooperative in nature: we therefore recommend policy studies and cost-benefit analyses of cooperative deployments in the context of current or future treaties and agreements.

3) Far-field (10-500 kilometer) monitoring of the presence or absence and operational status of 10 MWt reactors, and placing upper limits on plutonium production in such reactors, would require detectors at the 10,000 ton to 10,000,000 ton (10 Megaton) scale. For cost reasons, these would likely be composed of pure water doped with neutron capture agents. U.S. and international groups have proposed detectors of this kind at the 100,000 ton scale. These proposals are now in the Conceptual Design phase in with funding agencies, with the aim of achieving a variety of fundamental physics goals. We recommend that the technical nonproliferation community actively engage in these experimental and planning efforts. Such participation will help ensure that the best technologies from fundamental science are brought to bear on the nonproliferation problem. Similarly, the policy community as well as scientific and nonproliferation funding agencies, should analyze the consequences of the existence of such detectors, and in particular should consider planning scenarios and co-investment in projects involving joint physics and nonproliferation goals.

4) Detection of nuclear detonation offers the unique possibility of unambiguous remote confirmation of the nuclear nature of the event. Unfortunately, this capability is possibly the most challenging topic discussed in this white paper. As discussed in earlier unclassified reports[5], 10-100 kilometer range of foreseeable detectors for 1 kiloton yield fission explosions appears most suitable for cooperative monitoring of test sites in relatively proscribed circumstances. While there are differences in backgrounds due to the burst-like character of the fission bomb antineutrino pulse, much of the necessary detector development research will be accomplished in the course of R&D recommendations 2) and 3) above. In a policy context, we recommend analysis of the potential impact of various cooperative deployments on CTBT and other future test-ban verification regimes.

Table of Contents

## 2 Introduction

This white paper presents a comprehensive survey of applied antineutrino physics relevant for nonproliferation, summarizes recent advances in the field, describes the overlap of this nascent discipline with other ongoing fundamental and applied antineutrino research, and charts a course for research and development for future applications. It is intended as a resource for policymakers, researchers, and the wider nonproliferation community.

The paper is organized as follows.

In section 3 we give a general overview of the information that antineutrino detectors can provide for reactor monitoring, reactor finding and nuclear explosion detection, and provide illustrative examples of deployments.

In section 4 we describe the physics of antineutrino production and detection relevant for nuclear reactors and nuclear explosions in further detail.

In section 5, we consider 'near-field' (10 meters to 1 kilometer) applications of antineutrino detectors in further detail, with an emphasis on existing demonstrations of cooperative monitoring and safeguards of nuclear reactors. We review current IAEA safeguards practices at different reactors, consider possible benefits of antineutrino detectors in the context of safeguards, and describe the state of the art for antineutrino detection as applied to near-field monitoring. We also present a set of general requirements for near-field deployments, and describe research and development priorities for improved near field monitoring antineutrino detectors.

In section 6, we survey current and potential capabilities for finding operating reactors or excluding their presence in the 'mid-field' (1 to 10 kilometers standoff), and describe research and development priorities for detectors at this standoff distance.

In section 7, we survey capabilities and potential for 'far field' (10- 500 kilometer standoff) reactor finding and explosion detection, and enumerate research and development priorities for these very large scale detectors.

In section 8, we summarize ongoing fundamental physics research that is relevant for applied antineutrino physics.

We conclude with a discussion of the overlapping R&D priorities for applied and fundamental antineutrino physics.

## 3 Overview and Examples of Antineutrino-Based Measurements of Nonproliferation Interest

The possible utility of the antineutrino signal for nonproliferation is easily understood apart from detailed production and detection mechanisms, which are described in section 4. In descending order of accessibility with current technology, the measurements of interest for the applications discussed in this white paper are:

the antineutrino *rate* from a given source, whether a reactor or a fission bomb;

the antineutrino *energy spectrum*; and

the antineutrino *direction*.

*1. Information derived from antineutrino rate measurements.*

The emitted antineutrino rate from reactors depends on the thermal power and fissile isotopic content of the reactor. The antineutrino rate can therefore be used to measure the reactor operational status (off/on) and power continuously and in real time. If the reactor power and initial fuel loading are known by other means, and the antineutrino event rates are sufficiently high (roughly, hundreds or thousands of events per day or week) the antineutrino rate can be used to estimate the evolving amounts of fissile uranium and plutonium in the reactor core. For a given fuel type, the degree of neutron irradiation primarily determines these changing amounts of fissile material, and is referred to as the 'burnup'. The fuel burnup at discharge directly correlates with the amount of plutonium in spent fuel, and is an important parameter in the context of reactor safeguards.

Many experiments worldwide have performed antineutrino rate measurements at reactors[6,7]. These fundamental physics experiments are described in section 8.1. Two experiments, described in section 5.3, have been built specifically to demonstrate reactor monitoring capability in the context of nuclear safeguards.[1,2]

The antineutrino rate from nuclear explosions depends on the fission yield and fissile isotopic content of the explosion. The short burst of antineutrinos emitted by a nuclear weapon can therefore be used to measure the total fission yield of a nuclear explosion, as well is to confirm that the explosion is indeed due to fission. An earlier study[5] has provided a detailed examination of the utility, detector characteristics, and costs for fission explosion detectors.

*2. Information derived from the antineutrino energy spectrum*

Like the antineutrino rate, the antineutrino energy spectrum depends on the reactor power and fissile isotopic content. Estimates of both the reactor fissile isotopic content *and* its thermal power can be derived from spectral measurements, without the need for independent measurement of the reactor thermal power, provided sufficient statistics are available. As with rate measurements, antineutrino spectral measurements have been made in numerous fundamental physics experiments[6,7,8]. Theoretical estimates have also been made of reactor antineutrino spectra for the major fissile isotopes[9,10]. These are accurate to a few percent, and are derived primarily from experimentally measured fission product electron spectra of these isotopes[11,12]. In combination with models of the reactor fuel evolution, such as ORIGEN[13], the antineutrino spectrum of a given reactor can be predicted. Further information on antineutrino spectral analysis for reactor monitoring is provided in section 4.1.5.

3. Information derived from the antineutrino direction

Finally, the antineutrino direction can in principle be used to infer the location of a bomb or reactor. Directionality is difficult to measure, for reasons described in later chapters. However, the Chooz collaboration has successfully reconstructed the direction of antineutrinos from a known reactor source with a pointing accuracy of approximately 20 degrees[14].

These three types of measurements underlie all of the potential uses of antineutrino detectors that are described in the following sections. To further orient the reader to the possibilities, Figure 1 displays a number of possible deployment options, distinguished by range and by overburden. The sidebar presents examples of measurements that could be made currently or in the near term at three distance scales of interest. These ranges scales are:

1) *Near-field* 10 meters – 1 kilometer from a nuclear power or research reactor,

2) *Mid-field* 1-10 kilometers from a nuclear reactor, and

3) *Far-field* 10 km and beyond from a nuclear explosion or a nuclear reactor.

We will return to these examples, with more detail in specific cases, in the application sections of the paper.

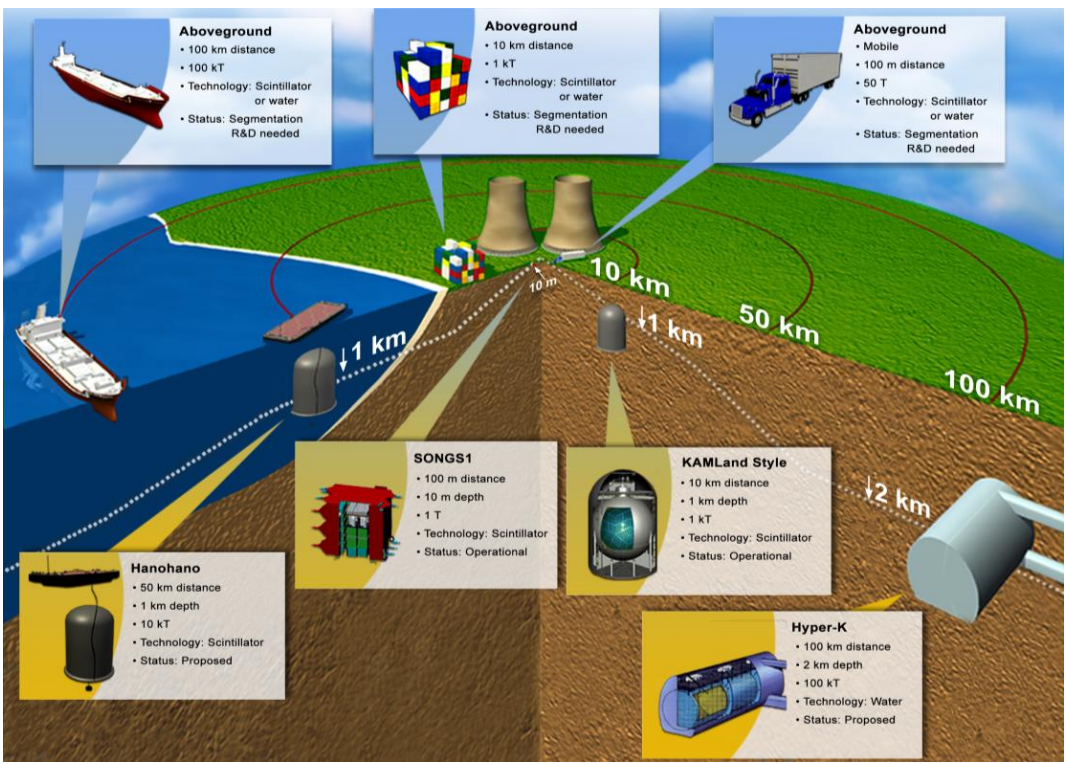

**Figure 1: Represenative deployment scenarios for antineutrino detectors. The 1 ton SONGS1 and 1000 ton KamLand detectors have been deployed and have operated for several years.**

Figure 2, is a summary graph of the size of the detector as a function of distance from a 10 MWt reactors. Because of the highly penetrating power of antineutrinos, shielding and attenuation calculations, such as those required for gamma and neutron detection, are irrelevant for the purposes of calculating signal rates. For a given interaction and detector type, the main variables influencing detector size are the distance from the source, the source strength, (fissile yield or reactor thermal power), and the amount of shielding against ambient backgrounds . (There is also a comparatively small correction factor due to antineutrino oscillations, described later, that depends on the ratio of the antineutrino energy and the reactor standoff distance.)
77

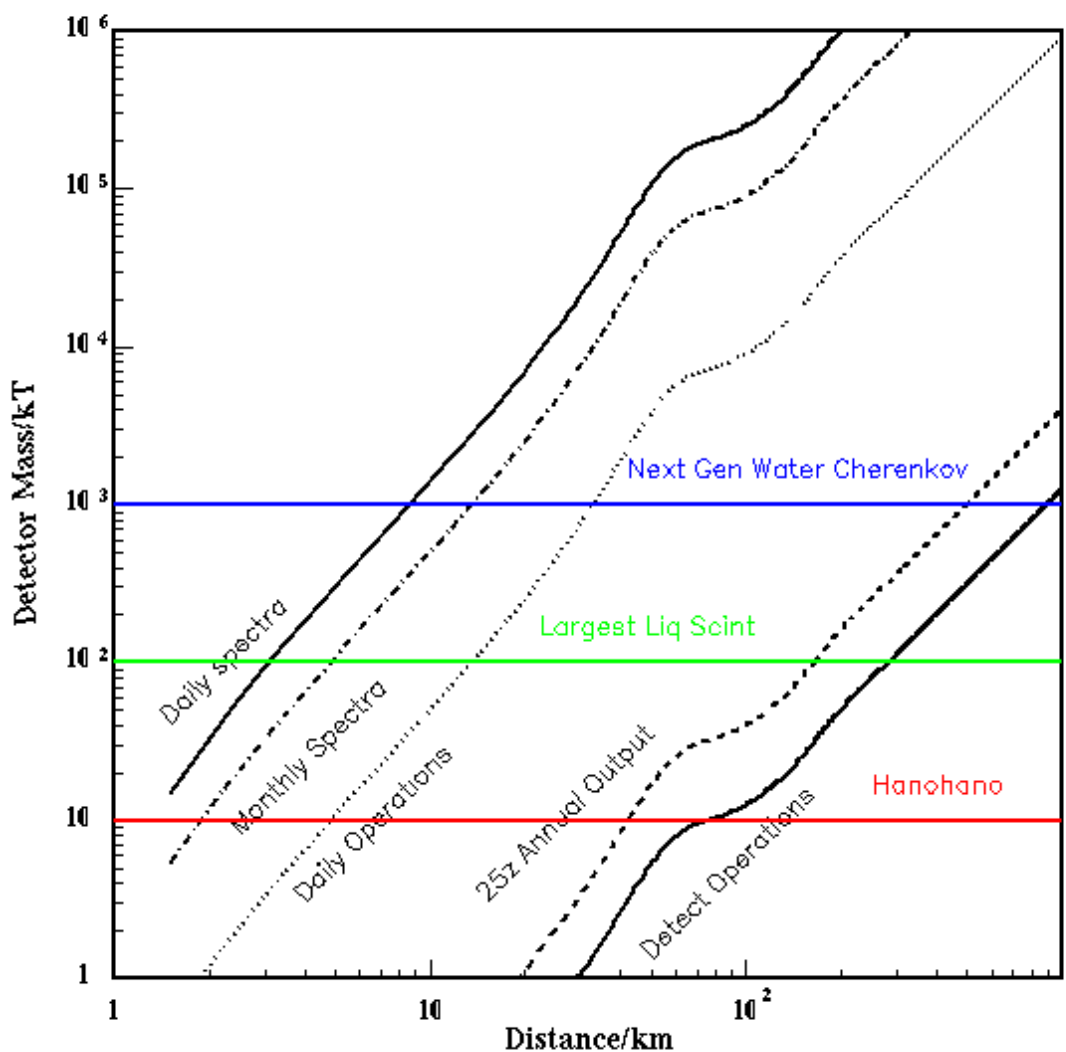

**Figure 2:** A plot of the required detector mass (in kilotons) for a given standoff distance (in km) and application, assuming zero background, and using the most commonly employed antineutrino detector interaction (inverse beta decay). Discovery refers to confirming the presence of a 10 MWt reactor within one year with >5 events. Measuring annual output with 16 measured events in 1 year would provide an estimate of the reactor thermal power with 25% precision. Measuring daily operations would require 3600 events/year, measuring monthly spectra would require 36K events per year, and daily spectra would require about 3000 events per day. Most of the scaling behavior is determined by the inverse square dependence of the antineutrino flux. The kinks in each curve are caused by the known effect of reactor antineutrino oscillations, first measured by the KamLAND experiment[8]. Further information about reactor antineutrino oscillations is found in section 8.1.

## 3.1 Representative Applications of Antineutrino Monitoring

The following table provides representative examples of nonproliferation applications of antineutrino detection in the near, middle and far field.

| |
|---|
| **Near-Field reactor monitoring** : <br> An antineutrino detector deployed at the San Onofre Nuclear Generating Station (SONGS) in Southern California.has been used to non-intrusively monitor the operational status, relative thermal power, and changes in fissile content of a 3.64 GWt reactor. The 640 kg (0.64 ton) detector operated at 24.5 meter standoff. The reactor operational status (on or off) is detected within 5 hours of shut down or start up at the 99% confidence level. The reactor power, relative to a known initial value, is measured with 3% accuracy in one week. With known constant reactor power and fuel loading, changes in fissile content corresponding to consumption of 500 kg of $^{235}$U and production of ~80 kg of Pu are observable in approximately 4 months. Further improvements can be expected with additional effort. Spectral measurements providing a more robust constraint on the reactor evolution are also possible. With approximately the same size detector at 5 meter standoff, similar precision for power and operational status could be obtained for a 100 MWt research reactor. <br><br> The deployment is described further in section 5. |
| **Midfield reactor operation/exclusion/discovery:** The largest currently operational reactor antineutrino detector is known as KamLAND (the Kamioka Large Antineutrino Detector) [8]. KamLAND is a 500 ton fiducial mass$^a$ (1000 ton total mass) liquid scintillator detector, operating at 1000 meters overburden. With this detector and overburden, the operation of a 10 MWt reactor could be excluded in a 10 kilometer radius in 3 months. <br><br> With no other reactors present, the measured antineutrino rate would be approximately 24 events per year for an 85% efficient detector such as KamLAND. With known KamLAND backgrounds of about 1 antineutrino-like event per month, 95% confidence of detection would be obtained in approximately 3 months. If a reactor is discovered, its power could be estimated to within 25% in approximately 8 months. Additional known or unknown reactors would change the background levels and alter the detector size requirements. Midfield applications are considered in section 6. |
| **Far-Field reactor exclusion/discovery:**: The operation of a 10 MWt reactor could be excluded within an 800 kilometer radius with a 1,000,000 ton (1 Megaton, fiducial mass) water Cerenkov antineutrino detector. Detectors of this scale are now being proposed by various physics collaborations in the United States and abroad. <br><br> The measured antineutrino rate would be approximately 5 events per year for a 85% efficient detector. The rate includes the effect of neutrino oscillations, described below, which must be taken into account at this distance. A key R&D consideration is that the dominant cosmic-ray backgrounds in the detector would have to be suppressed by a factor of about 30 - <u>per unit of detector mass</u> - relative to the KamLAND detector. 95% confidence of detection would be obtained in approximately 1 year. Again, additional known or unknown reactors would change the background levels and alter the detector size requirements. Far-field reactor applications and background considerations are examined in section 7. |
| **Far-field nuclear explosion detection**: the fissile nature and approximate yield of a 10 kton explosion at 250 kilometer standoff could be confirmed with a 1,000,000 ton water Cerenkov detector. An 85% efficient detector |

---

$^a$ The fiducial mass is defined as the fraction of the total detector mass within which valid antineutrino events are recorded. In the present context, the outermost thickness of the total mass of the detector is used as a shield against external backgrounds to create a central fiducial region.

would record 2-3 events in a few seconds from such an explosion. Oscillation effects are included in this rate estimate. Far-field nuclear explosion detection is considered in Section 7.5.

# 4 Production and Detection of Antineutrinos From Nuclear Reactors and Nuclear Explosi

## 4.1 Brief Description of Reactor Monitoring with Antineutrinos

In this section we provide a summary description of the behavior of the antineutrino signal emitted by the most common type of nuclear reactor, a Light Water Reactor (LWR) fueled with Low Enriched Uranium (LEU). This signal is modified significantly for different reactor types, such as fast fission reactors, and these modifications are considered in sections 4.1.5 and 5.4.

As an LWR proceeds through its irradiation cycle, the inventory of each fissioning isotope varies in time. Neutrons cause fission of uranium and plutonium, while the competing process of neutron capture on $^{238}$U produces plutonium. As a consequence, the relative fission rates of the isotopes vary significantly throughout the reactor cycle, even when constant power is maintained. This phenomenon is shown in Figure 3, a plot of fractional fission rates for a typical cycle of a Light Water Reactor (LWR), operating at constant power over 600 days. $^{235}$U and $^{239}$Pu contribute the most fissions: although it accounts for most of the mass of the reactor core, $^{238}$U can be fissioned only with MeV-scale or "fast" neutrons and contributes only about 10% of the total fissions in these reactors, in which the neutron population is mostly sub-eV scale or "thermal".

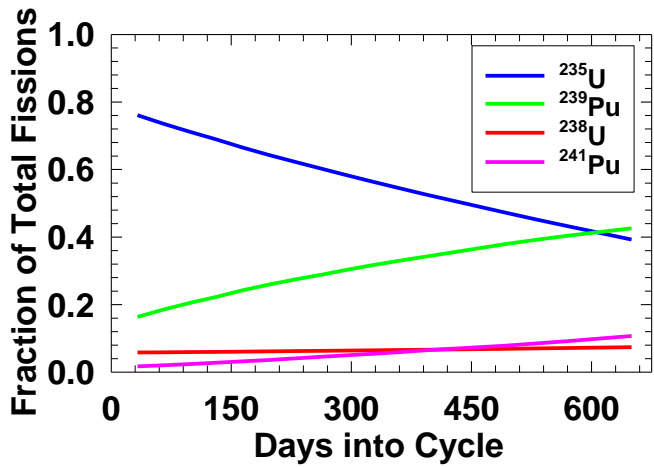

**Figure 3: The fractional fission rates of the main fissile isotopes in a Light Water Reactor plotted versus cycle day.**

This ongoing change in fissile content induces a systematic shift in the detected antineutrino flux and energy spectrum over the course of the cycle, known as the "burnup effect". Burnup, in units of Gigawatt-days per ton of heavy metal (GWd/THM), measures the integrated thermal power output of the fuel per unit of fuel mass. It effectively tracks total neutron exposure of the fuel, and therefore uranium consumption and plutonium production in the core.

The burnup effect has been measured in past reactor antineutrino experiments[1], and is an important consideration for near-field reactor monitoring applications. While the total antineutrino emission rate per fission is about the same for the main fissile isotopes, the emission rate per fission varies significantly among these isotopes in the energy regime (1.8-10 MeV) where reactor antineutrinos are most readily detected. Because of the measurable rate differences in detected number of events from the main fissile isotopes, the evolving state of the core is reflected in both the measured antineutrino rate and energy spectrum.

The size of this effect depends on the core type, fuel management strategy, and fuel composition. For example, in a standard 1.5 year cycle of operation of a Pressurized Water Reactor (PWR), the net decrease in $^{235}$U content and increase in fissile Pu content are roughly 1500 kg and 250 kg respectively.. This gradual consumption and ingrowth causes a gradual reduction in the antineutrino rate of about 12% over the same time period.

Figure 4 shows the size of the burnup effect relative to an initial value, as simulated for a LEU-fueled PWR. In Sections 5 and 8, we show that this predicted effect has been clearly confirmed by several reactor antineutrino experiments, can be seen even with detectors of simple design, and can be used to estimate the fissile content of the reactor throughout the cycle, including the burnup of discharged fuel.

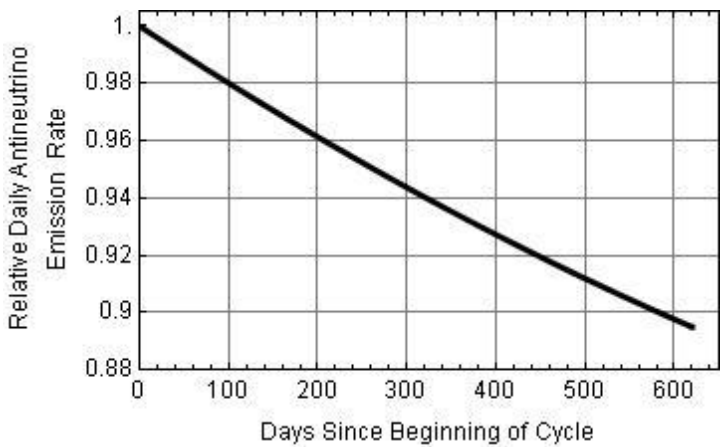

**Figure 4:** The change in the daily emitted antineutrino rate across a typical LEU fueled PWR cycle. The rate is normalized to 1 at the beginning of cycle, and the reactor operates at constant power throughout the cycle. In this example, the antineutrino rate changes by approximately 12% over the course of a 600 day cycle.

Summarizing points to be explained in more detail later, an estimate of the operational state of the reactor, the reactor power, and the reactor plutonium inventory can be obtained from either of two quantities measured by an antineutrino detector: the total antineutrino rate integrated over a broad energy window, as just described, or the energy spectrum of the detected antineutrinos. To further explain the connection between these measured quantities and nonproliferation metrics of interest, we describe reactor antineutrino production and detection processes in more detail.

### 4.1.1 Production of Antineutrinos in Reactors

Antineutrino emission in nuclear reactors arises from the β-decay of neutron-rich fragments produced in heavy element fissions. The average fission is followed by the production of about six antineutrinos, which corresponds to the average number of β-decays required for the fission daughters to reach stability. For a typical power reactor, the thermal power output is about 3 GWT, and the energy release per fission is about 200 MeV. For such reactors, therefore, the number of antineutrinos emitted from the core is approximately $10^{21}$ per second. These emerge from the core isotropically and without attenuation.

The antineutrino energy distribution contains spectral contributions from the dozens of beta-decaying fission daughters. Precise estimates of the distribution have been derived from beta spectrometry measurements[9,10,11,12], and validated by many reactor experiments[14,15,16]. An approximate formula for the antineutrino energy density per fission is

$$\frac{dN}{dE_{\bar{\nu}}} = \exp-(a + bE_{\bar{\nu}} + cE_{\bar{\nu}}^2),$$

where $E_{\bar{\nu}}$ is the energy of the antineutrino in MeV, and the coefficients a,b, and c are specific to each fissile isotope. The mean energy of the emitted antineutrinos is similar for all fissile isotopes, approximately 1.5 MeV.

### 4.1.2 Reactor Antineutrino Interaction Mechanisms

#### 4.1.2.1 Inverse Beta Decay

The most commonly used and practical method for detecting reactor antineutrinos is the well understood inverse beta decay interaction:

1) $\bar{\nu}_e + p \rightarrow e^+ + n$.

Here the antineutrino ($\bar{\nu}$) interacts with free protons (p) present in the detection medium. The numerical value of the cross section per target proton for this interaction is of order $10^{-43}$ cm$^2$, which is large in comparison to most other possible antineutrino interaction mechanisms, and which enables cubic meter scale detectors at tens of meter standoff from standard power reactors. The neutron (n) and positron (e+) are detected in close time coincidence, providing a dual-event signature that stands out strongly against backgrounds. In addition to the antineutrino flux, the antineutrino energy is accessible through the measured positron energy. The quantities are related by the formula:

$$2 \quad E_{\bar{\nu}} = E_e - M_p + M_n + m_e + O(\frac{m_e}{M_p})$$

where $E_{\bar{\nu}}$ is the antineutrino energy, $E_e$ is the positron kinetic energy, $M_p, M_n$ and $m_e$ are the proton, neutron and electron masses respectively, and $O(\frac{m_e}{M_p})$ are terms of order $\frac{m_e}{M_p}$ that mainly account for the nuclear recoil. The reaction in (1) has an energy-dependent cross-section and a threshold of ~1.81 MeV.

*4.1.2.2 Other Interaction Mechanisms*

Radiochemical methods, antineutrino-electron scattering and coherent antineutrino-nucleus scattering mechanisms are all possible methods for detecting reactor antineutrinos.

Radiochemical methods refers to a family of possible inverse beta decay interactions with various nuclei, following the generic process

3) $\bar{\nu}_e + (A,Z) + Q_\beta \to e^+ + (A,Z-1)$

Where $Q_\beta$ is the energy threshold for the interaction, and A and Z are respectively the mass number and proton number of the target nucleus. The interaction probabilities decrease with increasing Z, and energy thresholds vary with isotope[17]. Accounting for these effects, only a few targets have rates comparable with the inverse beta decay process on protons, most notably the transition which has a very low (0.018 MeV) energy threshold. Generally, these approaches are undesirable for detection on time scales of days to months, either because the materials are inconvenient or expensive is in the case of *He*, or because they involve radiochemical recovery of converted isotopes following long exposure times. Concepts of operation involving long acquisition times might still employ detectors of this kind.

$^3He \to {}^3H$,

Antineutrino-electron scattering involves the exchange of a neutral $Z_0$ gauge boson with an electron in a convenient target medium, with water being a common choice. It has an approximately 100–fold lower interaction probability per target than inverse beta decay, requiring correspondingly larger detector sizes, and is therefore not as readily applicable for security applications. A further limitation of this 'neutral current' interaction is that it is equally sensitive to antineutrinos and neutrinos, and so that solar neutrinos are a background process, unlike the inverse beta decay process on protons, which is sensitive to antineutrinos only.

Coherent antineutrino-nucleus scattering is collective neutral current interaction that occurs via coherent summing of $Z_0$ exchange amplitudes between the antineutrino and all the nucleons in any nucleus[18]. The coherent addition of nucleon amplitudes makes the interaction probability significantly *higher* than inverse beta decay on protons, by factors of 10-100. While theoretically very well motivated, the very low energy transfer in the process makes it extremely difficult to detect – indeed no experiment has yet succeeded in measuring coherent neutrino scattering. This high-rate but difficult to detect interaction may nonetheless be of interest in some practical applications, and is discussed in more detail in section 8.4. As with antineutrino-electron scattering, it suffers from an irreducible solar neutrino background. which for this interaction limits the useful standoff distance from reactors to a few kilometers, beyond which distance solar and reactor fluxes are comparable.

### 4.1.3 Reactor Antineutrino Detection and Background Rejection

Historically, organic liquid scintillator has been the most common choice of detection medium for reactor antineutrinos. It can be obtained in large quantities at low cost, has a high density of free proton targets to enable reaction (1), and it can be doped with different neutron capture elements to enhance sensitivity to the neutron in the final state of the antineutrino interaction.

Detection with liquid scintillator has been standard in nuclear physics ever since the early experiments leading to the discovery of the antineutrino. Modern liquid scintillator detectors, such as Rovno[2], SONGS1[19], Chooz[7] and Palo Verde[6], have fiducial masses of several tons and have run for a few years each, with total detector-related systematic errors on the absolute antineutrino count rate as low as 3%. These detectors have very good time stability, compatibility with plastic hardware and relatively modest health hazards.

In the process (1), both the positron and the neutron are detected in close time coincidence compared to other backgrounds. The positron and its annihilation gamma-rays produce scintillation signals within a few nanoseconds of the antineutrino interaction: this is collectively referred to as the prompt scintillation signal. This is followed by a second, delayed scintillation flash arising from the cascade of gamma-rays which come from decay of the excited nuclear state of the neutron-absorbing element following neutron capture. The scintillation light is recorded in photomultiplier tubes, with the number of scintillation photons proportional to the deposited positron and neutron-related energies, and the time of each deposition recorded. This time correlated pair of MeV-scale energy depositions, with the prompt and delayed signals separated by only a few tens or hundreds of microseconds, stands out strongly against backgrounds.

The neutron capture can occur on hydrogen, for which a 2.2 MeV gamma-ray is released. Often, a dopant with a high neutron-absorption cross-section is used, to improve the robustness of the signal. This brings the dual benefits of reducing the capture time and increasing the energy released in the gamma cascade following capture, compared to 2.2 MeV for hydrogen. For example, a 0.1% concentration of gadolinium, which has the highest neutron absorption cross-section of any element, reduces the capture time from about 200 microseconds to tens of microseconds relative to hydrogen, and increases to 8 MeV the neutron-related energy available for deposition in the detector (arising from neutron capture de-excitation gamma-rays).

For the relatively small detectors needed for near-field reactor safeguards it is also possible to use non-hazardous liquid scintillators, blocks of solid plastic scintillator coated with neutron capture agents, doped water Cerenkov detectors, and other approaches, all of which can improve deployability and ease of use of the system. These alternatives are also discussed in section 5.5. For mid-field detection, strategies such as segmentation and improved particle identification may be of use for the more stringent background rejection requirements imposed by the reduced flux available beyond 1 km. These alternatives are also discussed in section 6. For far-field detection of reactors (or nuclear explosions), only large homogeneous liquid scintillator or doped water Cerenkov detectors appear practical, as is discussed in section 7.

Aside from the neutrino oscillation effect that appear at very long ranges, the signal event rate falls off simply as the squared distance to the reactor or reactors, with no other attenuation effects even at Earth diameter standoff. However, as discussed later, relatively small systematic corrections to this oscillation, at the few percent level, may be revealed at the few kilometer standoff by a next generation of neutrino oscillation experiments.

Backgrounds depend in a more complicated way on overburden and nearby materials. Backgrounds are usually separated into 'correlated' and 'uncorrelated' types. Correlated backgrounds are those for which a single physical process is responsible for both the apparent positron and neutron signals, such as muon induced $^9$Li, or multiple neutron scattering. Uncorrelated backgrounds arise from two independent physical processes, such as two random gamma-ray interactions occurring within the time window that defines the antineutrino signal.

The cosmic ray muon flux, which is responsible for much of the correlated background, falls off exponentially with overburden. Overburden is usually expressed in meters of water equivalent (mwe). At distances relevant for near-field cooperative monitoring, out to approximately 1 km, overburdens ranging from 10-300 mwe have been shown to give sufficiently good signal to background ratios for precision measurements of the antineutrino flux at levels relevant for reactor monitoring applications. The KamLAND detector, currently the sole example of a dedicated far-field reactor antineutrino detector, with sensitivity to reactors at 200 km standoff and beyond, has an overburden of about 2700 mwe (or 1 kilometer of earth).

For both near and far field detection, an active muon 'veto' system is normally used to time-stamp the passing of muons through or near the main detector, allowing further rejection of muon-related backgrounds. These systems may consist for example of plastic or liquid scintillator read by photomultiplier tubes – several other options have also been used.

Various combinations of lead and/or steel for gamma-ray attenuation, and polyethylene, water and other materials for neutron attenuation are used as passive shields to reduce the uncorrelated backgrounds, which arise from local ambient neutrons and gamma-rays. The thickness of these shields are of the order 0.5-2 meters, depending on the signal strength and target signal to background ratio for the particular experiment. The practical considerations related to shielding and overburden in the context of cooperative monitoring, are described further in section 5.5, Since only a few experiments have been built specifically for cooperative monitoring, further optimization of shielding and overburden in the context of cooperative monitoring experiments is likely possible.

### 4.1.4 Rate Measurements and the Burnup Equation

Having discussed the production and interaction mechanisms, we turn to the detailed features of the antineutrino signal as measured through the inverse beta decay process. $N_{\bar{\nu}}$, the total rate of detected antineutrinos, is directly related to a sum over the fission rates of each fissile isotope. Following Klimov[2], this explicit dependence of the antineutrino rate on the individual isotopic fission rates can be factored as

$$4) \quad N_{\bar{\nu}} = \frac{N_p W \varepsilon}{4\pi R^2} \times \frac{\sigma_5 \left(1 + \sum_i \alpha_i(t)(\frac{\sigma_i}{\sigma_5} - 1)\right)}{E_5 \left(1 + \sum_i \alpha_i(t)(\frac{E_i}{E_5} - 1)\right)} \times P_{ee}(R, E_{\bar{\nu}})$$

For all parameters, the index $i=5,8,9,1$ extends over the four main fissioning isotopes $^{235}$U, $^{238}$U, $^{239}$Pu and $^{241}$Pu. The factorization allows for a simplified parametrization of the equation, to be described below, and emphasizes the dominant contribution from $^{235}$U ($i=5$) over most of the reactor cycle.

$N_p$ is the number of target protons, $W$ is the thermal power, $R$ the standoff distance and $\varepsilon$ is the detection efficiency. $\sigma_i$ is the cross section for the inverse beta interaction, averaged over the emitted antineutrino energy distribution for the isotope $i$:

$$\sigma_i = \int dE_{\bar{\nu}}^i \cdot \sigma_{\bar{\nu}p} \frac{dN_{\bar{\nu}}^i}{dE_{\bar{\nu}}^i}.$$

In the above, $\sigma_{\bar{\nu}p}$ is the inverse beta decay cross section, $\frac{dN_{\bar{\nu}}^i}{dE_{\bar{\nu}}^i}$ is the number of antineutrinos emitted by isotope $i$ in the energy range $dE_{\bar{\nu}}^i$, and the integral extends from the inverse beta decay reaction threshold of 1.8 MeV up to the maximum antineutrino energy of roughly 10 MeV.

$E_i$ is the energy release per fission. $t$ is time at which the rate is calculated, and $\alpha_i(t)$ is a dimensionless time dependent parameter, defined as the fraction of the total fission rate arising from isotope $i$, with the normalization

$$\sum_i \alpha_i = 1.$$

Throughout the cycle, the fractional fission rates $\alpha_i(t)$ change as $^{235}$U is consumed and $^{239}$Pu is produced. This introduces a time dependence to the antineutrino rate, as discussed below. Based on an ORIGEN simulation of a PWR, representative values of $\alpha_i$ at beginning and end of cycle are shown in Table 1 Also shown in the table are the fixed values of $\sigma_i$ and $E_i$.

| Isotope | Energy averaged cross-section per fission ($\sigma_i$, $10^{-43}$ cm$^2$) | Energy release per fission ($E_i$, MeV) | fission fractions ($\alpha_i$) beginning of cycle | fission fractions ($\alpha_i$), end of cycle |
|---|---|---|---|---|
| $^{235}$U | 6.38 | 201.7±0.6 | 0.763 | 0.423 |
| $^{239}$Pu | 4.18 | 210.0±0.9 | 0.162 | 0.397 |
| $^{238}$U | 8.89 | 205.0±0.9 | 0.0476 | 0.076 |
| $^{241}$Pu | 5.76 | 212.4±1.0 | 0.027 | 0.102 |

Table 1: The isotopic parameters used to calculate the antineutrino rate throughout the cycle. The fission fractions are derived from an ORIGEN simulation of a representative commercial PWR. The values for the energy release from fission and energy averaged cross sections are from references [20] and [2] respectively.

The final term, $P_{ee}(R, E_{\bar{\nu}})$ depends on the antineutrino energy and the standoff distance, and accounts for neutrino oscillations. Its value is unity for reactor antineutrinos for standoff distances of 500 m or less, but may be reduced by as much as several percent at 1 km, and is known to be approximately 60% at 200 km. Figure 5, produced by the KamLAND collaboration[21], shows the effect of the change in $P_{ee}$ on the measured antineutrino rate as a function of the standoff distance R, for a large set of reactor antineutrino experiments. Past and ongoing neutrino oscillation experiments, which have measured or will measure this correction at this distance and greater distances, are discussed further in section 8.1.

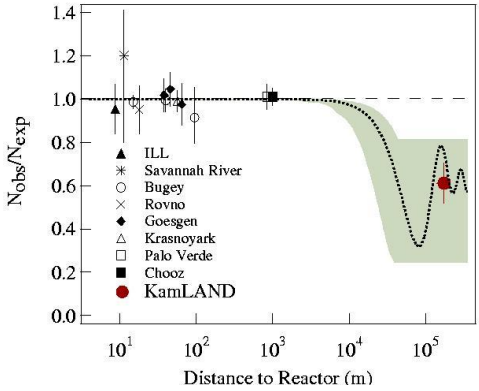

Figure 5: The ratio of observed to expected antineutrino events as a function of standoff from a reactor. This ratio is a measure of the strength of the oscillation term $P_{ee}(R, E_{\bar{\nu}})$ in Equation 4. Plot courtesy of the KamLAND collaboration[22].

Equation 4 can be simplified as:

5) $\quad N_{\bar{\nu}} = \gamma(1 + k(t))W(t)$.

$\gamma = \dfrac{\varepsilon N_p \sigma_5}{4\pi R^2 E_5}$ absorbs the constant parameters defined for Equation 4.

$$k(t) = \frac{\sum_{i=5,8,9,1} \alpha_i (\frac{\sigma_i}{\sigma_5} - \frac{E_i}{E_5})}{\sum_{i=5,8,9,1} \alpha_i (\frac{E_i}{E_5})}$$

is a ratio of sums over the time-varying relative fission rates of the individual isotopes. $W(t)$ is thermal power of the reactor with a possible time dependence now explicitly shown.

Equation 5 reveals the time/burnup dependent nature of the antineutrino rate. Using the parameters from Table 1, the equation shows that even for a reactor operating at constant thermal power, the parameter k(t) changes by about 10-12 percent throughout a typical modern LWR reactor cycle, as $^{235}$U is consumed and $^{239}$Pu is produced and consumed in the core. The result is specific to an LEU-fueled LWR. Other reactor types, such as Canadian Deuterium Uranium (CANDU) reactor, fast breeder reactors, or MOX-fueled reactors, will exhibit different variations in the antineutrino rate over the course of the fuel cycle.

In a 'typical' reactor antineutrino experiment, Equation 5 is used to predict the antineutrino rate, based on thermal power measurements provided by the reactor operator, and on a detailed simulation of the fissile isotopic content of the core as it changes over the course of the reactor fuel cycle. Proprietary codes provided by reactor operators, or publicly available simulation tools such as the ORIGEN/SCALE package[23] are used for this purpose. Predicted and measured rates are then compared to look for the anomalous behavior characteristic of antineutrino oscillations.

For reactor monitoring applications, the procedure used in the oscillation experiments is inverted: the reactor's fissile content is estimated or constrained using the measured antineutrino rate $N_{\bar{\nu}}$. If only rate information is available, additional inputs are required to extract such an estimate: the fuel geometry, the initial fuel enrichment and isotopics, the reactor thermal power $W$, the detection efficiency, and the predicted fissile isotopic evolution must all be known. Each of these inputs has associated uncertainties which limit the overall precision with which the fissile isotopic content can be estimated in an absolute sense. However, for safeguards purposes, a measurement made relative to a known startup fuel content may suffice, in which case many of these uncertainties are reduced or eliminated. This point is discussed further in 5.3, where experimental results from demonstration safeguards reactors are discussed in more detail.

Historically, the total integrated absolute flux of antineutrinos emitted by a thermal reactor has been shown to both predictable and measurable to an uncertainty of about 2%[26]. Recent work indicates that the uncertainty in the predicted flux can be further reduced to below 1%, based on improvements in the thermal power measurement methods available at some reactors, as well as a more comprehensive validation of the precision of the burnup simulation codes[24].

Proprietary simulation tools, which most accurately represent the isotopic evolution of the core, may in principle be available for safeguards inspectors, but their use represents an additional practical obstacle in a monitoring context. Members of the KamLAND collaboration has demonstrated that the evolution in the antineutrino rate for Japanese PWRs and BWRs can be described with simple four-parameter equations[25], one per reactor type, which reproduce the results of full reactor core evolution simulations to within 1%. This implies that detailed simulations of the fuel burnup may not be needed at each reactor in order to predict the antineutrino rate (or spectrum). Assuming this approach is extensible to other reactors, as is likely, it considerably simplifies the analysis of antineutrino data in a safeguards context.

*4.1.4.1 Statistical considerations for rate-based measurements of operational status, power and burnup*

The required event rates for rate-based measurements can be approximately estimated according to application. For simple determination of operational status on a daily or weekly basis, the change in the number of true antineutrino events from the reactor must be several standard deviations above background fluctuations within the desired period. This is easily achievable at ~10-100 meter standoff and modest overburden (10-20 mwe), in which circumstance detectable antineutrino rates are in the 100-1000 event per day per ton range, and ratios of signal to square root of background are ~10-50. This is shown by the prototype detectors described in 5.3. For near-field daily or weekly power or burnup monitoring in LWRs, relative to an initial value, enough data must be collected so that statistical uncertainty does not dominate the approximately 0.5% change per month in the fission rate induced by the burnup effect. This implies a data collection requirement of about 10000 events per month, or ~300 events per day. This also gives 1% precision on a relative thermal power estimate within a month, dominated by statistics. For absolute measurements of burnup, without reliance of operator declarations, the full antineutrino energy spectrum must be acquired, and the statistical requirements are therefore more stringent. This case is discussed in the following section. For longer term monitoring, current detectors can also achieve the required signal to background ratios in the mid-field, where rates are at the level of events per day or month in 100-1000 ton detectors, and backgrounds have been demonstrated to be suppressed to a factor of several below these values (see section 6).

### 4.1.5 Spectral Measurements

As long as flux alone is used to estimate plutonium inventory, both relative and absolute measurements would still require independent knowledge of the reactor thermal power, since it serves as a free parameter that could be used to tune the antineutrino rate and mask the burnup effect. By using the full antineutrino energy spectrum, fissile inventories can be independently confirmed, with little or no input from the reactor operator. This is a considerable advantage in a safeguards context compared to flux based measurements. However, measurement of a spectrum requires a more sophisticated detector and analysis than a simple counting detector. The trade-offs inherent in this approach are considered further in section 5.5.

The plot on the left in Figure 6 shows the representative energy spectra per fission of antineutrinos for the two most important fissile elements $^{239}$Pu and $^{235}$U. The rightmost plot shows same spectra after folding with the energy-dependent detection inverse beta cross section of Equation 1. The differences in both the emitted and detected spectra are apparent, amounting to 50% or more, especially at higher energies. While the mean emitted energy is about 1.5 MeV, the mean detected energy is higher, roughly 4 MeV, enhanced by the quadratic energy dependence of the inverse beta cross section.

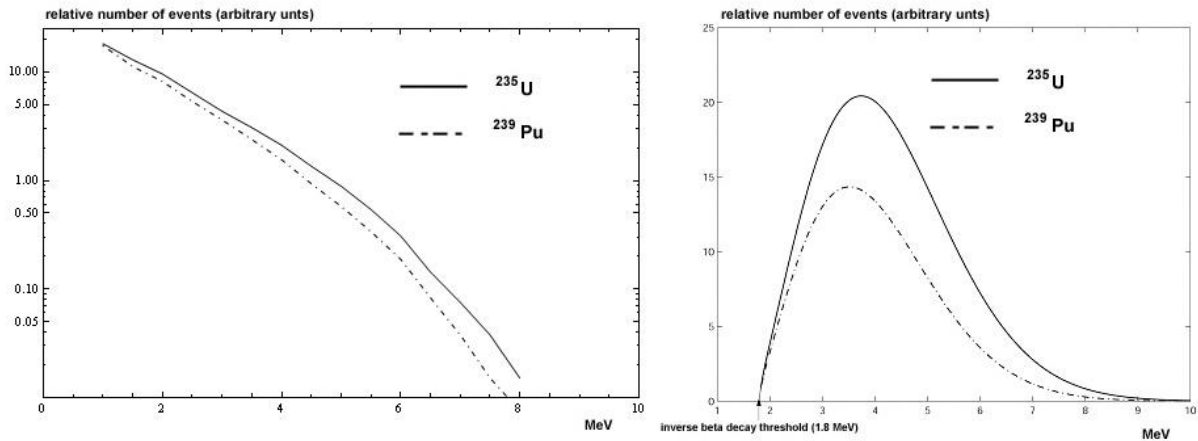

**Figure 6: Left: The emitted antineutrino spectra from $^{235}$U and $^{239}$Pu, as emitted by a LEU-fueled Pressurized Water Reactor. Right: the same spectra after convolution with the inverse beta decay cross section. Above approximately 3 MeV, the two spectra differ by 50% or greater. Only the sum spectrum is measured in an actual detector.**

Experimentally measured antineutrino spectra have absolute accuracies of about 2%[26], with errors dominated by uncertainties in the predicted reactor emission spectrum.

The relative contributions to fission of the different isotopes evolve over the course of the reactor cycle. This causes a change in the measured spectrum throughout the reactor cycle. A measurement of the energy spectrum is therefore sensitive to the evolving fissile isotopic composition of the core. Figure 7 shows the difference in the spectrum as simulated at the beginning and end of a representative 600 day LEU-fueled equilibrium LWR cycle, and the fractional change in the spectrum over this same cycle as a function of energy.

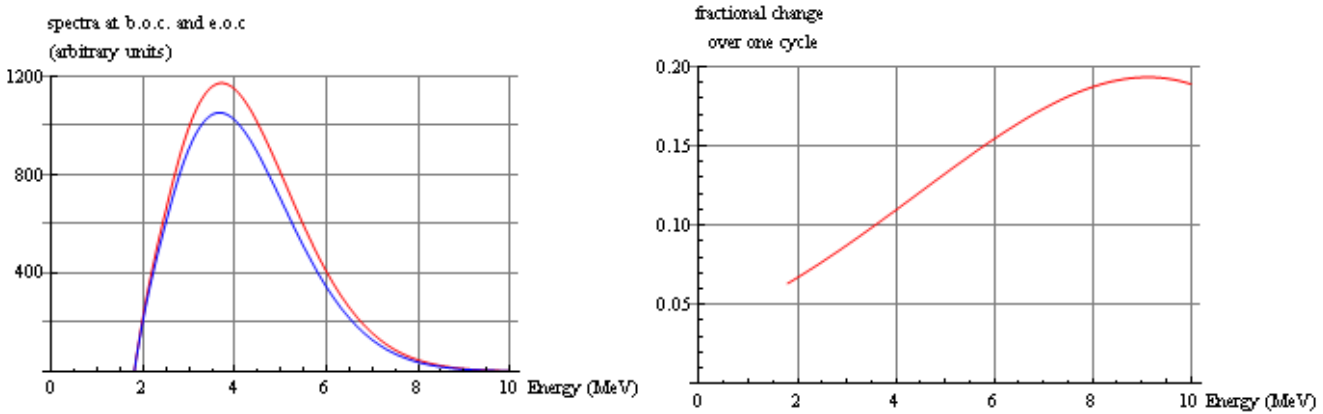

**Figure 7: Left: the antineutrino energy spectrum from a representative LEU-fueled LWR, at beginning of cycle (b.o.c, shown in blue) and end of cycle (e.o.c., shown in red). Right: The fractional change in antineutrino rate from beginning of cycle to end of cycle, plotted versus energy.**

The bin-to-bin differences between the beginning and end of cycle spectra range from 6-20%, with the most pronounced difference being at higher energies. Because these differences are larger than the known 2% uncertainties in the predicted spectra, they can be – and have been - experimentally observed with energy-resolving antineutrino detectors[16].

Plutonium and uranium isotopic content and power can be derived from a fit to an integral antineutrino energy spectrum, taking the unknown fissions fractions $\alpha_i$ and the thermal power $W$ as free parameters. Huber et. al.[27] have shown that the Pu content in an equilibrium cycle of an LWR can be estimated to within 10%, or about 40 kg of Pu ($^{239}$Pu+$^{241}$Pu), by measuring the antineutrino spectrum with $10^6$ events, assuming current 2% uncertainties in the antineutrino energy spectra, and a 0.6% uncertainty in detector response. A three-fold reduction in the spectral uncertainties would lead to a 20 kg estimate of total fissile Pu content. This method requires no prior assumptions about reactor power or the fission fractions. A further refinement is to constrain the fit by a direct comparison throughout the cycle with the expected spectra based on a detailed simulation of the reactor core evolution.

Improvements on this estimate of the plutonium inventory can be expected for two reasons. First, next-generation neutrino oscillation experiments such as Double Chooz, described in 8.1.1, will directly measure the antineutrino spectrum from a PWR, with an expected factor of two improvement in precision. Improved beta spectrometry measurements for fission daughters will also help improve the precision of the predicted antineutrino spectra. Perhaps more importantly, as already discussed in the context of rate measurements, safeguards regimes do not necessarily require an *absolute* estimate of the fissile inventory: instead a measurement may suffice of changes *relative* to a known inventory at beginning of cycle. For example, power and burnup information can be derived from declarations or other measurements, or an initial single cycle calibration period can be used to empirically correlate a known and measured plutonium inventory with the antineutrino spectrum (or rate) for the particular reactor in question in later cycles. Cycle-to-cycle consistency of this measurement would demonstrate that the reactor is operating according to declarations, without being affected by many of the systematic uncertainties in the emitted antineutrino flux. Such an approach would also eliminate many systematic effects imposed by the antineutrino detector. This relative analysis has been explored by the Rovno and SONGS1 experiment, and is discussed below in section 5.3. The approach is analogous to that adopted in some modern neutrino oscillation experiments such as Double Chooz[28], wherein the ratio is formed of the antineutrino spectra and rates as measured in two identical detectors at different standoff, in order to isolate the oscillation phenomenon and remove many reactor and detector related systematic effects. The difference in the current case is that this ratio is being formed using the same detector at different times during the cycle, rather than using two different detectors. For such relative measurements, the time stability of the detector response is therefore a key experimental criterion.

### *4.1.5.1 Statistical considerations for absolute spectral measurements of fissile content and power*

Summarizing results discussed above and derived in reference [27], about 100,000 events acquired over a several month period would be required in to achieve a 3% accurate absolute power measurement using a spectral analysis, assuming current spectral uncertainties. Approximately $10^{\wedge 6}$ events in a several month period would allow extraction of an estimate of the mass of the main fissile isotopes at an accuracy of 10% with existing spectral uncertainties, or 1-2% accuracy if these uncertainties can be reduced by a factor of 3-10.

### 4.1.6 Directionality

Directional information about antineutrinos is difficult to obtain. Using the inverse beta decay interaction, the Chooz collaboration has successfully reconstructed the direction of antineutrinos from a known reactor source with a pointing accuracy of approximately 20 degrees[14]. However, because the stochastic relation between the antineutrino direction and the measured quantities in this detector, thousands of events were required to achieve this pointing accuracy. This number of events is incompatible with the low event rates required for discovery of small reactors in mid-field and far-field applications. (1-500 km standoff distances). New methods of event-by-event reconstruction of the antineutrino direction are a key area of interest for long range applications.

Directional information can also be extracted from detectors that rely on Cerenkov light detection, such as the Super-Kamiokande detector[29]. The incident antineutrino direction is recovered by determining the apex and central angle of the reconstructed Cerenkov light cone. However, such detectors rely on antineutrino-electron scattering, which, as already discussed, is less suitable for security applications because of the smaller cross-section. This method also requires high statistics in order to reconstruct the source direction in an average sense.

## 4.2 Production of Antineutrinos in Fission Explosions

The burst of antineutrinos arising from a fission explosion arises from the same source as reactor antineutrinos: the chain of approximately six beta-decays that follow each fission. An unclassified picture of the antineutrino burst can be created using a few simple assumptions. First, unlike steady state reactors, all the fissions in the explosion are taken to occur within 1-10 microseconds. This must be true since criticality is maintained over time scales only of this order. For the same reason, the fissioning neutrons physically cannot have thermalized prior to the rapid disassembly (read explosion) of the weapon. Therefore, the population of fission daughters, and the antineutrino energies are both characteristic of those produced by fast neutron fission. These two facts suffice to crudely define the important features of the burst: its number, energy and time distributions.

The total number of antineutrinos in the burst is directly proportional to the explosive yield.

6) $\quad N_{\bar{\nu}} \cong 6 \cdot N_{fiss} \cdot Y$

with

$Y(kt) \qquad \rightarrow$ yield

$N_{fiss} = 1.45 \times 10^{23} \qquad \rightarrow$ number of fissions/kt

The antineutrino energy distribution is created by a fast neutron fission spectrum. Such a spectrum has been calculated theoretically by Vogel[10], (though not yet measured at a reactor or in an explosion). Figure 8 shows the energy distribution of the antineutrinos emitted by a fission explosion, which in this simple approximation is essentially that expected from a fast reactor.

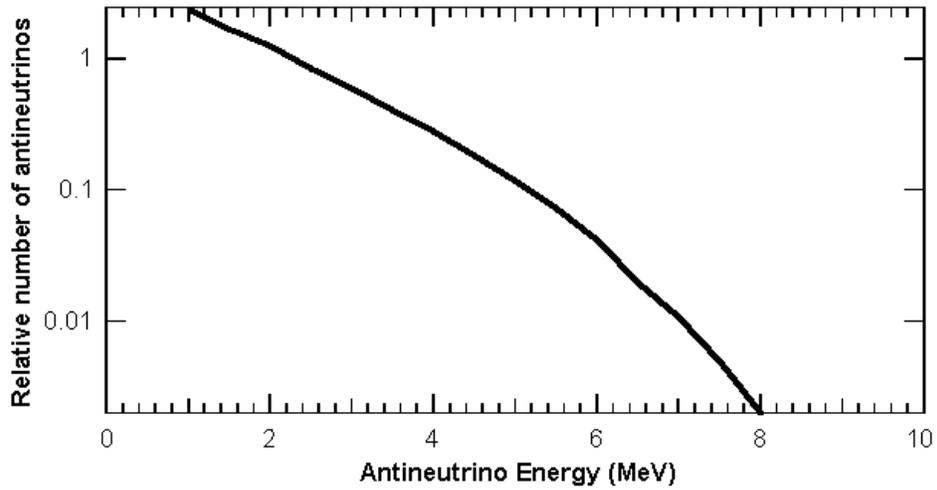

**Figure 8**: The approximate energy spectrum of antineutrinos produced in a U-235 fission explosion.

The time distribution of the antineutrino burst is set by the half-lives of the fission products. These peak at relatively short times, ranging from a few milliseconds to a few seconds, with a long tail extending out to very long half-lives. Figure 9 shows an approximate time distribution of the events, based on the lifetimes of the known most probable fission daughters. The mean lifetime is about 2.5 seconds.

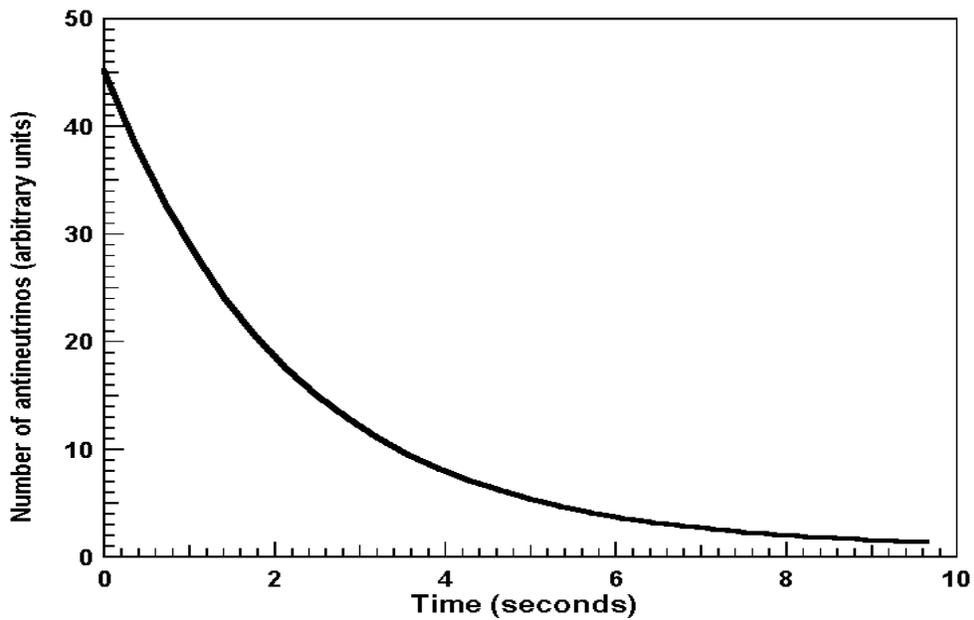

Figure 9: **The approximate time distribution of the antineutrinos produced in a $^{235}$U fission explosion.**

### 4.2.1 Detection of Antineutrinos from Fission Explosions

Summarizing the signal of interest, a detector must be capable of registering antineutrino events with energies up to about 8 MeV from a burst lasting a few seconds.

Accounting for energy threshold, the number of events in an inverse beta detector as a function of yield Y, distance D and detector mass M is given by:

7) $N_{\bar{\nu}} = 2.25 \text{ events} \times \left( \frac{Y}{\text{kTons}} \times \frac{M}{10^6 \text{Ton}} \times (\frac{100 \text{ kilometers}}{D})^2 \right)$

Unfortunately this is a dauntingly small number. Equation 7 reveals that only about 2 events would be detected in million ton detector at 100 kilometer standoff.

While the number of events is quite small, their burst-like nature is very effective for reducing backgrounds. However, as a result of this 'antineutrino-starved' circumstance, only detection is possible, and that, clearly only with enormous effort: no spectral (or temporal) distributions are likely to be measurable in any realistic scenario. In Section 7.5 we discuss the prospects and possible benefits of remote nuclear explosion monitoring with antineutrinos.

# 5 Near-Field Applications: Safeguards and Cooperative Monitoring

In this section we describe the goals and currently used protocols and technologies of the IAEA reactor safeguards regime in some detail, and then define possible roles that antineutrino detection may play in this regime. We also discuss the many historical examples of deployed antineutrino detectors. These deployments - including two that have been built for the purpose of testing and demonstrating cooperative reactor monitoring - provide valuable information on cost, intrusiveness, and ease of operation. Because of the historically limited budget of the IAEA, these practical considerations are decisive when considering adoption of the technology within the IAEA safeguards regime. Next, we discuss other operational concepts for cooperative monitoring beyond safeguards, and conclude this section with a discussion of the R&D needs for future antineutrino-based near-field monitoring applications.

## 5.1 Current Reactor Safeguards Methods

The objective of the IAEA safeguards regime is "… timely detection of diversion of significant quantities of nuclear material from peaceful nuclear activities to the manufacture of nuclear weapons..."[30] The regime applies to all civil nuclear infrastructure, including commercial and research reactors. Since antineutrinos are produced in practically measurable numbers only by critical or supercritical systems, reactors are the only part of the IAEA safeguards regime for which antineutrino detection is potentially relevant.

'Timely detection' and 'significant quantity' are essential elements of the IAEA safeguards regime. Both have formal definitions. Table 2 shows the IAEA timely detection goals. These goals are based on the 'conversion time', an estimate of the time needed to convert nuclear material of a given form into a weapon. A significant quantity (SQ) is defined as the amount of nuclear material for which the possibility of manufacturing a nuclear explosive device cannot be excluded. Table 3 shows the IAEA definition of a significant quantity.

| IAEA Timeliness goal | Material form |
| --- | --- |
| **One month** | unirradiated direct use material (e.g., Highly Enriched Uranium (HEU) (=> 20% uranium, enriched in $^{235}$U), separated plutonium, or Mixed Oxide (MOX) fuel) |
| **Three months** | irradiated direct use material, (e.g., plutonium (Pu) in spent or core fuel) |
| **One year** | indirect use material (e.g., Low Enriched Uranium (LEU) (< 20% uranium, enriched in $^{235}$U) or natural uranium) |

Table 2: IAEA timeliness goals for detection of diversion of nuclear material from reactors[2].

| IAEA Significant Quantity | Isotopic content |
| --- | --- |
| 25 kg of $^{235}$U in HEU | HEU is defined as uranium with $\geq$ 20% $^{235}$U content |
| 75 kg of $^{235}$U in LEU | LEU is defined as uranium with < 20% $^{235}$U content |
| **8 kg of elemental Pu** | Any isotopic mix of Pu except Pu with >20% $^{238}$Pu |

Table 3: IAEA definitions for Significant Quantities of nuclear material.

IAEA safeguards methods divide into two categories:

1. *Nuclear material accountancy*: counting, examination, and direct measurements which verify the quantities and continued integrity of declared nuclear materials.
2. *Containment and surveillance (C/S) measures*: C/S measures, such as video, tags and seals, and similar methods, are used to complement material accountancy methods.

Reactor safeguards activities vary according to reactor type, while possessing certain common features. At all reactors, safeguards begin with a Design Information Questionnaire (DIQ), provided by the State to the IAEA. The IAEA conducts an initial review and site inspection based on the DIQ information. The review consists of a comparison of the plant design documentation with the actual plant infrastructure relevant for safeguards. Throughout the lifetime of the reactor, the IAEA annually verifies the DIQ, in particular when the State reports any changes to reactor operations.

Once the reactor is online, the safeguards regime applies material accountancy and C/S measures to fresh fuel, in-core fuel and spent fuel. The essential accounting methods are audits of plant operating records, in which the records are checked for consistency and compared with earlier reports to IAEA.

The IAEA verifies that fresh fuel shipped from a fabrication facility is identical to that received at the reactor, by comparing serial numbers on received fresh fuel against records of fuel shipments from the fuel fabrication facility. A similar procedure is applied for all materials shipped from the site. Shipments of spent fuel are closely scrutinized: extra inspection trips are used if necessary to verify the fuel integrity and check serial numbers of casks of spent fuel being shipped offsite or to onsite dry storage.

Beyond these common features, accountancy and C/S measures vary with reactor type, as we now describe.

### 5.1.1 LWR safeguards

LWRs are distinguished by their fuel enrichment (typically LEU, 3-5% enriched) and by the property of off-line refueling. These factors mainly determine the IAEA safeguards protocols particular to LWRs. Off-line refueling means that the fissile material in the reactor is accessible only in reactor-off periods. These reactor outages are therefore the focus of much of IAEA verification activities at LWRs.

For fresh in-core and spent fuel, IAEA must verify every three months that no diversions of a SQ of Pu in core fuel and spent fuel have occurred, and must verify every year that no diversions of a SQ of LEU in fresh fuel, core or spent fuel have occurred. The IAEA conducts Interim Inventory Verification (IIV) inspections at three month intervals to meet its Pu timeliness goal, and conducts a Physical Inventory Verification (PIV) at LWRs once per year to meet its LEU timeliness goal. During the PIV, the IAEA verifies the presence of all nuclear materials onsite, and all shipments of nuclear material into and out of the site. Other inspections may occur during the year to verify movements of spent fuel to on-site wet or dry storage.

#### 5.1.1.1 Declarations of thermal power

As part of the inspection process, the IAEA receives from the State an official declaration of the total thermal power generated by the plant since the last inspection period. The IAEA uses the power history of the reactor to analyze Pu production in the core. Although reactor power monitoring systems are available, and used in other reactor safeguards regimes (such as at research reactors, see below) these systems have not been implemented at LWRs they have been found to be intrusive and difficult to maintain, with large cost and little benefit compared to C/S techniques[31]. Because of this expense and difficulty, the IAEA has no direct independent confirmation of the Pu ingrowth in the core beyond the internal consistency of the operator declarations.

#### 5.1.1.2 Fresh fuel and in-core fuel verification

During the Physical Inventory Verification, the IAEA verifies the continued presence of fuel in the core by using an underwater camera to check serial numbers on fuel assemblies. The IAEA also counts the number of fresh fuel items, verifies their serial numbers, and randomly verifies a number of fuel assemblies to confirm that they contain uranium, using an approved statistical sampling plan. A gamma spectrum is measured, within which the 185 keV peak of $^{235}$U is identified, in order to verify the uranium content of the fuel. Once the reactor core is verified the IAEA depends on containment and surveillance (C/S), using a prescribed combination of seals on strategic hatches and canals in the reactor pool and surveillance cameras, to reassure that no tampering with the core fuel occurs until the next PIV.

#### 5.1.1.3 Spent fuel verification

The IAEA also verifies and maintains Continuity of Knowledge (CofK) of spent fuel at LWRs, Spent fuel is the most attractive material from a proliferation standpoint since it contains the most Pu, is outside of the reactor and thus more susceptible to diversion.  During each Physical Inventory Verification visit, the IAEA verifies the continued presence of spent fuel in the cooling pond using a night vision device tuned to the Cerenkov light spectrum emitted by the spent fuel. For old and/or low burn-up fuel with reduced Cerenkov output, the presence of spent fuel items is verified by spectroscopic detectors which detect the 662 keV gamma ray line from $^{137}$Cs. Once the spent fuel pool is verified, the IAEA maintains CofK with surveillance cameras, and in some cases by sealing the cover plates over the spent fuel pond.

Every three months the IAEA verifies the spent fuel for timeliness by verifying the functionality and integrity of surveillance equipment and seals.  If no C/S system exists at the spent fuel pond, the IAEA must reverify the entire spent fuel pond every three months using Cerenkov or gamma-ray detectors.

### 5.1.2  CANDU Safeguards

At CANDUs, the main focus of safeguards is on accountability for spent fuel.  As in LWRs, accounting records are audited during each visit and compared with past reports to IAEA. In addition, CANDUs are subjected to the following safeguards measures:

Verification of fuel in the core by continuously monitoring spent fuel discharges using a Core Discharge Monitor;

Counting and monitoring of spent fuel bundles as they are transferred from the vault to the spent fuel bay using Spent Fuel Bundle Counters.  Where required, other vault penetrations are monitored to verify that all spent fuel is transferred to the spent fuel bay via the route containing the bundle counter;

Surveillance in the spent fuel bay.  In addition, at some stations tamper-indicating containers of fuel are closed with AECL Random Coil seals as a complement to the surveillance system.  At other stations, non-destructive assay instruments are used as a back-up for the surveillance system.

Since re-fueling of a CANDU occurs almost daily, permanently installed instrumentation is used to continuously monitor and track spent fuel movements.  In contrast, because LWR cores are not usually opened more than once per year it is usually possible to apply IAEA safeguards seals to verify that the reactor pressure head remains closed between IAEA inspections.

### 5.1.3  Research Reactor Safeguards

Safeguards practices at research reactors varies considerably because of the  diversity of core types and operational characteristics. Audits and C/S measures are pursued in a manner broadly similar to CANDUs and LWRs. However, a significant difference in some research reactor safeguards protocols is the use of Reactor

Power Monitor (REPM), which monitors neutron levels, and the Advanced Thermo-hydraulic Power Monitoring (ATPM) which measures the flow and temperature differential on the primary coolant loop to verify core thermal performance. Since fuel in research reactors can be shuffled easily and targets positioned without any IAEA knowledge, there is potential in large (>25 MWth) research reactors of unreported Pu production. This led to the creation and use of the REPM and ATPM. ATPM systems must be introduced into the coolant loop in the reactor in order to function.

## 5.2 Antineutrino-Based Reactor Safeguards Methods

The protocols just described rely in large part on C/S measures and on so-called 'item accountancy', which in this context refers to the counting of fuel assemblies or fuel rods. By contrast, antineutrino detection is a type of 'bulk accountancy' method for directly estimating the amounts and rates of consumption and production of fissile material in the reactor core, in a manner that is nonintrusive to reactor operations, outside of containment, and under control of the safeguards inspectorate. These features provide a complementary set of capabilities that can further enhance the reactor safeguards regime.

As discussed earlier in section 4.1.5 and shown in the examples introduced below, the current state of the art in antineutrino detection is such that the measurement, when used alone and independent of any other safeguards measure, is sensitive at the one sigma level to changes of roughly 50 kg or 6 significant quantities of Pu *and* 300 kg or 4 significant quantities of $^{235}$U in LEU in reactor fuel, within the IAEA timeliness criterion (respectively 3 months and 1 year) set for these materials. These values can be expected to be improved upon by further development of antineutrino detectors and data analysis methods. Moreover, antineutrino-based safeguards methods are not intended be used alone, but in conjunction with other safeguards measures, improving IAEA's overall ability for timely detection of significant quantities of fuel, possibly reducing the number of inspector days at the site, and providing information that is completely controlled by the inspection agency. For example, a fault-tree-based analysis of one diversion scenario, has shown that an independent antineutrino-based measurement of the reactor thermal power can provide a three-fold improvement in ability to detect diversion of fuel compared to a regime without such a measurement[32]. The complementary, non-intrusive nature of antineutrino-based methods, as well as the robustness and ease of use of the underlying detectors, point to its strong potential as a new safeguards tool.

The antineutrino rate and spectral measurements described earlier can be used to extract information about the operational status, thermal power and fissile inventory of reactors. The following examples are suggestive but not exhaustive of how this information would be used in a safeguards context. In practice, each reactor type – LWR, fast reactors, CANDUs, and others – will require a specific safeguards analysis to establish the utility of antineutrino detectors within a particular safeguards protocol.

**Operational status**
One purpose of tracking the reactor power is to ensure that the reactor was not stopped and started with unusual frequency, or otherwise operated in a suspicious manner. For example, frequent shutdowns early in the fuel cycle could be indicative of attempts to recover low-burnup fuel, from which plutonium would be easier to recover. (The isotopics of such low burnup plutonium are also somewhat more favorable for bomb design.) This points to one possible benefit from antineutrino measurements: they can continuously and independently confirm the correctness of operator declarations of the reactor operating history between inspection periods. This type of measurement is of particular interest for off-line refueling reactors, such as LWRs or many research reactors. For example, the SONGS1 prototype, deployed at an LWR, has shown that shutdowns can be discovered within five hours, and that a 20% change in the thermal power can be seen within about 15-20 hours.

**Power**

IAEA does not itself currently use power monitoring technologies at power reactors because of the intrusiveness and cost of these systems. Instead, the agency relies on the reactor operators' formal declarations of power throughout the fuel cycle to estimate core fissile inventories at the end of cycle. By contrast, antineutrino rate and spectral measurements offer a means for the IAEA to independently check operator power declarations. As described in section 4.1.4, antineutrino rate measurements provide a non-intrusive, real-time estimate of instantaneous and integrated thermal power, folded with a correction term that explicitly depends on the burnup. As shown below for the SONGS1 experiment, the precision on a *relative* measurement of thermal power - that is, relative to an power measurement initially verified by other means – has been demonstrated at the 3% level with a simple detector, limited only by counting statistics and after correcting for the predictable effect of burnup. Additional improvement on the precision of the relative power measurement, approaching the 1-2% level, could be obtained with larger or more efficient detectors than the prototypes so far deployed.

**Isotopic content**

Estimates of the isotopic content of in-core fuel are directly accessible in real-time through measurement of either the antineutrino rate or energy spectrum. This allows a more stringent consistency check to be enforced on operator declarations: the declared burnup of both fresh fuel and spent fuel must be reconciled with the estimate of isotopic content provided by the antineutrino rate or spectral measurement. As discussed in section 4.1.4, rate-based measurements require knowledge of the reactor power and initial fuel loading, while measurement of an energy spectrum would allow absolute and independent estimates to be made of both power and burnup at few percent level, *without* additional input about reactor operational parameters. This requires a more sophisticated detector and analysis compared to a simple rate measurement. Alternatively, rate-based measurements could be used in conjunction with other safeguards measurements to more rigorously constrain the fissile content, approaching the significant quantity level.

## 5.3 Antineutrino Detector Deployments Relevant for Reactor Safeguards

As seen in Table 4, a large number of experiments have successfully measured antineutrino events near nuclear reactors. This rich experimental history was inaugurated with the discovery of the antineutrino interaction at the Savannah River Site in the 1950s[33], using the same organic scintillator technology that has become the standard for reactor antineutrino experiments.

| Experiment | Power (GW) | mass(ton) | Distance (m) | Depth (mwe) | Detector | μ-rate($s^{-1}$) | Deadtime (%) |
|---|---|---|---|---|---|---|---|
| ILL[34] | 0.057 | 0.32 | 8.76 | 7 | $^3$He scint. PSD | 250 | 8 |
| Gosgen[16] | 2.8 | 0.32 | 38/46/65 | 9 | $^3$He scint. PSD | 260-340 | 8 |
| Krasnoyarsk[35] | 1.6 | 0.46 | 57/231 | 600 | $^3$He only | - | - |
| Bugey 3 [36,37] | 2.8 | 1.67 | 15/40/95 | 23/15/23 | 6Li scint. PSD | - | 2 |
| SavannahRiver[38] | 2.2 | 0.25 | 18.2/23.8 | ~10 | Gd scint. PSD | - | - |
| CHOOZ[7] | 4.4 | 5 | 1000 | 300 | Gd scint. | 1 | 2 |
| Palo Verde[6] | 11.6 | 11.3 | 800 | 32 | Gd scint. | 2000 | 2 |
| KamLAND[8] | ~80 | 408 | ~180,000 | 2700 | Gd scint. | 0.3 | 11 |
| Rovno[39] | 1.4 | 0.43 | 18 | | Gd scint | 350 | 7 |
| SONGS1[19,40,41] | 3.4 | 0.64 | 24.5 | 10 | Gd scint. | 600 | 10 |

**Table 4** The characteristics of several previous reactor antineutrino detection experiments. The detector column includes the technique used, along with whether or not there was Pulse Shape Discrimination (PSD) in the electronics.

Following the discovery of the antineutrino, and as discussed further in section 8.1, most of these experiments were built to search for evidence of neutrino oscillations, which appear as a regular, repeated deviation of the antineutrino flux from inverse square behavior, as a function of distance. Since there was little theoretical guidance on the oscillation spatial frequency, experiments were done at a range of different distances, as seen in Figure 5. The large number of experiments has provided a solid foundation of theory and experiment which can be applied to safeguards applications. Perhaps most important for near-field applications, the experiments have shown that oscillations do not affect the spectrum or inverse square behavior of the antineutrino flux out to approximately 1 km. This non-deviation is measured with the absolute precision on the reactor antineutrino flux and spectra of the various experiments to the level of 2%. Experiments now being built, such as Double Chooz[28] and Daya Bay[42], will improve the precision of flux and spectral measurements at these distance ranges. Additionally, the effect of isotopics on the antineutrino rate and spectrum has been clearly demonstrated by many of the earlier experiments. In some experiments, such as the first Chooz deployment[14], the direction of the antineutrino has been successfully reconstructed.

Results from many of the oscillation experiments have provided valuable information for the design of detectors for non-proliferation purposes. For example, experience with changes in the optical properties of scintillator in CHOOZ and Palo Verde point out the importance of careful handling of the scintillator during production to minimize any contaminants which might cause the Gd to come out of solution. This experience has resulted in more stable operations and modified scintillator handling procedures that benefit safeguards deployments. A wide range of calibration methods, degrees of overburdens, and active and passive shielding methods have been studied in these experiments, lating the groundwork for applications.

Beyond the oscillation experiments, The Russian deployment at the Rovno complex in Ukraine[2] and the U.S. deployment at the San Onofre Nuclear Generating Station in Southern California[19] were deployed explicitly to demonstrate the feasibility of practical monitoring of reactors in a safeguards context with relatively small (cubic meter scale) antineutrino detectors.

### 5.3.1 Deployment Considerations

The overall goals for antineutrino detection for safeguards fall into two conflicting categories. First, the detectors must be designed, deployed and used to optimally extract measurements relevant for safeguards. These will consist of one or more of the operational status, thermal power and isotopic content measurements described earlier, using either rate or spectral information (or both). Second, the detectors must meet additional criteria that are generally not a priority in physics deployments, and which may conflict with physics goals. These criteria are imposed by IAEA needs for low cost, robust, reliable, easy-to-use, and largely unattended monitoring systems, and by the strong desire of reactor operators for minimal impact on plant systems and minimal disruption of plant activities.

In the first category fall basic detector design criteria including standoff distance, the detector target mass and local gamma/neutron shielding, the amount of overburden, the need for active muon background rejection, the stability in time of the detector response, and the precision with which the detector can be calibrated. The second set of criteria include practical considerations such as reducing the detector footprint to the extent possible, minimizing energy consumption and replacement of parts, construction from non-hazardous and low cost materials, and ability to operate with minimal or no overburden. As with many applied physics efforts, the tension between these practical concerns and the physics requirements for extraction of a reliable signal are the focus of current R&D.

The state of the art is that both the Rovno and SONGS deployments have successfully demonstrated that antineutrino detectors at shallow depths and with 2-3 meter linear dimensions (including shielding materials) can stably track operational status, thermal power and fissile content in real time in LWRs over the course of several years. We next describe the two experiments and summarize the significance of their results for safeguards applications.

### 5.3.2 Rovno

Russian physicists appear to have been first worldwide to recognize and exploit antineutrino detection as a tool for reactor monitoring[43]. Multiple basic and applied experiments were conducted over the course of many years, beginning in 1982, at the Rovno Atomic Energy Station in Kuznetsovsk, Ukraine. The work summarized here is from an experimental deployment that focused specifically on measurements relevant for safeguards applications[2].

The detector consisted of 1050 liters of Gd-doped organic liquid scintillator, viewed through light guides by 84 photomultiplier tubes. A 510 liter central volume was used as the primary target, with a 540 liter surrounding volume, separated by a light reflecting surface, employed as shield against external gammas, and as a capture volume for gamma-rays emitted respectively by the positron annihilation and neutron capture. The deployed location of the detector was 18 meters vertically below the reactor core, providing substantial overburden for screening of muons. An active muon rejection system was apparently not used in this deployment.

The antineutrino source was a Russian VVER-440 pressurized water reactor, loaded with LEU fuel, with a nominal power of 440 MWt. The gross average daily antineutrino-like event rate was 909 ± 6 per day, with a reactor-off background rate of 149±4 events per day (i.e. a net antineutrino rate of about 760±7 per day). The intrinsic efficiency of the Rovno detector was approximately 30%.

The Rovno deployment clearly demonstrated sensitivity to both the power and the isotopic content of the core, based on rate and spectral measurements. Figure 10 reveals the expected variation in the antineutrino count rate due to the burnup effect. The 5-6% change in rate is well matched to a prediction based on a model of the fuel isotopic evolution in the core, combined with the theoretically predicted antineutrino emission spectra for each isotope.

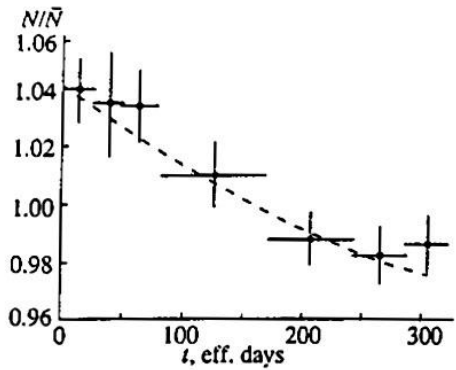

Figure 10: The antineutrino rate N, normalized to the time average of the rate over the course of the cycle, plotted versus effective cycle day, for the Rovno experiment[44]. The points with error bars are the data, the curve is a prediction based on a simulation of the reactor core evolution.

The absolute precision on the reconstructed thermal power of the reactor is 2%, with the largest uncertainties arising from imperfect knowledge of the detection efficiency and detector volume. For comparison, direct thermal power measurements by the most accurate methods used by reactor operators range have precision ranging from 0.5-1.5% depending on the method.

The change in the antineutrino spectrum over the course of the fuel cycle is also visible in the Rovno data. Figure 11 shows the ratio of the reconstructed antineutrino energy spectra at the beginning and end of the fuel cycle. The variation in spectra is most pronounced at the highest energies, consistent with predictions, and is caused by the net consumption of 521 kg of fissile material (both plutonium and uranium) over the course of the fuel cycle. While not directly quoting an uncertainty on this value derived from the antineutrino measurement, the Rovno group independently estimated fuel consumption from the reactor's thermal power records, and find a value of 525±14 kg, close to the antineutrino derived value.

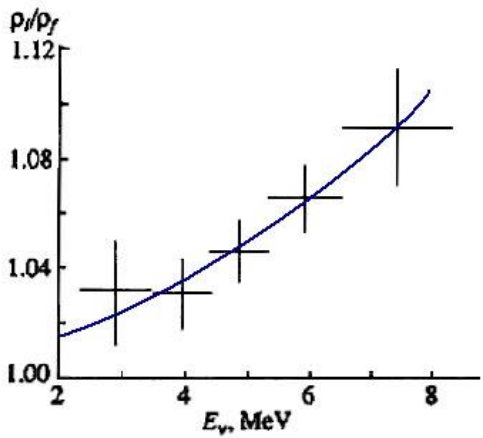

Figure 11: The ratio of beginning-of-cycle to end-of-cycle antineutrino spectra, as measured at the Rovno reactor. The points with error bars are the data, the curve is a prediction based on a simulation of the reactor core evolution.

### 5.3.3 San Onofre

The SONGS deployments in the United States were performed independently of the earlier Russian experiments, for the express purpose of demonstrating the feasibility of antineutrino detection in the context of IAEA safeguards. In a report following a 2003 experts meeting at IAEA headquarters in Vienna[45], the IAEA requested study of the feasibility of "confirmation of the absence of unrecorded production of fissile material in declared reactors" with antineutrino detectors, and further stated:

"The appropriate starting point for this scenario is a representative PWR. For this reactor type, simulations of the evolution of the antineutrino flux and spectrum over time should be provided, and the required precision of the antineutrino detector and independent power measurements should be estimated. This effort should be coupled with the already ongoing prototype detector development."

The SONGS1 detector responded to this request, and sought to demonstrate that antineutrino detectors could meet other anticipated IAEA requirements. In addition to simulating the antineutrino signal and confirming its sensitivity to thermal power and fissile content, the SONGS detector demonstrated stable long-term unattended operation, using a simple, low channel count detector design, non-intrusiveness to reactor site operations at a commercial power plant for several years, and remote and automatic collection of antineutrino data and detector state of health information.

The SONGS1 detector at the San Onofre Unit 2 Nuclear Reactor has been operating since 2002, with the full detector volume operational continuously from 2006 through summer 2008. SONGS1 has an approximately cubic meter central target, containing 0.64 tons of gadolinium (Gd) loaded liquid scintillator contained in four stainless steel cells, each read by two Photomultiplier Tubes (PMTs). As seen in Figure 12, a six-sided water/polyethylene shield of average 0.5 meter thickness is used for passive shielding of neutrons and gamma-rays, and a 5-sided muon detector for tagging and vetoing muon-related backgrounds. A total of 24 PMTs were used to read out both the muon veto and main detector.

The detector was deployed in the Unit 2 'tendon gallery', an annular room that lies directly under the containment dome, and which allows access to steel reinforcement cables that extend through the containment structure. The gallery is 25 meters from reactor core center. Many commercial reactors have tendon galleries. They are well suited for deployment of an antineutrino detector because the large, vacant space is rarely accessed by plant personnel, and because of the muon-screening effect of approximate 10 mwe earth and concrete overburden. (The SONGS overburden is 10 mwe – other sites may differ but are of the same depth scale.) At the SONGS location, background muon rates are reduced by a factor of approximately 5 compared to above-ground backgrounds.

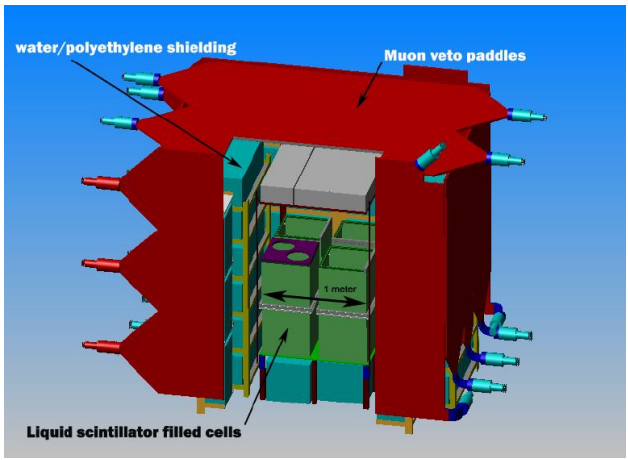

**Figure 12: A cut away diagram of the SONGS1 detector, showing the scale and major subsystems.**

The measured antineutrino rate (signal plus background) in the SONGS detector, which has roughly 60% of the mass of the ROVNO detector, was 564 ± 13 events per day at full reactor power, with a measured background rate of 105 ± 9 events per day at zero reactor power. The intrinsic efficiency of the SONGS detector is approximately 10%.

SONGS1 demonstrated sensitivity to the three antineutrino-based safeguards metrics introduced in section 5.2: operational status, power and fissile content.

Figure 13 shows a short-term excursion of the reactor power as reflected in the antineutrino rate. The upper plot is the hourly antineutrino count rate plotted versus hour. As described in detail in [40], a cumulative statistical test statistic, known as the Sequential Probabilistic Ratio Test, is used to determine to a desired level of confidence the operational status of the reactor. The evolution of this statistic is shown in the lower plot. In this example, confirmation of a change in the reactor operational status is determined within 4 hours.

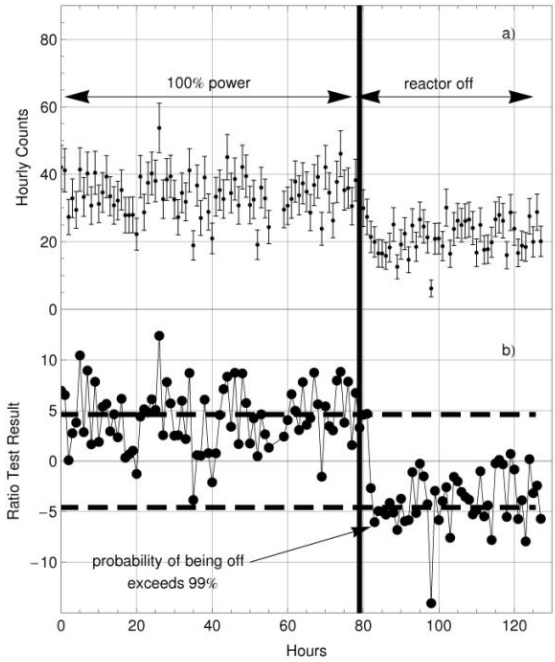

**Figure 13: (a):** the hourly number of antineutrino-like events, plotted versus hour, through a reactor outage. **(b):** the value of the test statistic plotted versus hour over the same time range. In both plots, the vertical line indicates the hour in which the reactor shutdown occurred. The dashed lines in (b) are the 99% confidence level values of the test statistic, known as the Sequential Probabilistic Ratio Test. For this data set, these values are obtained only during the appropriate period (on or off), meaning that no false positives or negatives occurred.

Figure 14 is a histogram of the weekly antineutrino rate, normalized to average weekly rate for the prior 4 weeks. This metric provides a relative estimate of the reactor power on a weekly basis, accurate to 3%. This limitation is imposed only by the counting statistics for data accumulated over a week. Further precision can be obtained with a larger or more efficient detector.

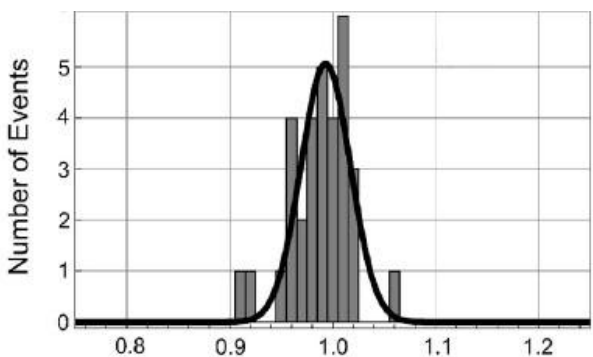

**Figure 14:** Histogram of the weekly detected antineutrino rate divided by the average of this quantity for the prior four weeks. This relative metric gives a 3% accurate measurement of the reactor thermal power, limited only by counting statistics.

Figure 15 shows the long term change in the antineutrino rate measured in SONGS1 over the course of a full reactor cycle. The total change in the antineutrino rate is approximately 12% over the entire cycle, and this change is predicted by a detailed simulation of the core isotopic evolution. The measured change in rate can be used to roughly estimate the sensitivity to changes in fissile content. In a SONGS refueling outage, approximately 250 kg of $^{239}$Pu and $^{241}$Pu is removed and 1500 kg of $^{235}$U is added in fresh fuel. This leads to the measured 12% change in antineutrino rate. At the one standard deviation level in this prototype, this corresponds to the ability to detect a reduction in total Pu content of 40 kg and a simultaneous increase of $^{235}$U content of about 250 kg, using only the antineutrino rate. Further improvement in this sensitivity can be expected from a higher statistics measurement (increased detector size or efficiency) or through spectral analysis.

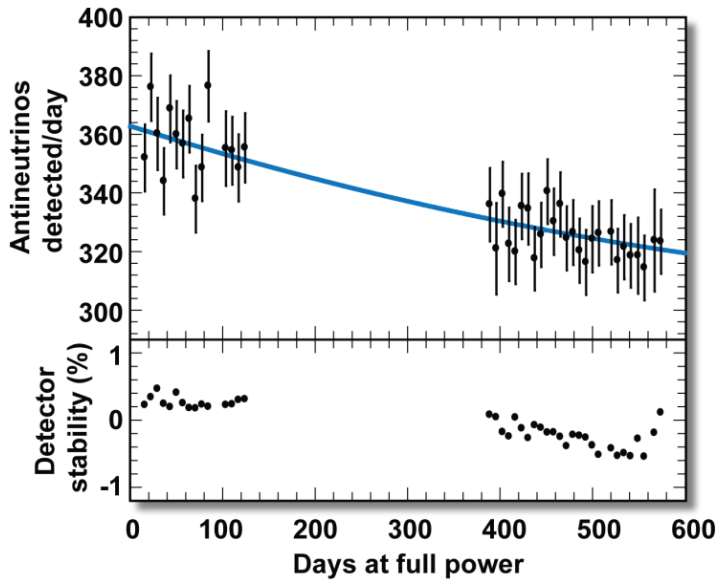

Figure 15: The daily background subtracted antineutrino rate over a 600 day cycle of the SONGS Unit 2 reactor. Data from two successive cycles were combined to create this plot. The solid line shows the rate change predicted by a simulation of the reactor isotopic evolution. The lower plot shows a measure of the stability of the detector response over this period. The response is stable in a relative sense to better than 1% over the entire data-taking period, meaning that the measured rate change is not due to changes in the detector response.

In an actual safeguards regime, information derived from the antineutrino detector will be used in conjunction with other safeguards methods to determine the absence of diversion of a significant quantity of fissile material. To determine the marginal utility added by antineutrino monitoring in this context, it is useful to evaluate how the presence or absence of antineutrino rate information affects the ability to detect diversion in specific scenarios. One such study[46] has shown a 3-fold improvement in the probability of detecting diversion at a reprocessing plant, when inventory information derived from a SONGS1-style antineutrino detector is used in conjunction with other safeguards information, compared to a scenario in which this information is not available. In the example scenario, the antineutrino detector measured a 5% power increase, which resulted in the presence of an otherwise unaccounted for significant quantity of plutonium (8 kg), assumed diverted at a downstream reprocessing facility.

Beyond the safeguards metrics just described, SONGS1 addressed important practical considerations relevant to safeguards deployments. SONGS1 collected and analyzed data continuously, remotely and in real time, using a local computer and telephone modem uplink to a laboratory in Northern California. The detector location was completely removed from daily reactor operations, and no maintenance of the detector by site personnel was required. Only occasional detector maintenance was required by LLNL and SNL deployment team, even for this non-optimized prototype.

An important shortcoming of many liquid organic scintillators, including the material used in SONGS1, is their low flashpoint, high flammability and relatively high toxicity. Though SONGS1 has demonstrated non-intrusive and safe operation, the use of the flammable liquid does impose some safety burden on the operator and inspector that it is preferable to avoid. A more recent effort at SONGS has demonstrated antineutrino sensitivity with a non-toxic plastic scintillator based detector. One of the two identical modules of this detector, SONGS2, is depicted in the cutaway rendering in Figure 16. Each module consists of 24 2 cm x 0.75 m x 1m plastic scintillator slabs, interleaved with thin Mylar sheets painted with Gd-doped paint. Each module is read by a total of 4 PMTs, and is placed in an aluminum framework for portability. The total active volume of this detector was 0.4 cubic meters. The plot to the right in Figure 16 shows the gross daily antineutrino count rate through a short outage, revealing clear sensitivity to the antineutrino signal. While preliminary, these results show a short term monitoring capability similar to that of the SONGS1 liquid scintillator detector. Further analysis is needed to demonstrate long term sensitivity to burnup.

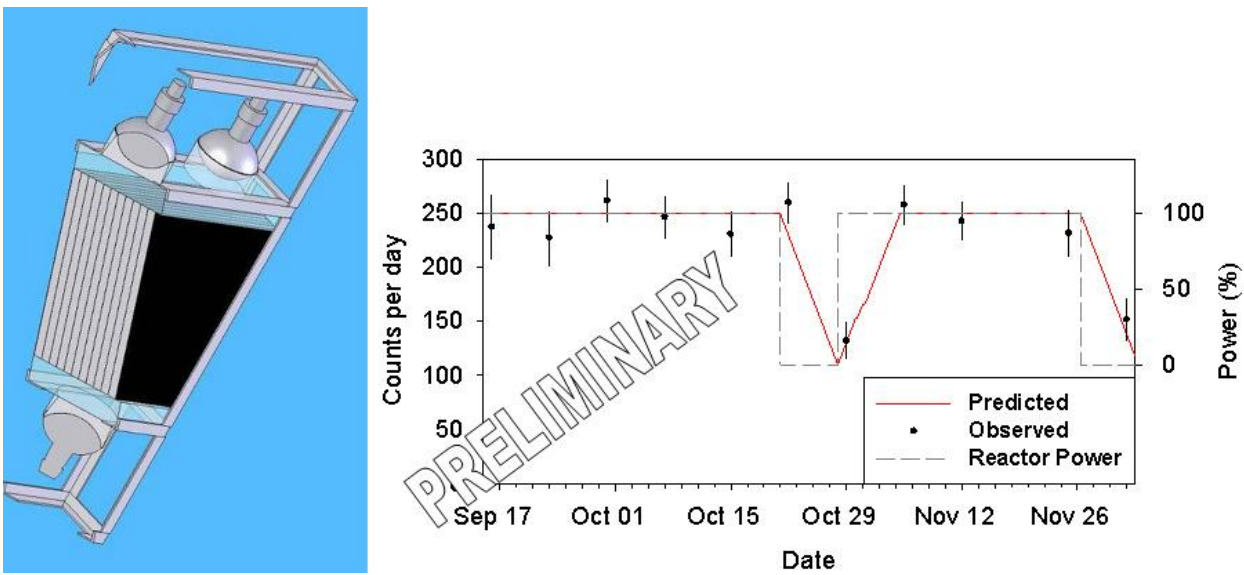

**Figure 16: Left: a cutaway view of the plastic scintillator/Mylar sandwich detector SONGS2. Mylar sheets, covered with a Gd-doped paint, are interleaved between the 2 centimeter thick plastic scintillator blocks. Light collection is accomplished with 4 photomultiplier tubes. Right: initial data from the plastic detector showing sensitivity to the antineutrino signal.**

The advantages of this approach are several. Most importantly the flammable, toxic, and carcinogenic liquid organic scintillator is eliminated. In addition, the detector can be fully assembled offsite, rather than built in place, as is often the case with the liquid detectors. It also can be transported to the site without using resorting to the more expensive shipping methods and onerous transport regulations required for hazardous materials. The disadvantages are some reduction in overall detection efficiency due to the detector design and the plastic scintillator proton density.

### 5.3.4 Beyond Rovno and SONGS: A Demonstration Safeguards Project

The Rovno and SONGS1 deployments both clearly demonstrate many of the expected requirements for antineutrino-based safeguards. They show that antineutrino detectors can extract measurements of direct safeguards interest for years at a time, without affecting plant infrastructure or interfering with plant personnel activities. The relatively simple design of the detectors, with their low channel counts, readily available raw materials, and low maintenance requirement, demonstrates that IAEA criteria for low cost, simple unattended remote monitoring capabilities can all be met. While absolute detection of significant quantities of fissile material have not yet been achieved with antineutrino detectors alone, initial studies indicate that the additional information provided by antineutrino detectors, in conjunction with other safeguards information, can constrain fissile content at the few dozen kg level, and improve ability to detect diversion in specific scenarios. In section 5.5, we discuss additional R&D that would be useful to further improve the utility of antineutrino detectors for safeguards.

A logical next step from these efforts would be a deployment of a detector at an IAEA safeguarded facility in a non-nuclear weapons state, preferably with the direct involvement of IAEA. Such a deployment would represent an important advance beyond the earlier safeguards demonstrations, by serving as a pilot project and an example for IAEA safeguarded reactor facilities worldwide. Valuable information on integration of antineutrino detectors into the modern reactor safeguards regime would be gained from such a test deployment.

### 5.4 Beyond Safeguards

Aside from the existing IAEA safeguards regime, other existing or future cooperative near-field monitoring regimes might benefit from antineutrino detectors. Examples include tracking the progress of plutonium disposition in reactors, verifying the cessation of fissile material production at a previously active reactor site for the Fissile Material Cutoff Treaty or similar regimes, or verifying core conversion in plutonium production reactors in weapons states. Here we provide one example of a verification problem beyond safeguards that might be addressed with antineutrino detectors.

'Plutonium disposition' refers to the management of separated weapons-grade plutonium inventories declared excess to military needs by the United States and Russia. As defined in a 1994-1995 National Academy of Sciences study[47], an important goal for any plutonium disposition program is to comply with the 'Spent Fuel Standard'. The Spent Fuel Standard requires that separated surplus military weapons-grade plutonium is converted into a physical form from which it is as difficult to recover plutonium as from ordinary commercial reactor spent fuel[b]. One proposed way to meet this Standard is to convert the weapons-grade plutonium into mixed-oxide (MOX) fuel, where the mixture typically contains 5% Pu and 95% natural or depleted uranium, and to irradiate this fuel in a commercial reactor. In this context, antineutrino detectors could be used to verify that a reactor is actually burning weapons plutonium and not a substitute such as LEU fuel, or other separated MOX fuel not derived from weapons material. Figure 17 shows the predicted variation in antineutrino rate for a PWR operated with LEU, and with MOX fuel composed of either reactor-grade and weapons-grade plutonium. The weapons-grade plutonium used in this simulation contains 94% of the fissile isotope $^{239}$Pu, and the reactor grade plutonium contains 52% $^{239}$Pu. A detailed simulation of the evolution of the reactor fuel composition was used to predict the variation in antineutrino rate[48]. Since the absolute errors on rate are at the 2% level, changes between LEU and MOX should be clearly visible in an antineutrino detector, while changes in the isotopic mix of plutonium are at the current limit of sensitivity for antineutrino detectors.

---

[b] Briefly, the logic of the Spent Fuel Standard is that converting separated military plutonium stockpiles into a form similar to spent fuel, which contains most of the world's plutonium, is both necessary and sufficient after consideration of nonproliferation, cost, speed, and technical factors. A more resistant form would be too expensive and slow to produce; a less resistant form would be too attractive from the standpoint of a proliferator. (We note that since no significant disposition of plutonium has taken place in the one and one half decades since the study was completed, choosing methods based on the estimated time to completion seems less important than was previously thought.)

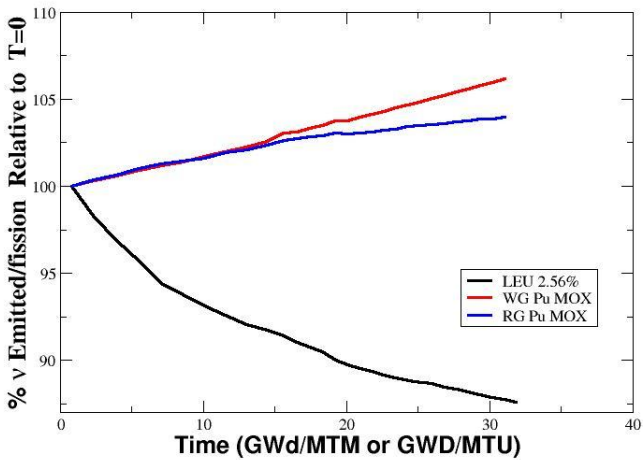

Figure 17: The variation in the emitted antineutrino flux from a PWR, according to fuel composition. The abscissa units are Gigawatt-Days per Metric Ton of Heavy Metal GWD/MTM or per Metric Ton Uranium (MTU)). The difference between LEU-fueled (black) and MOX-fueled (blue and red) cores is evident in the changing antineutrino rate. The distinction between weapons-grade (WG) and reactor-grade (RG) fuels may just be visible in high burnup conditions (> 20 GWD/MTM). The plot is from reference 48.

## 5.5 R&D Needs For Near-Field Antineutrino Detectors

Several demonstrations have explicitly shown that current generation antineutrino detectors offer new, practically achievable capabilities for IAEA reactor safeguards. In this section we will discuss research and development paths for improving the effectiveness of antineutrino detectors in near-field applications. R&D needs divide into three broad categories:

- improving the precision with which diversion of significant quantities of fissile material can be detected.
- improving the ease of deployment and operation of safeguards antineutrino detectors.
- performing additional systems analyses for IAEA safeguards and other cooperative monitoring applications.

These R&D paths are summarized here.

### 5.5.1 Increasing Sensitivity to Diversion of Material

Since the goal of current IAEA reactor safeguards is to detect diversion of significant quantities of material from reactors, an important class of R&D problem relates to improving the precision of estimates of fissile inventories derived from antineutrino detectors. (A closely related problem is improving our understanding of the synergies between antineutrino-based monitoring and other elements of the safeguards regime, which is considered in section 5.5.4.)

Little improvement is required in the precision of measurements of reactor operational status, which alert inspectors to unusual operating conditions and constrain the total amount of material that may have been generated in a given time period for a known reactor. Existing near-field antineutrino detectors have already demonstrated detection of shut-downs or tens of percent changes in reactor power within a few hours. With timely response by the inspecting agency, this reaction time is sufficient in the context of IAEA safeguards.

Some enhancement is both necessary and achievable in the precision of thermal power and fuel inventory estimates. Relating these quantities to the measured antineutrino rate and spectrum involves three steps, each amenable to further improvement. First the initial fuel loading and the thermal power profile of the reactor are used as inputs to predict the evolving fissile content at any time during the reactor cycle. Typically a reactor simulation package such as ORIGEN is used for this purpose. Second, the neutrino spectrum per fissile isotope is folded with the varying isotopic information. Third, the emitted antineutrino spectrum is related to the measured positron spectrum in the antineutrino detector, often by means of a detector response Monte Carlo code. Here we provide a discussion of R&D priorities that could lead to improvement in each area.

### 5.5.1.1 Predicted isotopic inventories and fission product yields

In the context of verifying declarations, simulations of the predicted evolution of the reactor fissile isotopic inventories play an important role. A recent review of the benchmarking of the precision of simulation codes has shown that for PWRs and BWRs, isotopics are successfully predicted by ORIGEN and other simulation packages with a mean precision below 1%, but a spread in precision across many experiments of several percent. This translates into an uncertainty in the predicted antineutrino spectrum of below 1%[24].

In this area, there are two main useful R&D paths in the context of safeguards. The first is to develop a comprehensive code package that combines all the elements needed for complete simulation of antineutrino spectra. Currently, the isotopic evolution simulation packages are normally decoupled from codes that reproduce antineutrino spectra. A more comprenhensive and integrated treatment, as outlined and initiated by the MCNP Utility for Reactor Evolution (MURE) project[49], is to build the spectrum on a per-fission basis, accounting for each individual fission products contribution to the total spectrum. The second area of need is to further extend the simulation base to a wider range of reactors, including CANDUs, fast reactors, and other reactor types of interest for safeguards.

### 5.5.1.2 Predicted antineutrino spectra

Distinct from the isotopic evolution, the absolute uncertainties in the predicted spectra of antineutrinos emitted per fission by each isotope can be further reduced beyond their current 2% level. An analysis by Huber[27] has shown that a three-fold reduction in these uncertainties would result in a 20 kg (2.5 significant quantity) uncertainty on the absolute quantity of fissile plutonium in a PWR as derived from a spectral analysis. In the limit, with no uncertainty on the spectra of individual isotopes, the Pu content could be determined to the significant quantity level, using only antineutrino information. While the absolute uncertainty in the spectra will never be zero, we note that a relative cycle-to-cycle comparison of directly measured antineutrino spectra at a given reactor could reduce or even eliminate these systematic effects. Moreover, antineutrino-derived information is likely to be used in conjunction with other data to meet safeguards goals.

One approach to predicting antineutrino spectra is to use *electron* spectra measured directly from fission. When the electron spectrum is suitably binned in energy, the intensity and energy values of each bin - along with the known analytic form of the underlying beta spectra - are used to predict the endpoint energies and intensities of each beta branch. The individual beta branches are constrained to sum to the original experimentally measured electron spectrum[c]. Once the endpoint energy and intensity of each underlying branch is estimated, this information can be analytically converted to an antineutrino spectrum arising from the same fissile isotope. The most recent efforts using this approach show a 1% uncertainty in converting electron to antineutrino spectra[50]. Needed R&D in this area lies in possible improved analytic or Monte Carlo treatment of the conversion process, and improvement in the experimentally measured electron spectra from fission from which the antineutrino spectra are derived.

A second approach is to precisely measure individual fission-fragment yields of each isotope, and fold these yields with the calculated individual antineutrino spectra, with endpoint energies now derived from electron spectra measurements on individual fragments rather than on a per fission basis. The precision of this approach could be improved with additional experiments that measure missing electron spectra from short-lived high Q-value decays.

### 5.5.2 Improving Safeguards Antineutrino Detector Performance

Current detectors will work for many safeguards applications. Still, further improvements would be useful and can be anticipated in the performance of antineutrino detectors for safeguards. With a wide range of possible designs available, the central challenge lies in balancing the need for the greatest possible sensitivity with ease of deployment and operation. Here we consider key performance requirements, and the R&D paths of greatest interest for improving the performance of cooperative monitoring antineutrino detectors. Improved deployment characteristics are considered in the following section.

Statistical considerations place the first constraint on detector performance. Several hundred events per day are needed for short-term (1-2 hours) shutdown information at commercial PWRs, and to achieve weekly power measurements at the few percent level. These capabilities have already been demonstrated by the cubic meter scale Rovno and SONGS detectors, at 20-25 meter standoff from few GWt reactors. The same event rates have also been shown by both Rovno and SONGS to suffice for detecting the 5-10% per cycle burnup effect in PWRs.

---

[c] In this prescription, the number of derived endpoints and intensities is equal to the number of bins.

Research reactors have powers ranging from a few kW up to a maximum of several hundred MWt. A scaling example is instructive to consider the requirements imposed by lower power reactors. For a 100 MWt reactor, a 3 ton detector with the same efficiency (30%) and standoff (18 meters) as the Rovno detector, would have an event rate of over 300 events per day, still sufficient for monitoring short term shutdowns and thermal power to the few percent level on a weekly basis. Improvements in detection efficiency, discussed below, may change the conclusion, but given the current state of the art, practical detectors probably cannot reach safeguards sensitivity goals for reactors much below 100 MWt.

Spectral analysis imposes a more stringent constraint on the required statistics. As already discussed, about $10^6$ total antineutrino events are required for an *absolute* measurement of fissile Pu inventories accurate to the 10% level with no prior assumptions about reactor operating parameters, with current emitted flux uncertainties, and with an assumed detector uncertainty on the spectral measurement of 0.6%[27]. This represents about 1300 days of running of the cubic meter Rovno detector, too long to be relevant for safeguards, as well as requiring a lower uncertainty on detector response than has been achieved to date. To reach this same absolute sensitivity in the several month time period consistent with IAEA safeguards needs would require approximately 11,000 events per day. This level might be reached at SONGS standoff with a 2.75 ton detector with 70% intrinsic efficiency, roughly similar to the Chooz detector parameters. Though this detector operated successfully for a period of a year or more, and its efficiency substantially exceeds the 10-30% efficiency of the safeguards demonstration detectors at SONGS and Rovno, the Chooz installation was significantly more complex and expensive than either of the safeguards detectors.

It is important to add that the requirements for a spectral measurement to be initially useful in a safeguards context are considerably less stringent than this, if independent constraints on reactor operations are available from other parts of the safeguards process. Indeed, with a spectral analysis combined with knowledge of the reactor power output, the Rovno experiment has already shown sensitivity at roughly the few tens of kg level to the total amount of fissile material generated in a single cycle, using just 760 net antineutrino events per day. This demonstrates that currently available detectors can be used for highly sensitive relative safeguards spectral measurements.

To meet the more stringent event rate goals imposed by smaller reactors and by absolute spectral measurements at large reactors, increases in event rates could come from the simplest expedient of increasing detector size. Reactor operator and IAEA needs for relatively small and non-intrusive detectors clearly limit this most obvious approach. However, we note that in future reactor construction, it may be possible to deliberately allocate enough space close to the core (but outside of containment) to accommodate multi-ton detectors. A 100 ton detector (including shielding) such as Chooz would fit in a cubic footprint 5 meters on a side. This is no larger than many stand-alone infrastructure elements already in place at reactor sites, such as water storage tanks.

Another way to increase event rates is to improve the intrinsic efficiency of the detector beyond the 30% level achieved in a practical detector by the Rovno group. The Chooz intrinsic efficiency of 70% came at the price of a large detector (a 5 ton target and a 100 ton shield), needed to fully contain the spatially diffuse gamma shower from capture of the neutron on the Gd dopant. Alternatives to Gd-capture are an interesting albeit challenging R&D path. For example, alpha emitting neutron capture agents will have a sub-millimeter extent of energy deposition following neutron capture, so that the total spatial extent of the antineutrino interaction (including displacement of the positron and neutron signals but not the 511 keV positron annihilation gamma-rays) is of the order several centimeters. Among other problems, this approach suffers from the fact that high linear energy transfer particles such as alphas have reduced light output relative to the Compton electrons created by gamma-interactions, so that energy thresholds for such detectors would have to be reduced, potentially increasing background rates. Nonetheless this is an important area of R&D for safeguards.

Another detector parameter affecting safeguards measurements is the overall systematic uncertainty in detector response. In an LEU-fueled PWR, the antineutrino rate changes systematically by about 0.6-1% per month in response to an ingrowth in plutonium of roughly 20 kg per month (assuming linear ingrowth for simplicity). An absolute detector response systematic uncertainty within a factor of a few of this value should suffice for safeguards measurements that meet IAEA timelinesss criteria. The total systematic detector response uncertainty in the Chooz experiment in France was 1.5%. Moreover, making use of a relative measurement will further reduce several systematic errors, including important detector-related uncertainties such as the uncertainty in the number of target protons. The forthcoming Double Chooz experiment has a set a target single-detector absolute systematic uncertainty of 0.6%[28].

### 5.5.3 Improving the Ease of Deployment and Operation

The SONGS and Rovno experiments have already demonstrated practical capabilities suitable for near term use in a safeguards context. Nonetheless, there are a variety of methods for improving the ease of deployment and operation, differing in impact and difficulty. Here we focus on the following areas that we judge to have the greatest likely impact for safeguards, ranked in order of difficulty:

- development of detectors with improved safety and deployment characteristics compared to current liquid scintillator.
- operating detectors at sea-level rather than underground, and
- shrinking the overall detector footprint including shielding.

*5.5.3.1 Improved safety characteristics*

SONGS2, described in section 5.3.3, has already demonstrated short-term monitoring capability using a plastic detector with improved safety characteristics suitable for safeguards applications. One area of useful further development is to improve the overall efficiency and energy resolution of plastic detectors. For example, it may be possible to dope plastic directly with Gd, allowing for homogeneous detectors and likely increasing the neutron capture efficiency relative to a sandwich design[51]. The key breakthrough needed is to maintain a long light attenuation length in the plastic scintillator while attaining the 0.1% Gd concentrations of interest for neutron capture. Alternative dopants such as Boron might also be considered.

Current liquid scintillator options still retain certain physics advantages relative to a sandwich design plastic detector like SONGS2. These include better neutron detection efficiency, lower cost per unit volume and long attenuations lengths. For this reason, in addition to plastic scintillator development, it is also useful to explore non-toxic replacements for the older generation scintillators. Recently, Gd-doped liquid scintillators such as [52] and linear alkyl benzene[53] with high flashpoints and reduced toxicity have been demonstrated by various groups. Their scintillation yield, attenuation lengths and long-term stability all appear suitable for use in safeguards applications, and development of safeguards detectors using these scintillators should also be pursued.

### 5.5.3.2 Above ground deployment

A more challenging technical problem arises when considering expanded deployment options. Not all reactors have underground galleries with sufficient overburden to accommodate an antineutrino detector. Above ground deployment therefore brings obvious advantages compared to previous reactor antineutrino detector installations. Detectors can be placed in a wider range of locations, and this freedom will make acceptance of the device easier for reactor operators, and can improve access to and control of the detector for safeguards inspectors. Further development along this path is both important for, and probably a unique requirement of, antineutrino-based cooperative reactor monitoring.

Above-ground detection is a challenge because of the backgrounds induced by cosmic rays.
The composition of the cosmic ray components at the earths surface (at sea level elevation) is shown in Table 5[54].

| Total Flux | Muons | Secondary Neutrons | Electrons | Protons, pions |
|---|---|---|---|---|
| $3 \times 10^{-2}$ cm$^{-2}$s$^{-1}$ | 63% | 21% | 15% | <1% |

**Table 5: Composition of the cosmic ray background at sea level. From [54].**

For inverse-beta detectors, the correlated event rate at sea-level increases substantially relative to a shallow detector. This is due in part to the presence of the strongly interacting hadronic component of the cosmic ray flux, which is removed with just a few meters of shielding. Furthermore, even at the relatively shallow deployment depths of the SONGS and Rovno experiments, the more penetrating muon fluxes are themselves reduced by factors of 5-10 relative to above ground cosmic ray backgrounds. Fundamental physics experiments have typically sought the conceptually simplest solution of reducing these backgrounds by burying the detector. In a safeguards context, the additional effort at above-ground background rejection, while a significant challenge, may be amply repaid by its deployment advantages.

For inverse beta detectors, above ground detection requires developing more sophisticated means for rejecting or screening out backgrounds, without inducing unacceptably high detector deadtime. A straightforward scaling of only muon rates from the SONGS1 detector 30 mwe depth to sea-level, would result in an unsatisfactory 40-50% detector deadtime. This scaling does not include the additional hadronic component present at sea-level.

The muon-induced veto rate could be reduced by shrinking the overall detector footprint. Since the same active detector volume is desired to maintain the signal rate, this implies a reduction in the passive shield. This might be possible by using segmentation, pulse shape discrimination, or other methods within the active volume to directly reject backgrounds in lieu of shielding. For example, reducing the shielding thickness of the SONGS1 detector from 0.5 meters to 0.25 meters would keep the sea-level muon trigger rates equal to those of the buried detector.

While space does not permit discussion of all approaches, less familiar designs, such as water-Cerenkov detectors and coherent scatter detectors, may also enable above-ground detection, by reducing sensitivity to backgrounds, or strongly boosting sensitivity to the antineutrino signal. We refer the reader to conferences[55,56,57] and literature[58,59] for further discussion.

### 5.5.4 Systems Analysis Needs for IAEA Safeguards and Other Applications

Worldwide, most of the research on antineutrino based safeguards is focused on detector development and improved understanding of the emitted reactor antineutrino signal. Less common but no less important is the analysis of the performance of antineutrino detectors in the context of IAEA safeguards and for other applications.

The study of specific diversion scenarios is a common methodological framework for evaluating the effectiveness of safeguards techniques. While some work as already begun in this area, especially for PWRs, much more study is needed to assess the benefits of antineutrino detectors for all cases of interest to IAEA. Performance against CANDUs, breeder reactors, research reactors of various design must be evaluated, as well as the effect of combining antineutrino-based metrics with other safeguards information. Understanding of the safeguards regime, along with a thorough command of the performance and limitations of antineutrino detectors is required for such analyses. Because of the newness of this technology to cooperative monitoring applications, antineutrino researchers and IAEA experts must work together to develop a more mature set of analytical tools and personnel capable of using these tools. A similar analytical framework is required to examine possible additional uses of antineutrino detectors outside of the current IAEA safeguards regime, for applications such as plutonium disposition, verification of a Fissile Material Cutoff Treaty, and others.

### 5.5.5 Worldwide Efforts To Develop Safeguards Antineutrino Detectors

25 years after the Russsian demonstrations at Rovno, and four years after the first IAEA experts meeting on this topic, there are now many efforts underway around the world to explore the potential of antineutrino based reactor safeguards. The evolution of these efforts is summarized in the agendas of annual Applied Antineutrino Physics (AAP) Workshops[55,56,57] At present, work is funded by a variety of national agencies acting independently, though there is frequent communication between the physicists involved at the AAP meetings. We conclude this section by summarizing worldwide activities in this burgeoning field.

*5.5.5.1 Effort in Russia*

As mentioned above, the concept of using antineutrinos to monitor reactor was first proposed by Mikaelyan, and the Rovno experiment[2] was among the first to demonstrate the correlation between the reactor antineutrino flux, thermal power, and fuel burnup. Several members of the original Rovno group continue to develop antineutrino detection technology, e.g. developing new Gd liquid scintillator using the LAB solvent. They now propose to build an improved cubic meter scale detector specifically for reactor safeguards[60], and to deploy it at a reactor in Russia.

*5.5.5.2 Effort in the U.S.A.*

A collaboration between the Sandia National Laboratories (SNL) and the Lawrence Livermore National Laboratory (LLNL) has been developing antineutrino detectors for reactor safeguards since about 2000. This group's focus is on demonstrating the feasibility of antineutrino based monitoring to both the physics and safeguards communities. This involves developing detectors that are simple to construct, operate, and maintain, and that are sufficiently robust and utilize materials suitable for a commercial reactor environment, while maintaining a useful sensitivity to reactor operating parameters.

Inspired by the GADZOOKS concept[61], the LLNL/SNL group is now investigating the use of Gd-doped water as an antineutrino detection medium. This method should be largely insensitive to the correlated background produced by cosmogenic fast neutrons that recoil from a proton and then capture. A 250 liter tank of purified water containing 0.1% Gd by weight was built to test this concept. An above-ground calibration with neutrons has clearly demonstrated the required sensitivity to neutrons, and to correlated events[58]. The detector has also been deployed in the SONGS tendon gallery. Data analysis in unshielded and passively shielded configurations is underway.

The coherent scatter process, described above, also holds promise for reactor monitoring since it has cross-section several orders of magnitude higher than that for inverse beta decay, which could eventually yield significantly smaller monitoring detectors. To explore this process, LLNL/SNL is currently collaborating with the Collar group of the University of Chicago in deploying an ultra-low threshold Ge crystal at SONGS, as well as investigating the potential of dual phase argon detectors for coherent scatter detection.[62]

*5.5.5.3 Effort in France*

**5.5.5.3.1 Double Chooz**

The Double Chooz collaboration plans to use the Double Chooz near detector ( about 400 meters from the two Chooz reactors, for a precision non-proliferation measurement[28]. The Double Chooz detectors will represent the state-of-the-art in antineutrino detection, and will be able to make a benchmark measurement of the antineutrino energy spectrum emitted by a commercial PWR. The MURE effort, described earlier, is being led by Double Chooz to improve the reactor simulations used to predict reactor fission rates and the measurements of the antineutrino energy spectrum emitted by the important fissioning isotopes. This work is necessary for the physics goals of Double Chooz, and will also improve the precision with which the fuel evolution of a reactor can be predicted.

#### 5.5.5.3.2 Nucifer

The Double Chooz near detector design is too complex and costly for widespread safeguards use. Therefore, the Double Chooz groups in CEA/Saclay, IN2P3-Subatech, and APC plan to apply the technology developed for Double Chooz, in particular detector simulation capabilities and high flash-point liquid scintillator, to the development of a compact antineutrino detector for safeguards named Nucifer[63]. The emphasis of this design will be on maintaining high detection efficiency (~55%), good energy resolution and background rejection. Nucifer will be commissioned against research reactors in France during 2009-2010, including the 70 MWt OSIRIS reactor at Saclay in 2009 and possibly the ILL reactor in Grenoble. This last location is particularly interesting as the fuel used in the ILL is 97% $^{235}$U. Following the commissioning phase, Nucifer will be deployed against a commercial PWR, where it is planned to measure reactor fuel evolution using the antineutrino energy spectrum.

### 5.5.5.4 Effort in Brazil

An effort to develop a compact antineutrino detector for reactor safeguards is also underway in Brazil, at the Angra dos Rios Nuclear Power Plant[64]. This may also be a precursor to a second generation theta-13 experiment.

Several deployment sites near the larger of the two reactors at Angra have been negotiated with the plant operator and detector design is well underway. The Brazilian work is particularly interesting, since a third reactor is being built at Angra, at which space may be reserved specifically for an antineutrino monitoring detector, and because of the regional safeguards presence (ABBAC, the Agencia Brasileiro-Argentina de Contabilidade e Controle de Materials Nucleares) in addition to the IAEA. Regional agencies such as ABACC often take the lead in the development and testing of new safeguards technologies.

### 5.5.5.5 Effort in Japan

A prototype detector for the KASKA theta-13 experiment has been deployed at the Joyo fast research reactor in Japan[65]. This effort is notable, since it is an attempt to observe antineutrinos with a compact detector at a small research reactor in a deployment location with little overburden, addressing two of the declared areas of major R&D interest for antineutrino monitoring.

# 6 Mid-Field Applications: Detecting and Monitoring Reactors From 1-10 km

## 6.1 Introduction

Detection and monitoring of reactors in the mid-field, between 1 and 10 km, is complicated primarily by the substantially reduced antineutrino flux compared to near-field monitoring. Depending on the outcome of current experiments[28,42] the phenomenon of neutrino oscillations may also affect mid-field cooperative monitoring, through a modest additional reduction in the observable electron antineutrino flux. Past oscillation experiments leave open the possibility of a systematic flux deficit due to oscillations as large as ~15% relative to the predicted flux assuming no oscillations[66].

In the mid-field, as for the near-field, we assume a cooperative monitoring regime which permits deployment of a detector at a suspect site. The aim may be to demonstrate that a country is not operating unknown illicit reactors, or that a known reactor or set of reactors is non-operational. Based on event rates and statistical considerations, these capabilities are the likely main focus of any future mid-field monitoring regimes. As shown below, precision power and fissile content monitoring require higher event rates than are likely to be accessible with practical detectors.

Of course, other methods exist for remotely verifying the operation or non-operation of known reactors, including thermal and visible wavelength satellite surveillance[67], monitoring of tritium releases and other radionuclides[68], and even actual destruction of infrastructure, such as was performed recently (and cooperatively) in North Korea[69]. Satellites and air or water borne radionuclides may also be used to search for unknown reactors. In these contexts, the advantages of antineutrino based monitoring in the mid-field are similar to those claimed for near-field monitoring: the signal is an inevitable and unique indicator of the presence of an operating reactor, and can't be masked except by other reactors, nor imitated by any source other than reactors. Therefore, aside from destroying the reactor, the other approaches mentioned are more susceptible to masking or spoofing than antineutrino-based monitoring. For example, a determined proliferator could divert heat from a small reactor with underground cooling or other means, while radionuclide releases are susceptible to weather patterns, or might be captured to frustrate detection.

In this section we consider detector characteristics and performance at the two extremes of the defined mid-field deployment range - one kilometer and ten kilometer standoff. We set signal and background performance targets and compare these with results achieved in existing experiments. With sufficient overburden, we show that existing technology allows deployment of detectors sensitive to 10 MWt reactors at one to ten kilometer standoff distances. We then discuss possible improvements in background rejection capability that might be obtained with more sophisticated detection methods.

## 6.2   Reactor Signal and Background in the Mid-Field

Following the prescription set forth in 3.1, we assume a reactor power of 10 MWt. For simplicity, we also assume that no other reactors are contributing to the background, although the analysis is easily modified to incorporate this possibility. At one kilometer standoff from a 10 MWt reactor, a 100 ton fiducial mass detector would detect approximately one event per day. At 10 kilometer standoff from a 10 MWt reactor, a KamLAND-like antineutrino detector[8] (1000 ton total mass, and 408 ton fiducial mass) would detect about one event per month. The Kamland detector is shown in Figure 18.

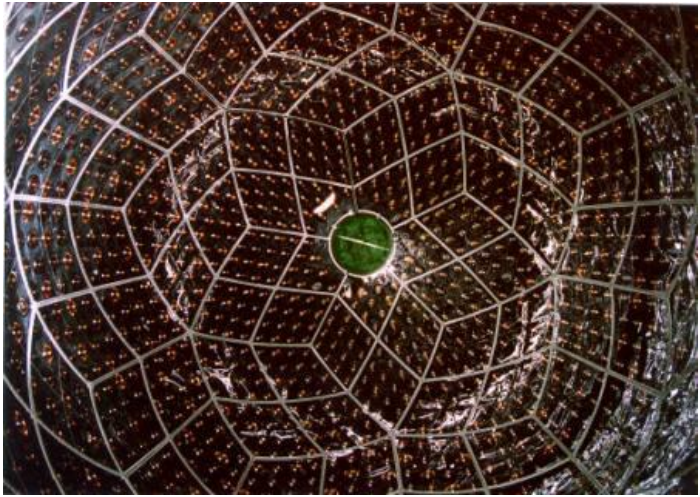

**Figure 18: The Kamioka Liquid Scintillator Anti-Neutrino Detector (KamLAND):  Buried 1 kilometer under earth, in an old mine northwest of Tokyo, KamLAND (Kamioka Liquid Scintillator Anti-Neutrino Detector) is the largest scintillation detector ever constructed. KamLAND is highly instructive for remote monitoring applications, since it unambiguously demonstrates long range reactor monitoring - albeit with reactor powers higher than those of likely interest in nonproliferation contexts. In particular, KamLAND clearly measures antineutrino flux from reactors in Japan and Korea at hundreds  of kilometer standoff.**

For both standoff examples, the intrinsic detection efficiency for antineutrinos is assumed to be 85%, similar to the demonstrated efficiency of the KamLAND detector. Assuming no observed events, a background-free detector with an expected signal rate of 1 event per day or month, would allow exclusion at the 95% confidence level of a reactor at one and ten kilometer standoff distances within three days or three months respectively. This conclusion remains valid as long as the expected background rate is comparatively low – for example, close to 95% confidence would still be achieved in the same time windows with 0.5 or fewer background events per day or month at 1 and 10 kilometer standoff respectively. The KamLAND reactor antineutrino experiment demonstrates that detectors of the necessary size and background rate for the entire range of mid-field detection can be built with current technology.

To further understand the performance and scaling properties of mid-field reactor monitoring detectors, it is useful to consider the KamLAND backgrounds in more detail. KamLAND background rates from two different analyses are summarized in Table 6. The CHOOZ detector background rates are also shown for comparison[7].

| Detector | Fiducial Mass (ton) | Overburden (mwe) | Total background rate | Correlated muogenic background rate | Accidental background rate |
|---|---|---|---|---|---|
| KamLAND '03[8] | 408 | 2700 | 0.4 per month | 0.4 per month | 0.002 per month |
| KamLAND '06[70] | 706 | 2700 | 5.5 per month | 4 per month | 1.6 per month |
| CHOOZ | 5.5 ton | 300 | 2 per day | 1.8 per day | 0.5 per day |

**Table 6: Background rates achieved in the CHOOZ and KamLAND detectors.**

As shown in Table 6, backgrounds may be divided into correlated (muogenic) and uncorrelated or accidental backgrounds, with the accidental backgrounds arising primarily from radioactive elements within and surrounding the detector. At KamLAND depths, backgrounds are seen to be dominated by the correlated muogenic events, which include spallation neutrons and various long-lived activation products. The relative suppression of accidental backgrounds is especially pronounced if a severe fiducial mass cut is made, as was done in the first (2003) KamLAND analysis. Below a few meters of overburden, after which the hadronic components of the cosmic ray background are screened out, backgrounds scale primarily with the underlying muon rate. Since higher backgrounds than were achieved in KamLAND can be tolerated at kilometer standoff distances, the most expedient simplification relative to the KamLAND design is to bury the detector at a shallower depth, in order to reduce excavation costs.

For example, in order to achieve a target background suitable for 1 km monitoring, of approximately 0.5 event per 100 ton mass per day, the required depth is about 600 meters (1600 mwe), which is about the depth of the neutrino experimental halls at the Waste Isolation Pilot Project plant (WIPP)[71] site in Carlsbad, NM. Figure 19 shows the scaling of muon flux for various experiments as a function of depth.

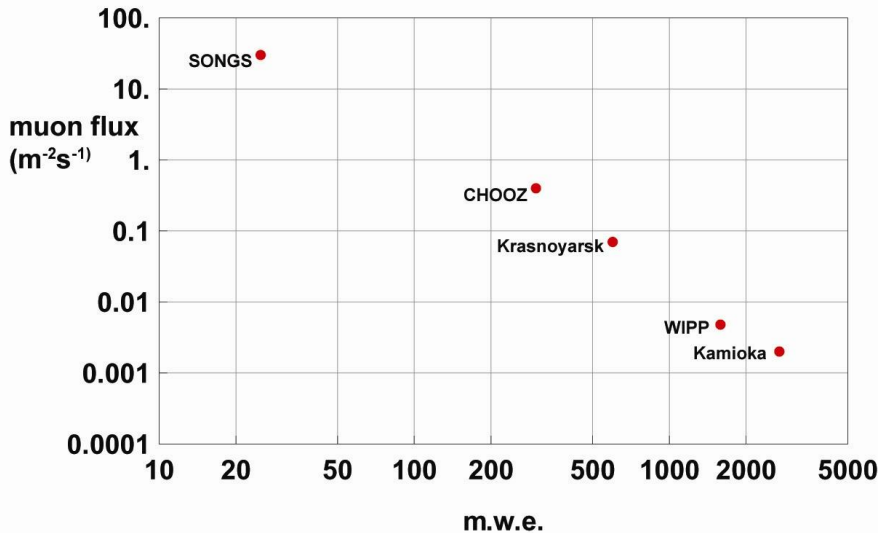

**Figure 19: The muon intensity per square meter and second at various underground experimental sites worldwide, plotted versus the depth of the experiment in m.w.e.. Kamioka is the site of the KamLAND antineutrino detector. WIPP is the Waste Isolation Pilot Project underground site in Carlsbad, N.M. Further information on the other experiments can be found in Table 4.**

## 6.3 The Effect of Reduced Overburden on Backgrounds and Detector Performance

Throughout the 1-10 km mid-field range of interest, existing and proven designs such as KamLAND could be adapted for cooperative monitoring purposes with little or no modification, provided only that suitable underground deployment sites were available. However, excavation costs makes up a large or even dominant fraction of the total cost of buried detectors. For example, the Braidwood neutrino oscillation experiment estimated a 23 M$ construction cost of four detectors comprising 260 tons fiducial mass, compared with a 35 M$ cost for two shafts and experimental halls at a depth of 450 mwe[72]. Therefore, for reasons of expense and ease of deployment, it would be preferable to deploy on or near the Earth's surface. In shallow or surface deployments, the detector would have to be designed to allow rejection of the large backgrounds due to cosmic radiation. This problem of above-ground or shallow-depth detection is similar to that discussed for the near-field in Section 5.5.3.2, but is made considerably more difficult by the larger detector size and reduced event rate compared to near-field deployments. It is also important to note that while reducing overburden is a strong cost consideration, the channel count and complexity of the detector can also considerably raise construction and operating costs. The following discussion is meant to explore some of the detector-related considerations in this trade-off, rather than prescribing a specific path to deployment. In some cases, burial will remain the simplest and most cost-effective approach.

The background radiation at or near the surface of the earth is composed of cosmic and terrestrial components. The terrestrial component, arising primarily from decays of uranium, thorium and potassium and other radionuclides, depends strongly on the composition of local materials, but does not differ much from that in existing underground antineutrino detectors. For inverse-beta detectors, the relative rates from KamLAND and Chooz in Table 6 show that internal backgrounds can be kept at or below the level required for mid-field monitoring. While care must be taken to ensure pure materials and to control contaminants such as radon, the ambient radioactive component of the background does not present any unsolved technical challenge, and we need not consider this background further. The muonic component of the cosmic background at shallow depths is increased by about 3-4 orders of magnitude relative to the KamLAND depth. This steep dependence on overburden is seen in Figure 19. In addition to the increased muon rate, a significant change in the character of the cosmogenic backgrounds occurs in the last 5-10 mwe of overburden near the Earth's surface. As shown earlier, only about 63% of the cosmic ray flux at the earths surface comes from muons – the rest comes from neutrons, electrons/gamma-rays, and other hadrons. Below roughly 5 mwe, the muonic component of the backgrounds dominates the total flux, since most of the hadronic flux has been screened out. In this circumstance, the cosmogenic antineutrino-like background arises primarily from secondary fast neutrons induced by these muons as they pass through the detector and nearby materials. Above 5 mwe, temporally complex and spatially extended cosmic ray showers, arising from hadronic interactions further complicate the background rejection problem.

In small detectors, the charged components of the cosmic ray background (muons, electrons, protons and pions) are relatively easy to suppress using thick passive shields or veto techniques, while maintaining good signal efficiency. For 100 ton or 1000 ton scale detectors, however, the need to veto signals arising from typical cosmic ray fluxes of $1-2 \times 10^2$ m$^{-2}$ s$^{-1}$ , with veto windows having typical durations of 100 microseconds or more, quickly increases dead-time to intolerable levels. For example, a cubic 100 ton scintillator detector at the Earth's surface would have a 65% dead-time arising from muons alone, assuming a 100 microsecond veto window. A cubic 1000 ton detector would have 100% dead-time with the same veto window.

Part of the background in a surface detector arises from the secondary neutrons produced by hadronic interactions in the atmosphere, and by muon interactions with surrounding materials. At sea level, the cosmogenic secondary neutron flux is some 100 times greater than it would be after only about 4 m of rock overburden. These high background rates confirm that the leading problem in designing a surface or near-surface midfield detector is to suppress cosmogenic backgrounds in the detector, and that simple veto strategies will not suffice due to the dead-time these strategies incur.

## 6.4 R&D needs for mid-field detectors

Here we describe some possible approaches for reducing backgrounds at shallow depths, while maintaining tolerable deadtime. The most promising directions are:

- reducing dead-time by vetoing only segments of the total detector rather than the entire detector.
- improved identification of the particle types for the initial and final state positron and neutron.

One way to achieve both goals is with a highly segmented detector, and the use of two types of scintillator – the first sensitive primarily to neutrons, and the other to electromagnetic interactions. Segmentation keeps the cosmic veto rate local and manageable, thereby reducing the dead time effects, since only small sections of the detector need to be vetoed following a background interaction. Judicious choice of segmentation will also allow rejection of the dominant fast neutron signals in the detector. The use of two scintillators allows event by event particle identification, providing more specific identification of the positron-neutron pair that characterizes the antineutrino interaction. Below we provide one possible design that serves to illuminate some R&D directions of interest.

### 6.4.1 A possible detector design for improved background rejection

One possible design for a segmented detector is shown in Figure 20.

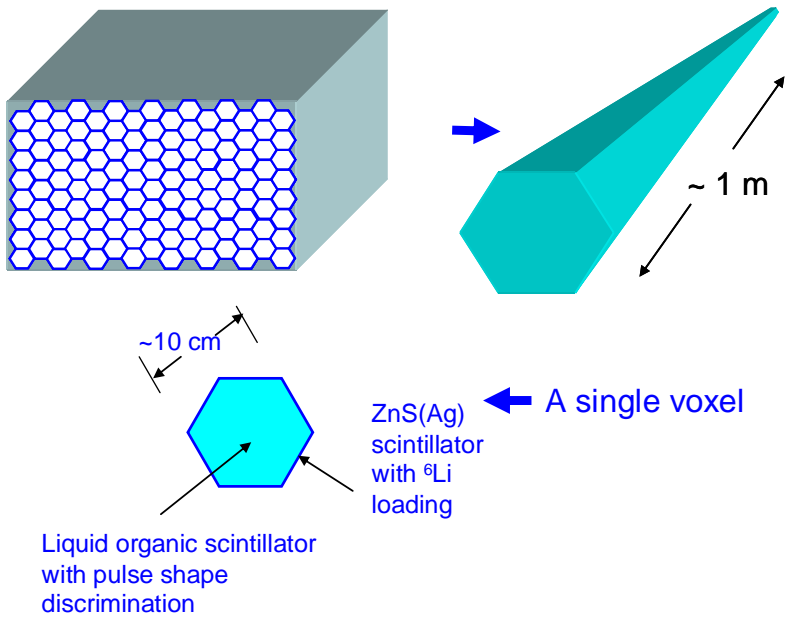

**Figure 20: A possible design for a segmented detector element incorporating particle identification features.**

The design consists of a series of liquid scintillator cells, whose walls are coated with a neutron-sensitive scintillator such as $^6$LiZnS(Ag). Because ZnS is an inorganic scintillator, the signal produced by neutron absorption (a triton, $^3$H, and alpha particle with a total Q value of 4.8 MeV) is not heavily quenched, as it would be in an organic scintillator. As a result, the light signal produced by neutron interaction is very large. Conversely, the range in the scintillator of the alpha and triton are very small (< 10 microns), and it is possible to dope ZnS with sufficient concentrations of $^6$Li such that efficient thermal neutron absorption will occur in relatively thin layers. Since it would be nearly impossible for a gamma-ray to deposit more than a few 100 keV in a thin (< 1mm) ZnS layer, the net effect is that ZnS is a neutron-only detector. Furthermore, it is a very efficient neutron only detector, since all of the reaction products of neutron interaction are contained in a thin layer. This situation contrasts with that for the more common homogeneous gadolinium-based antineutrino detectors, in which the neutron-induced signal, consisting of a gamma-ray cascade following capture, is not specific to particle type. Moreover, unliked gadolinium-doped detectors, the localized nature of the interaction in the inorganic scintillator allows for complete containment of the event within a single voxel.

### *6.4.1.1 The antineutrino signal*

The basic principle behind this segmented design is to create distinct and physically localized sets of signals for the antineutrino interaction and for each type of background. We first describe the essential features of the antineutrino interaction. When an antineutrino interacts via inverse beta decay with one of the protons in the liquid scintillator it will produce a positron (with a stopping distance of approximately 1 centimeter or less in the liquid) and a neutron. With a judicious choice of voxel dimensions (say 5 centimeter in diameter), the positron will slow and annihilate in a single voxel. The positron will produce a fast decaying "electron-like" light pulse in the liquid scintillator which is distinguishable via pulse shape discrimination from the slower decaying light pulse produced fast neutron recoils. The antineutrino-induced neutron will slow and be captured by the $^6$Li in the ZnS scintillator on the walls of the very same voxel in which positron energy deposition occurred, because the length scale for the slowing of a positron is comparable to thermalization length for the neutrons produced in inverse beta decay. Thus, the unique signature of an antineutrino interaction in this detector is an electron-like event (as determined by pulse shape discrimination) followed by the capture of a neutron in an adjacent detector wall.

### *6.4.1.2 Fast neutron interactions*

In a conventional homogenous scintillator detectors, fast neutrons induce proton recoils in the organic scintillator, then thermalize and capture with nearly the same statistical behavior in time as the neutron produced in inverse beta decay. Thus, the signature of a fast neutron may closely mimic an anti-neutrino event in such detectors. However use of pulse shape discrimination in a ~5 cm voxel should allow reliable differentiation between proton recoils and electron-like events, with a resolving power of about 1 in $10^4$ in the best liquid scintillators. Furthermore, it is very unlikely that a fast neutron will lose all (or even most of its energy) in a single scattering event, so that most neutrons will undergo scattering events above threshold before thermalizing over a length scale of tens of cm. Thus, segmentation combined with pulse shape discrimination should effectively discriminate fast neutron interactions from true inverse beta decay antineutrino interactions.

### *6.4.1.3 Muon and fast electron interactions*

In our proposed detector design, muons (and other minimum ionizing events), would be identified by the topology of their energy deposition within the detector. We would expect muons to travel relatively undeflected through a large number of detector segments. As noted earlier, we would deliberately avoid use of a conventional muon veto system where a few large paddles provide muon rejection, because such a system would have intolerable dead time effects in a near surface installation. Instead, only a subset of voxels would be paralyzed during the passage of a muon.

### *6.4.1.4 Long-lived cosmic-ray activation products*

As occurred in the KamLAND detector, $^9$Li and $^8$He can be generated by cosmic rays, then beta-decay in associated with a delayed neutron emission from the beta delayed daughter. The decay time constants are 178 ms and 119 ms respectively. The combination of MeV-scale beta emission with the time-correlated neutron can produce an antineutrino-like signal. These backgrounds require further study, but might be rejected without fully vetoing the detector because they will occur within a few centimeters and milliseconds of the associated reconstructed muon track in the detector.

### *6.4.1.5 Gamma-ray backgrounds*

As mentioned earlier, ambient gamma-ray backgrounds are similar for a near surface detector as they are for an underground detector and do not present a more serious background at the surface than underground. Nonetheless, gamma-rays in combination with the high neutron background at the surface could present special difficulties due to accidental coincidences between these independent event classes. Further study of these rates in specific detector designs is required.

## 6.5   Summary for Mid-Field Applications

In this section we reviewed some of the technical issues associated with deploying a large inverse beta decay detector at or near the surface that would be useful in a mid-field detection scenario (1-10 km). We conclude that existing detectors could be used for discovery or exclusion of small reactors throughout the mid-field, provided they are relatively deeply buried (100s of mwe).  To reduce costs associated with excavation, we considered detectors with improved background rejection capabilities for the dominant cosmogenic backgrounds, which increase quickly towards shallower depths. Since at these depths local detector shielding alone is insufficient for reasons of deadtime, segmentation techniques appear to offer promise for improving specificity for the antineutrino signal, reducing deadtime and backgrounds, and thereby reducing overburden requirements in the mid-field. While segmentation offers a logical and promising R&D direction, we note that we have not examined other background rejection techniques in detail, such as development of new materials or readout systems capable of improving particle identification in unsegmented detectors.

# 7  Far-Field Applications: Detecting Reactors and Explosions at 10 – 500 km

The distinguishing features of far-field applications are:

1. Detector sizes are tens of kilotons for tens of kilometer reactor standoff distances, and tens of megatons for hundred kilometer distances, and 100 megatons for 500 kilometer distances;
2. Event rates at the level of a few per month for the small reactors of likely interest, even in very large detectors;
3. Neutrino oscillations must be taken into account;
4. Detector related backgrounds, and real antineutrino backgrounds from other reactors play a more important role; and,
5. Unlike near-field and mid-field applications in which there are examples of detection capability down to 10 MWt reactor power, no antineutrino detectors have been built larger than the 1000 ton KamLAND detector, nor neutrino detectors larger than then 50 kiloton Super-Kamiokande detector[d]. However, instruments at the 100 kiloton to megaton scale are now being considered or proposed to study a wide range of fundamental physics topics.

The overlap of this work with fundamental neutrino and dark matter physics is discussed in Section 8. Here we discuss the state of the art in antineutrino detection relevant for far-field monitoring, and consider specific monitoring examples. We then examine the prospects for fission explosion monitoring, and conclude with a review of the necessary R&D paths for far-field nonproliferation applications.

---

[d] As discussed in this section, large Water Cerenkov neutrino detectors such as Superkamiokande must be modified to make them sensitive to the antineutrino signature.

| Goal | Required Antineutrino Event Rate | #/Yr | Detector Mass | | |
|---|---|---|---|---|---|
| | | | *10 KT* | *1MT* | *100 MT* |
| **Detect Operation ~1yr** | ~5 /yr | ~5 | 70 km | 800 km | >>1000 km |
| **25% accurate estimate of total annual energy** | 16 /yr | 16 | 35 km | 400 km | >>1000 km |
| **Detect reactor shutdown within ~1 week** | > 10/day | 3600 | 6 km | 60 km | 600 km |

**Table 7:** *Required fiducial detector masses for several possible remote reactor monitoring goals for 10 MWt reactors. 1000 ton scale liquid scintillator antineutrino detectors (KamLAND, Borexino) exist now; 10,000 ton (10 Kton) detectors are a straightforward extrapolation of this technology and are now being developed by several groups worldwide. 1,000,000 ton ( 1 Megaton) Water Cerenkov detectors are also being considered by several groups for fundamental physics studies. 100 Megaton detectors are not currently being considered. For the first two options, the background is assumed to be 1 event per year from all sources. Relative to the KamLAND detector, this would require background suppression by factors of 10, 100 and 1000, increasing with detector size. Since cosmogenic backgrounds dominate, suppression could be achieved by burying the detectors at 3 km water equivalent, compared to the 2 kilometers water equivalent depth of the KamLAND experiment.*

We begin our discussion with reactor monitoring. The example introduced in Section 3.1 sets the required detector scale: exclusion of the presence of a 10 MWt reactor within an 800 kilometer radius, *with no other reactors present*, would require a one megaton water Cherenkov or liquid scintillator detector, with backgrounds suppressed by a factor of 100 compared to the KamLAND detector. As seen in Table 6 and discussed below, it is important to note that the dominant source of backgrounds derives from cosmogenic activation, which can be suppressed by additional overburden. While large, this detector mass could be achieved by building about three modules of about 600 kilotons, as presently proposed for various physics purposes (and discussed in Section 8). A Super-Kamiokande[29] sized detector (50 kTon), modified to be sensitive to antineutrinos, could measure the integrated power output of a 10 MWt reactor with 25% statistical precision, out to a distance of about 100 km.

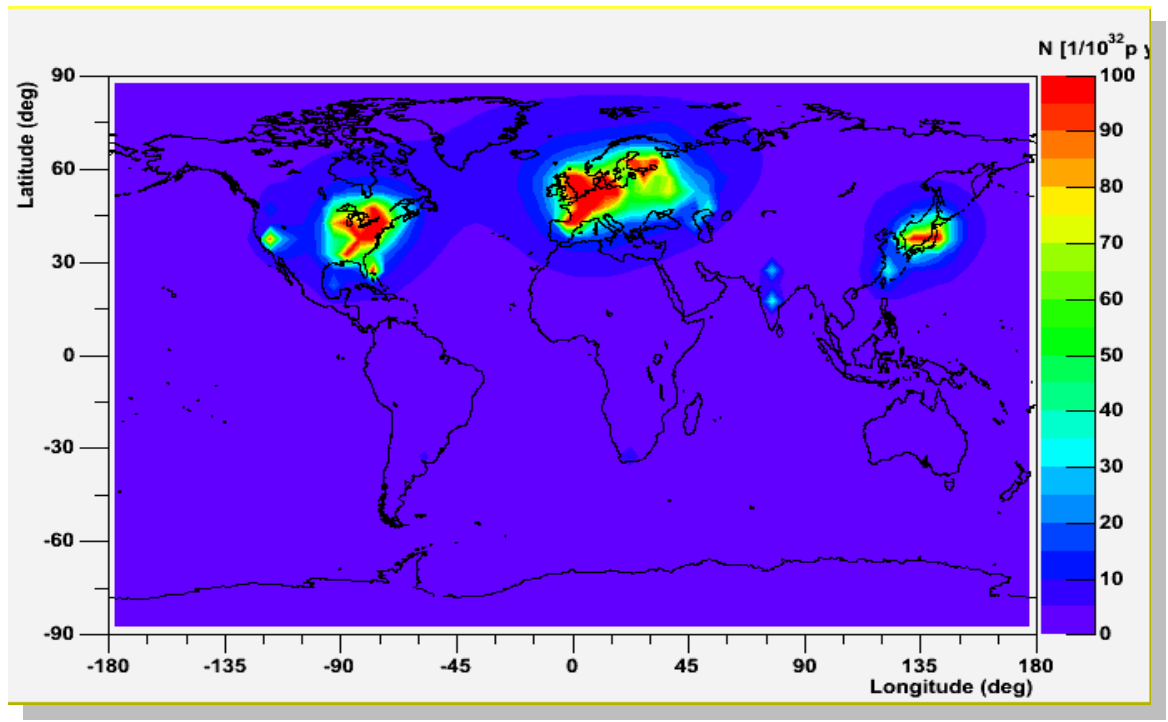

**Figure 21: The *predicted* average antineutrino flux worldwide arising from known nuclear reactors. The color contours show the antineutrino interaction rate per $10^{32}$ proton per year, or about that of a 1000-ton detector per year. The red indicates areas where there are dense concentrations of reactors in France, the eastern United States, and Japan. Data are modeled based on known reactor powers, and are not from antineutrino flux measurements.**

The large scale examples above (1 and 100 Megatons) assume successful suppression of non-antineutrino backgrounds by 2-3 orders of magnitude relative to the current state of the art, *and* no reactor antineutrino backgrounds. In fact, reactor antineutrino backgrounds vary widely across the globe. Figure 21 shows the global antineutrino fluxes from all known reactors worldwide. These fluxes are an additional background beyond the detector-related backgrounds discussed in Section 6.2. Absent a directionally capable detector (discussed below in 7.6), they are irreducible without resorting to multiple detectors. An exception to this rule discussed in arises if the number of collected events in the detector is sufficiently high, in which case the specific signature due to neutrino oscillations can be used to provide some range information for a distributed set of reactors.

North Korea, South Korea, Japan and France would clearly be difficult locations for reactor monitoring due to backgrounds from large numbers of power reactors. However, monitoring in developing countries may have no significant local reactor antineutrino contributions (as for example in Africa and generally in the Southern hemisphere.) The reactor-related backgrounds can thus be quite small out to standoff ranges of hundreds of kilometers. Even where these backgrounds are considerable, under many circumstances subtraction of known reactor signals is possible with a few percent accuracy, Power declarations for known reactors are currently required for IAEA safeguards. The same accounting could be performed more directly and with increased reliability if local antineutrino monitoring were available at each reactor.

For most of our long range monitoring examples, the detector mass has been set to provide the minimum number of events possible to determine reactor presence or operational status. With the higher event rates made possible by very large detectors (tens of megatons) in the few hundred kilometer range, we can in principle take advantage of neutrino oscillation phenomena to separate the signature of closer target reactors from those background power reactors at greater distances. In the next section, this idea is illustrated with the hypothetical, topical and difficult example of a hidden reactor in North Korea.

## 7.1 Discovery of an undeclared reactor in North Korea, including backgrounds from reactors in South Korea, using antineutrino rate information

Having introduced an example in the previous section incorporating only detector-related backgrounds, we now consider remote monitoring of North Korea. North Korea is chosen as an enduring and well known proliferation problem, and as an especially vivid indication of the confounding problem of reactor related backgrounds. We emphasize that many places in the world would allow deployment of the few hundred kiloton to megaton scale detectors discussed above, due to the substantially reduced reactor-related antineutrino backgrounds and more favorable siting requirements. The example described here concerns only standoff, reactor power and detector performance. We make no assumptions about cooperation from North Korea or neighboring states nor the many other practical obstacles that would have to be overcome for such a deployment to occur. We assume that cosmogenic backgrounds are reduced by burial of the detector at 50% greater depth than currently deployed KamLAND detector, and that either liquid scintillator technology could be used, or that background suppression in water Cerenkov detectors is achievable at the same level as KamLAND. The latter assumption in particular is far from being demonstrated.

To reduce as much as possible backgrounds from the large installed capacity in South Korea, we assume that a large and deeply buried detector could be built in southern China, near the North Korean border. The deployment scenario is shown in Figure 22. The detector size of 10 megatons allows 8 standard deviation confidence of detection of a sole 10 MWt reactor in Yongbyon above backgrounds, where both detector-related *and* other reactor backgrounds are accounted for. Reactor backgrounds are assumed to be fully reported by neighboring cooperative states, in this case primarily South Korea (with a small contribution from Japan). Exploitation of correlations amont rates in a smaller (Megaton scale) array of detectors would actually serve the purpose even better. The detector related backgrounds are scaled from the KamLAND experiment (2003 analysis, see Table 6 above), The background estimate includes the suppression effects of recent purification efforts at KamLAND, which reduce radon and associated backgrounds and also assume 50% greater depth (3 km water equivalent, or about 1 km of rock, compared to the actual KamLAND depth of 2 km water equivalent). Total backgrounds are estimated at about 1% of the remote reactor signals. The estimated power reactor generated backgrounds are based on known South Korean reactor thermal power ratings. One may also assume cooperative reporting of daily power production by these reactors, which aids in background subtraction.Table 8 shows the expected signal and background rates.

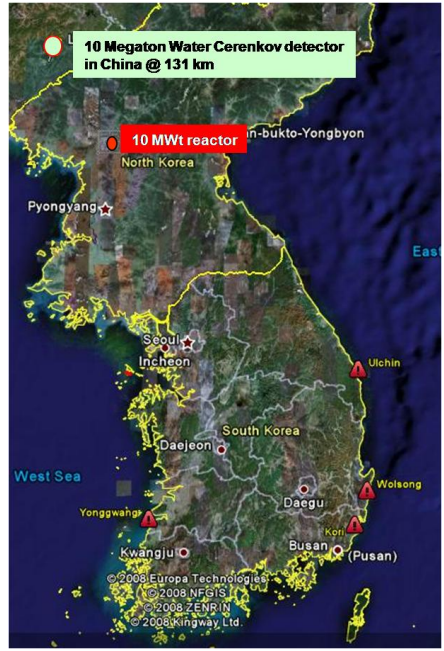

**Figure 22: A hypothetical deployment sensitive to a 10 MWt in Yongbyon, North Korea. The size of the detector (or array of detectors) is 10 Megaton. A one year dwell time would achieve an 8 sigma measurement of a 10 MWt reactor at the indicated location. Included in the calculation are backgrounds from all South Korean reactors, with sites in South Korea indicated by the orange triangles, as well as detector related internal and external backgrounds, using a rate per unit mass extrapolated from the KamLAND experiment.**

|  | *Annual rate in a 10 megaton detector at 4 kmwe depth* |
|---|---|
| Yongbyon 10 MWt reactor | **1900 events** |
| background from ~38 GWt of S.K. reactors | 185,000 events |
| Cosmogenic and internal backgrounds | 12000 events |
| Fluctuations in total background | **450 events** |
| Statistical significance after one year | **~ 4 S.D.** |

**Table 8: The signal and background rates in a hypothetical 10 Megaton detector deployed at 131 km standoff from an unacknowledged 10MWt reactor in Yongbyon, North Korea. The depth is chosen to make cosmogenic backgrounds negligible compared to reactor backgrounds. The statistical significance is determined solely by counting statistics. Antineutrino oscillations provide greater resolution, as discussed in the text.**

Not included in this rate analysis is the time variation of known backgrounds, such as the South Korean power reactors. In the KamLAND experiment, the predicted neutrino flux varied by about a factor of two over several years timescale, as reactors went down for service and due to problems. This temporal signature further strengthens the analysis, but has not been employed in the present simulations.

## 7.2 Discovery of an undeclared reactor in North Korea, including backgrounds from reactors in South Korea, using spectral information

Neutrino oscillations, which change the type or 'flavor' of the antineutrino as a function of distance and energy, have been observed with reactor antineutrinos and are discussed in Chapter 8. When one has an expected signal in the range of a thousand or so events, employment of neutrino oscillations provides a powerful tool to further distinguish signal from backgrounds, even when the total number of signal events is small (~ percent) compared to background. The detector, still 10 Megatons in this example, would have to achieve energy resolution comparable to the KamLAND detector, so that a water based system would probably not be feasible. Nonetheless it is interesting to consider the spectral analysis technique in the far field.

There are many ways to do this analysis. Here we employ optimal filtering in the energy distribution. With a known reactor antineutrino background, including the now well measured oscillations which distort the antineutrino energy spectrum, an optical filter or a correlation function can be convolved with the observed spectrum. This filter will yield a peak at the distance of the "unknown" reactor, with an amplitude proportional to the reactor power.

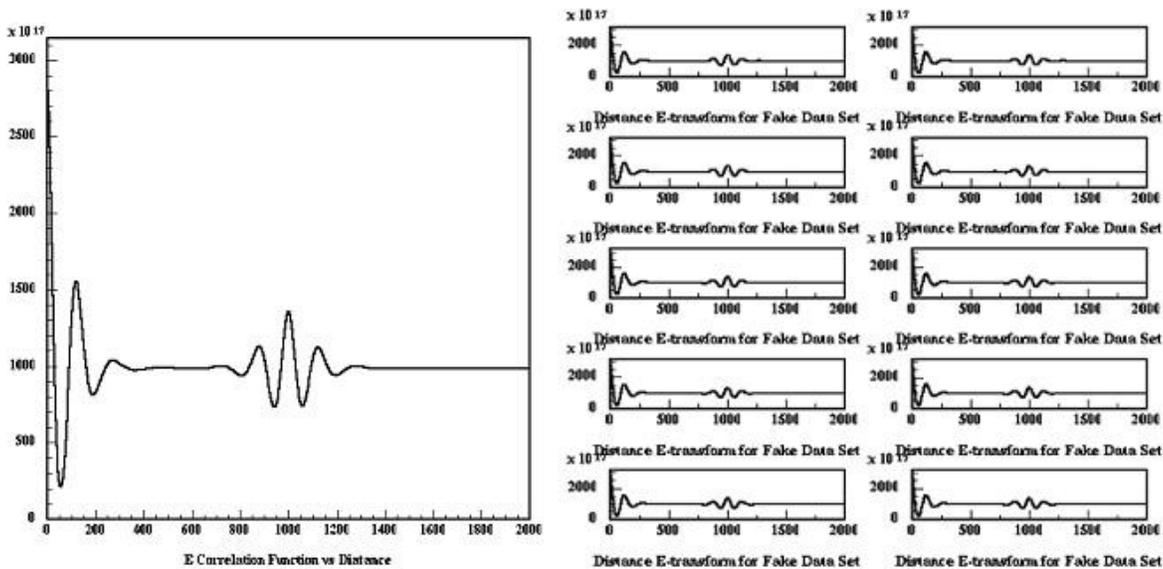

**Figure 23: Left:** *Correlation function versus distance for the hypothetical example of a 10 MWt reactor at 131 km distance and 10 GWt background reactors at 1000 km distance. One clearly sees the signatures of both background reactors and the "hidden" reactor at the unknown 131 km distance, with the ranging accurate to a few kilometers. Right: Ten examples of Monte Carlo simulations of one year of data in the example above. This plot demonstrates that with these high count rates, statistical fluctuations are negligible.*

These correlation functions can be further improved. The shape of the signature versus distance is exactly equivalent to a point spread function (PSF) in, for example, radio astronomy. Techniques have been refined to deconvolve the PSF, using for example the 'maximum entropy method'[73] or the 'CLEAN' algorithm[74]. The latter seems particularly attractive for this application as it allows sequential subtraction of the peaks, revealing underlying detail.

The reactor distribution in this example is slightly simplified compared to reality, since South Korean reactors are distributed in at four main distances, rather than a single distance, as discussed here. Nonetheless the example illustrates the additional power provided by spectral filtering. Not yet exploited are the opportunities for inclusion of directional information (discussed in Section 3.1.6). While we consider these various effects here separately (counting statistics, backgrounds with time dependence, range dependent spectral distortions, and directionality), a real monitoring program would employ all simultaneously in a Maximum Likelihood approach, squeezing as much information as possible from the data set. It is difficult at this time to say how much this will add in analysis power, but 40-50% improvement seems reasonable.

In summation we have shown that with next generation scientifically motivated instruments in the one megaton class (see Fig. 2), one could in principle monitor small reactors in the 10 MWt class out to ranges of order of 800 km, and at ranges of a few hundred kilometers make more detailed assessments of operations. These statements assume that the ambitious energy resolution, background rejection and underground mine engineering goals of these detectors are all realized. The ongoing Conceptual Designs efforts related to these detectors will aid further and more detailed assessments of prospects for reactor monitoring. In the following sections, we discuss planned detectors that begin to approach scales of interest for far-field monitoring.

## 7.3 Next Generation Large Liquid Scintillator Detectors

In Chapter 6, we introduced the 1000 ton KamLAND liquid scintillator detector, in which current background levels allow sensitivity to 10 MWt power reactors throughout the entire 1-10 km mid-field range, even extending partially into the far-field (out to 30 km). Next generation scintillator antineutrino detectors will likely be built on a scale roughly 10 times bigger. An example[e] is the proposed 30,000 ton submersible Hanohano detector[75], a schematic of which is shown in Figure 24. Assuming a fiducial volume of 10,000 tons, (extrapolated from the KamLAND self-shielding radius) and scaling rates from KamLAND estimates, Hanohano would have a 95% confidence exclusion range of about 70 km for a 10 MWt reactor in one year, assuming total *non-reactor* backgrounds comparable to KamLAND (including removal of radon in the latter experiment). However, in these deeply buried detectors, the non-reactor backgrounds depend primarily upon depth. Backgrounds from cosmic rays will be totally negligible below 4 km depth, and may be manageable at shallower depths. For example, the KamLAND detector operates at about 2.7 km water equivalent depth.

The difficulty of background rejection increases at shallower depths, and this is an important element of future research and development, as discussed below.

---

[e] There are several multi-kiloton scale beingproposed antineutrino detectors worldwide. Hano-hano is unique in that it is submersible rather than being deployed in a deep mine.

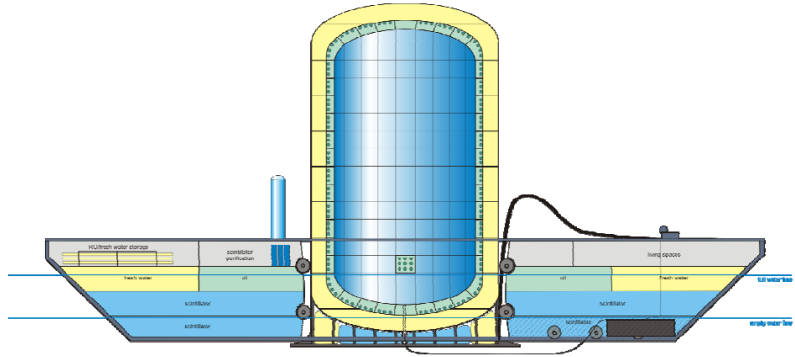

**Figure 24:** The proposed ocean-going 10 kiloton liquid scintillator detector Hanohano. Shown in cutaway profile, the cylindrical detector will be transported by a 100 meter long barge and lowered to depths up to 5 km for data taking. The detector is designed for study of neutrinos from reactors and measurements of geoneutrinos from the Earth's mantle. It is mineral oil based, and is surrounded with large photo-detectors. While the ocean-going engineering presents challenges, the detection technology is a straightforward extrapolation from the operating KamLAND instrument in Japan, and other detectors.

As discussed in Section 8, the Hanohano detector is to be built primarily for antineutrino physics, geophysics and astrophysics studies. It may be the first experiment after KamLAND to demonstrate remote detection and monitoring capability for a single reactor, including measurement of the reactor operational status and power at distances of 50-100 kilometers. The Hanohano design has particular interest for nonproliferation applications because of its flexible deployment platform, which allows submersion of the detector at various locations and depths in the world's oceans. It will also give experience in the sort of large detectors that will ultimately be needed for nonproliferation, and will help further develop the expertise and the scientific personnel to carry out remote monitoring tasks of the future.

An example deployment would be 55 km offshore of the San Onofre Nuclear Generating Station. (This distance was chosen to allow maximum sensitivity to neutrino oscillation parameters in a fundamental physics experiment.) At this distance, Hanohano would collect roughly 25 events per day for both reactors, which would easily allow determination of the operational status of either reactor within one day. While of little utility for nonproliferation, the example illustrates that relatively high statistics are achievable even at tens of kilometer standoff in next generation detectors.

## 7.4 Next Generation Large Doped Water Cerenkov Detectors

Beyond distances of a few tens of kilometers, and above detector masses of the order of 100 kilotons, the detection technology must almost certainly change from liquid scintillator to using a water base. The liquid scintillation medium has the advantages of being able to work with lower energy neutrinos (lowering the threshold from roughly 4-5 MeV in water Cherenkov detectors down to the inverse beta decay kinematic limit of 1.8 MeV), and to have roughly ten-fold better energy resolution. One central consideration is cost:

photomultiplier tube coverage costs scale with the linear dimension of the detector squared, while the proton target mass goes as the cube. Hence the cost of the organic liquid scintillator alone (at roughly $1/kg) will begin to dominate and eventually become prohibitive as one scales the size upwards. The cost of scintillator for a 100 kiloton detector is about 100 million dollars. Assuming 40% PMT coverage, and a cost of 10,000 dollars per square meter of PMT, the PMT cost for the same detector would be about 40 million dollars. Therefore, water (at about 1% of that cost of scintillator) is the preferable target material for very large detectors.

While large water Cherenkov detectors have been built and detected MeV-scale neutrinos, no large water-based detector has yet demonstrated reactor antineutrino detection. The largest presently operational low energy *neutrino* detector based on water Cherenkov technology (leaving aside ice Cherenkov technology for the moment, see below) is Super-Kamiokande[29], shown in Figure 25. It has 50 kilotons gross and 22 kilotons fiducial volume and a present energy threshold of about 4.5 MeV. It has been operating for over a decade (since 1996) and its operation is well understood. Studies of water Cerenkov detectors on megaton mass scales have been put forward in Japan[76], the United States[77] and Europe[78]. These or others may be constructed for neutrino physics studies in the next decade, with projected costs in the range of $500 – 1000 M.

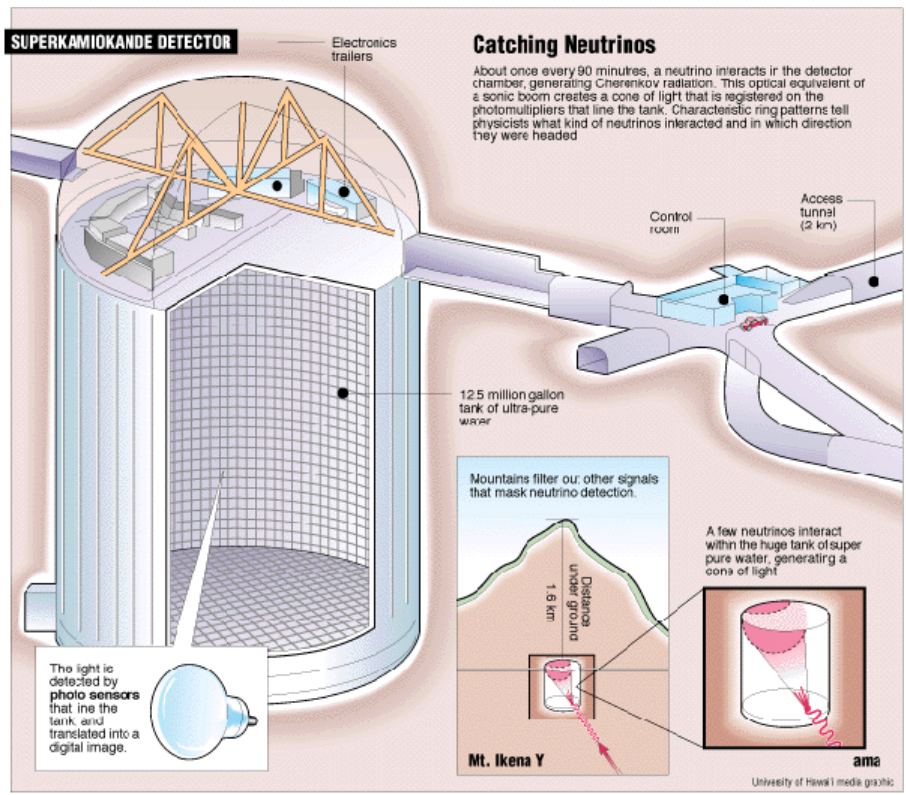

**Figure 25 Super-Kamiokande is a joint Japan-US large underground water Cherenkov detector operating in Japan. The observatory was designed to search for proton decay, study solar and atmospheric neutrinos, and keep watch for supernovas in the Milky Way Galaxy. Super-Kamiokande is located 1,000 m underground in the Mozumi Mine (Kamioka Mining and Smelting Co.), Gifu Prefecture, Japan. The detector consists of a cylindrical stainless steel tank 41.4 m tall and 39.3 m in diameter enclosing 50,000 tons of ultra-purified water, with 13,000 large photomultiplier tubes.**

To be relevant for far-field reactor monitoring, water Cherenkov detectors must be made explicitly sensitive to antineutrinos, using the inverse beta decay interaction of Equation (1). This requires the ability to detect neutrons in the water, in order to exploit the two-fold time-coincident signal generated by the positron and neutron, which signature is needed to distinguish the anti-neutrinos from many other sources which cause a single flash of light (particularly solar neutrinos). Doping with gadolinium (at the 0.1% scale) of large scale water detectors has been proposed as one way to make water Cherenkov detectors sensitive to neutrons. A detailed study of this option has been performed in consideration of a possible upgrade to the Super-Kamiokande detector, known as the GADZOOKS detector[79]. As a small scale proof of the detection principle, an LLNL group has demonstrated water-based neutron detection, and sensitivity to inter-event time correlations identical to those produced by antineutrinos, using a 250 kg Gd-doped water detector[80]. Other schemes have been proposed for neutron detection and for enhancing the light output in water Cherenkov detectors, and these are matters of active study at present. As demonstrated by Superkamiokande[81], the achieved limit for light attenuation in purified water is roughly 80-100 meters, so that megaton scale detectors, with 100 meter linear dimensions, are not greatly affected by attenuation. However, it remains to be demonstrated that attenuation lengths are not affected by wavelength shifters, or, crucially, gadolinium doping. Moreover, larger detectors than 1 megaton would require modular construction or other expedients. Beyond the question of gadolinium doping, many other obstacles remain for scaling the technique to the megaton scale and beyond. The relevant research and development paths are discussed below in Section 7.6.

On another front, a billion ton detector, the cubic kilometer IceCube instrument is nearing completion at the South Pole by a US-Europe collaboration[82]. This instrument trades off sensitivity for size, with a neutrino detection threshold of a few GeV instead of our required energy down to a few MeV, and no current sensitivity to the inverse beta decay reaction. Despite its attractive dimensions, IceCube is unlikely to play a role for reactor monitoring applications. However, we return to the example of IceCube in section 7.5, with a discussion of its potential sensitivity to burst-like antineutrino phenomena in the few MeV range.

## 7.5 Detection of Nuclear Explosions

Nuclear explosion monitoring is an important element of the proposed Comprehensive Test Ban Treaty (CTBT)[83] and similar regimes. A variety of on-site and remote sensor technologies can be and are already used for verification within such regimes. The technologies can be used to detect nuclear explosions conducted in the atmosphere, underwater, underground, or outer space. Depending on the geographical location and site access, explosive yields approaching 1 kiloton can be measured in some, though not all circumstances. Given this technology base, and the likely achievable capabilities of antineutrino detectors for explosion detection, antineutrino detection is likely to play at best a supplementary role in nuclear explosion detection in the near and medium term, most likely in cooperative contexts and as a confidence building measure. Its main possible advantages are unambiguous evidence of a fissile character of the explosion (rather than some other explosive or seismic event), and the ability to provide a competitive estimate of the device yield. In the discussion below, we briefly summarize the state of the art for remote nuclear explosion detection technologies, for the purpose of comparison with antineutrino based explosion detection. A more detailed comparative study may be found in [84].

### 7.5.1 Atmospheric and Exoatmospheric Explosion Detection Technologies

Several sensor types can be used to record the nuclear signature of an atmospheric or exoatmospheric explosion, seconds to days after the event. Bhangmeters, X-ray, gamma ray, neutron, and EMP detectors are space-based and designed to see the radiation from the explosion. Radionuclide sensors can detect the long-lived radioactive gases and particulates from the explosion. Since the Cold war era of intensive nuclear test activities, proliferant nations have avoided above-ground tests and instead pursued underground testing. The following technologies, as well as antineutrino detection, are suitable for detection of subterranean and underwater explosions.

### 7.5.2 Underwater and Underground Technologies

Within the global CTBT regime, there are five remote sensor technologies used for detecting and identifying underwater or underground nuclear tests: radionuclide, seismic, infrasound, hydroacoustic, and satellite imaging. Only radionuclide sensors can see an intrinsically nuclear signature from a fission explosion, and then only if the contained explosion accidentally vents into the atmosphere. The other four technologies can see various blast effects, but not direct nuclear radiation effects. Consequently, by utilizing evasion tactics such as decoupling, deeper burial, camouflage, concealment, and deception, it may be possible to conduct low-yield nuclear explosions underground that would escape detection altogether, or be incorrectly identified as non-nuclear phenomena (e.g., earthquakes, rockbursts, and chemical explosions).[85]

There have been at least three documented cases in which natural earthquakes were incorrectly identified as nuclear explosions: two in January and August of 1996 near the Russian nuclear test site at Novaya Zemlya, and a third by Pakistan on April 28, 1991.[86] In addition to the occasional occurrence of false positives, false negatives are also possible - the incorrect classification of low yield underground nuclear explosions as innocuous events.[87] Although seismic sensors have the best chance of detecting a low yield underground test, the relatively weak signal is not inherently nuclear and would be just one of the thousand estimated ambiguous seismic signals produced by earthquakes worldwide each year.[88] As a result, the rare occurrence of a clandestine, underground nuclear test could be lost in the noise.

### 7.5.3 A Role for Neutrinos

The potential for ambiguity in seismic monitoring raises the question of whether antineutrino detectors might give useful supplementary information. The required detector size for a given standoff and number of events are shown in Figure 26. For example, detection of ten events from a 10 kiloton explosion at 200 kilometers would require a three megaton detector.

Detection of even one antineutrino in coincidence with a blast, located and time-tagged by other means, would have important implications: one would know first that the blast was certainly nuclear, and one would know the yield within a factor of several. With ten neutrino events, the yield would be determined within thirty percent, better than is typically achieved seismically. Unfortunately the detector sizes needed for standoff detection are daunting: roughly 10-100 times larger than the largest detectors now being proposed, and 1000 to 100,000 times larger than state of the art antineutrino detector (KamLAND). A distributed array of such large anti-neutrino detectors offers the additional possibility of explosion location, as well as improved yield estimates. For completeness Table 9 summarizes a set of possible nuclear explosion monitoring goals that could be enabled with such detectors.

Of course, while assumptions differ slightly, the numbers of events expected and required have not changed since the 2001 study[84] (leaving aside the relatively small correction now known to come from electron antineutrino oscillations, which is not included in our tabulated examples). The main advance compared with the earlier study is that megaton scale detectors, required for the simplest standoff detection goals for kiloton scale explosions, are now being considered by various groups for fundamental physics studies, which was not the case a decade ago. If such detectors are developed, antineutrino detection might someday provide a transformative additional capability for CTBT verification.

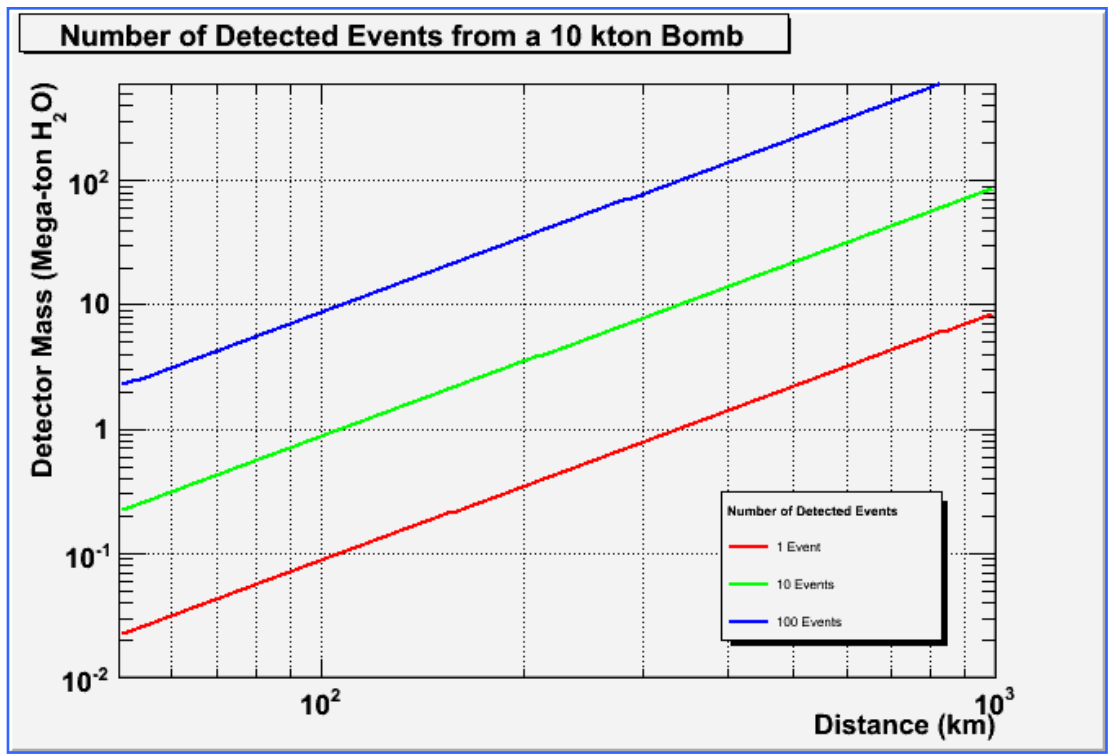

**Figure 26:** The number of events which would be detected on average for a given detector mass, versus the distance from the explosion of a nominal 10 kiloton-TNT equivalent nuclear weapon. Oscillation effects are not included in this crude estimate. The water based detector has an assumed energy cut of 3.8 to 6.0 MeV. The lower line would permit confirmation of the nuclear nature of the blast and a crude estimate of yield. The green line, corresponding to 10 detected events, would allow autonomous detection and a 30% yield estimate. The upper blue line for 100 events would permit a 10% yield estimate, and extraction of other information.[89]

| *Goal for a hypothetical 100 MT instrument* | *Number of Events* | *Range* |
|---|---|---|
| **Detect explosion seen by other means** | >1 evt in coincidence | 1500 km |
| **Estimate yield to 30%** | 10 events | 500 km |
| **Detect otherwise undetected explosion** | 5 events in 4 sec | 700 km |
| **Estimate range for known yield, via number of events** | 10 events | +/- 15% |
| **Precision range, via oscillations** | 100 - 1000 events | +/- 1 % |
| **Location with 2 detectors** | 2 time >10 events | <250 km^2 |
| **Details of explosion (using time & spectrum)** | 1000 events | 210 km |

Table 9: Possible goals for detection of nuclear explosions with a hypothetical 100 megaton detector. Detectors on this scale are for now beyond the level now being proposed for large antineutrino detectors in the fundamental science community.

A final consideration in the context of explosion detector are very large detectors such as Ice-Cube[90]. As currently operated, these detectors are relatively sparsely instrumented with PMTs, and are therefore sensitive only to GeV-scale or greater antineutrino energies. However, large detectors of this kind have been considered for MeV scale antineutrino detection, in the context of supernovae and gamma ray burst detection[91]. The essential idea is to search for very small but well-correlated fluctuations in response among many PMTs, induced by the pulse of antineutrinos engendered by some astrophysical event. This interesting idea merits further consideration in a nonproliferation context, since it might someday allow these very large scale detectors to be adapted to nuclear explosion detection.

## 7.6 R&D Needs for Far Field Detectors

The principal research and development avenues for far-field detection divide into issues related to sensitivity and those related to cost. Perhaps the most important advance, doping of water based detectors with gadolinium compounds, has already been discussed in section 7.4. Other important areas of research are summarized below.

### 7.6.1 Background Suppression:

By various methods, such as those discussed in Section 6.4, far-field detection requires suppression of backgrounds in large scale water Cerenkov and liquid scintillator detectors to levels comparable to or exceeding those achieved in the KamLAND experiment. Methods have been developed for purification of scintillator liquids from radioactive materials (the most notorious being radon and decay daughters). KamLAND, Borexino[92] and SNO+[93] groups have all made great progress in bringing high radiopurity to levels thought to be unachievable within the last two decades. Backgrounds from detector boundaries and rock or water surroundings become less of a problem as detector size scales up, but material cleanliness remains an issue. Though well handled in present detectors, work is needed to make this a matter of industrialization in future huge instruments.

Cosmic ray muons and muogenic neutrons and isotopes decrease rapidly with depth. As stated earlier, below about 3 km water equivalent depth these are not a problem. At lesser depths they are more so, and from experience we know that they are manageable at depth of 2 kmwe. Even shallower depths may be practical depending upon the size of the instrument and the target signal. Further study is needed to determine just how shallow is tolerable.

### 7.6.2 Development of event-by-event direction reconstruction for the antineutrino signal:

Antineutrino directionality is an extremely difficult problem to solve, but could transform concepts of operation by allowing real-time location of reactors and much improved background rejection. A simple qualitative analysis of the physics reveals the problem. The challenge lies in the fact that the direction of the recoiling neutron and positron is only loosely correlated with the direction of the incoming antineutrino. When directionality has been achieved, most notably by the Chooz experiment with a reconstructed half-angle of about 20 degrees[14], this means that dozens or hundreds of events are required to extract the average direction of the antineutrino signal. For long range detection in which only a few events are available, this is obviously of no use. However, since momentum and energy are conserved in each individual event, it is possible *in principle* to reconstruct the antineutrino direction on an event-by-event basis. This essentially requires direct reconstruction of the positron and neutron momenta within the first few scatters following the antineutrino interaction, which in water and liquid scintillator occur within mere millimeters to a few centimeters of the event vertex. Thus what is required is a detector with a linear dimension of 10-100 meters, and position sensitivity on the scale of millimeters. Possible solutions included virtual segmentation via time projection chambers, or devices based on laser illumination and reconstruction of tracks, but the problem is clearly enormously difficult and no solution appears imminent.

Concerning cost, the two dominant factors, at least for water detectors, are PMTs and excavation or, in the ocean, containment. This suggests a focus on the two following R&D paths:

### 7.6.3 Low cost photodetection:

At present all large instruments use PMTs. These large (Super-Kamiokande uses 20 inch diameter tubes) are beautiful devices, low noise and have high photon detection efficiency. But they involve significant mechanical complications, and the number required for a 10 Megaton detector would be very difficult to manufacture and handle. The megaton class detectors mentioned above primarily anticipate using existing technology because of the long timelines for new devices. The cost of the present devices is also intimidating, at about $1/cm$^2$. For example the cost of photomultipliers for a 10 megaton instrument would be around $1.1B. Progress is possible if increased funding and effort is directed towards advanced photodetection technology. Indeed there is a lot of commercial activity in this area, often directed towards small pixel sizes – for example the tremendous revolution in CCD cameras. The precision of dimensions required has been demonstrated by flexible circuit manufacturers, with the prospect for being able to make the needed acres of photodetector in rolls like wall paper. There are however many concerns about manufacturing, noise levels, lifetime, and other factors.

Ultimate manufactured costs have been estimated to be as low as a few cents per cm$^2$, down by a factor of 10-100 below those of present photomultipliers. Both within and beyond the neutrino detection community, there is great motivation to pursue this development.

### 7.6.4 Ocean and Mine Tunnel Engineering:

While the problems are different , deployments of large detectors in mines and the ocean will require significant engineering R&D to proceed. In the case of mine based instruments, many of the problems are already under active study in Japan (HyperK), Europe (LENA) and the United States (DUSEL). There surely is a strong limitation on size of such detectors with reasonable cost, the present limitation being thought to be in the scale of one megaton. Even at that size the excavation time becomes a problem (approaching a decade timescale). For the grand visions of detectors beyond the megaton scale it seems inevitable that such instruments be placed in the ocean. There have been very preliminary studies of engineering at these sizes, but beyond about 1 megaton (the mass of the world's largest oil tanker) is new territory. We will not list the technical problems here, but simply state that while no in-principle problems exist, there are formidable engineering challenges including construction of giant bags, massive water purification, transportation, and other problems. One possible synergy comes from the large investments in underwater technology made by oil companies in recent years.

### 7.6.5 Reducing overburden requirements:
These approaches were discussed in Section 6.3 above.

## 7.7 Summary of Far-Field Applications

In this section we reviewed some of the technical and cost issues associated with deploying a large inverse beta decay detector at ten to several hundred kilometer standoff distance from reactors. We conclude that the several next generation detectors being proposed for physics experimentation might be useful for discovery or exclusion of small reactors in the far-field in some areas of the world. Of particular moment is the considerable and natural overlap in detector technology between the physics and nonproliferation applications. Sensitivity goals for hundreds of kilometer distant monitoring of small reactors with no other reactors present are currently beyond the state of the art, with the required detector masses roughly a factor of ten beyond the current state of the art. Portable next generation liquid scintillation detectors such as the proposed ten kiloton Hanohano can pursue fundamental physics topics while demonstrating and developing the technologies that move this area ahead. While affordability and allocations of national budgets ultimately relate to the desirability of the nonproliferation outcome, use of water Cerenkov technology, coupled with breakthroughs in the area of low-cost photodetection, appears to be the most cost effective approach.

# 8   Fundamental Physics and Reactor Antineutrino Detection

The fundamental science developed with nuclear reactor antineutrino sources has provided many of the ideas, and much of the research and development funding that has made nonproliferation-related reactor monitoring possible. Given the natural connection between the two fields at the level of technology, it can be expected that cross-pollination of this kind will increase over the next decade as new, more sensitive, detectors are built. Moreover, in a kind of reverse spin-off from international security applications to science, widespread deployment of antineutrino detectors for reactor monitoring could provide an important additional global resource for studying the fundamental properties of neutrinos. Here we summarize current and projected activities in the area of fundamental neutrino physics using reactor sources.

Over the last ten years there has been considerable progress in understanding the physical properties of neutrinos. It is now known that neutrinos have mass, and that the three types of neutrinos ($\nu_e$, $\nu_\mu$, $\nu_\tau$) - paired with the lepton (e,μ,τ postulated by the Standard Model (SM) of particle physics do not correspond to the three types of neutrinos found in nature. Physical neutrinos propagate through space according to their mass. These physical neutrinos ($\nu_1$, $\nu_2$, $\nu_3$) are now known to be quantum mixtures of the three types postulated by the SM. Thus they travel through space as ($\nu_1$, $\nu_2$, $\nu_3$), but interact with matter as ($\nu_e$, $\nu_\mu$, $\nu_\tau$). Establishing this experimental fact was the culmination of over twenty year of measurements of neutrinos from the sun, and from cosmic ray interactions in the Earth's atmosphere. These pioneering (and somewhat astonishing) results have since been verified in detail with experiments using neutrinos from the sun, cosmic rays nuclear reactors and accelerators. Current research that uses reactors as neutrino sources centers on measuring the amount of mixing between neutrino types, and in studying two other fundamental properties of neutrinos: their collective interactions with nucleons in nuclei, known as coherent scattering, and their magnetic moments. Here we introduce all three areas in turn, and discuss ongoing research in each, as it impacts nonproliferation missions. The list of ongoing experiments here is not exhaustive but is illustrative of the overlap in research and development between nonproliferation and fundamental physics applications. More comprehensive surveys of various neutrino experiments worldwide are available[94,]

## 8.1 Neutrino Mixing

Neutrino mixing, also known as neutrino oscillations, can be described as a mathematical transformation of vectors corresponding to the three different neutrino types or flavors. It occurs for neutrino flavors *and* for antineutrino flavors separately, so that the phenomenon can be observed with reactor sources. In the following paragraph, the word antineutrino can be globally substituted for the word neutrino.

Physical neutrinos that interact with matter - including detectors - can be represented as the three components of a vector. If we let the three axes of a coordinate system represent the three types of neutrinos in the SM ($\nu_e$, $\nu_\mu$, $\nu_\tau$), neutrino mixing can be thought of as a rotation of the physical neutrino vector in space. Geometrically, three angles can be used to describe such rotations, which are conventionally labeled $\theta_{12}$, $\theta_{23}$, and $\theta_{13}$. The first two of these angles have been measured while the third is as yet unknown. Thus there is currently a program to increase the measurement precision for the first two angles and make an initial measurement of the third angle, $\theta_{13}$. Some of this work has and will be done at nuclear reactors, and is described below. The experiments all use the inverse-beta reaction – as do proposed reactor monitoring detectors. In addition, much of this program requires a high precision knowledge of the antineutrino reactor flux and its variation with reactor type, fuel loading, and operational history – essentially the same types of measurements needed for non-proliferation monitoring.

### 8.1.1 Neutrino Mixing: θ₁₂

This neutrino mixing angle has been measured both using solar neutrinos and at nuclear reactors via *neutrino oscillations*. In this process, the physical neutrino flavors travel at different speeds through space (since they have slightly different masses) with the result that the admixture of flavors *varies* at every point along the path of the neutrino beam. Such a process has been dramatically demonstrated by the KamLAND detector in Japan, first introduced in Section 6. This detector used the combined flux from many Japanese nuclear reactors at a distance of roughly 200 km to show that such oscillations were indeed taking place. They showed that the mixture was a function of energy consistent with the form predicted by neutrino oscillations.

The detector consists of 1,000 tons of organic liquid scintillator viewed by roughly 2,000 phototubes (figure 1). The detector is surrounded by an active water-based detector as a cosmic ray veto – and the whole assembly is located 1 km under a mountain at the Kamioka Observatory in central Japan.

Figure 2 shows the resulting best measurement[95] of what is essentially $\theta_{12}^{\phi}$ based on the KamLAND reactor measurement and the solar neutrino data. The y-axis is the mass difference between the main two neutrinos involved. The mass difference is known to about 7% and the angle (32 degrees) is known to about 9%.

KamLAND is still operating, but it is not ideally placed[g] to make the most precise parameter measurement and the main focus is now on neutrino measurements relevant to solar astrophysics. Thus there may be "medium baseline" experiments for precision measurements of $\theta_{12}$. Hanohano, discussed in 7.3, is one example.

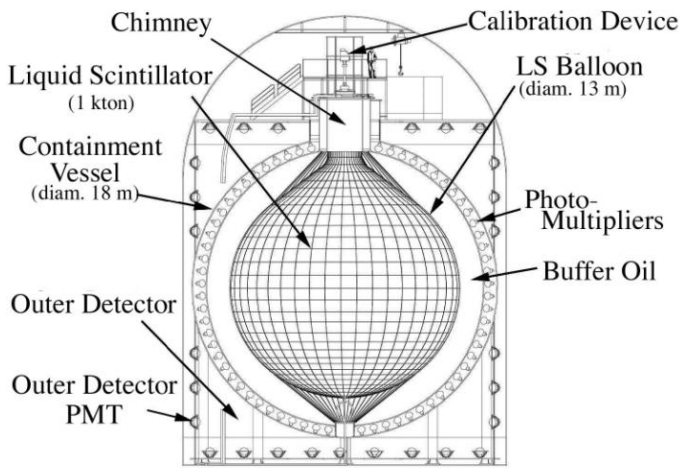

**Figure 27: The KamLAND detector for reactor antineutrinos. It has a fiducial mass in the 1,000 ton range and can measure the summed neutrino signal from multiple commercial power plants several hundred kilometers distant[95].**

---

[f] There are small corrections due to the influence of the other mixing angles.
[g] KamLAND is actually too far away – about 200 km on average whereas the ideal location is about half that distance.

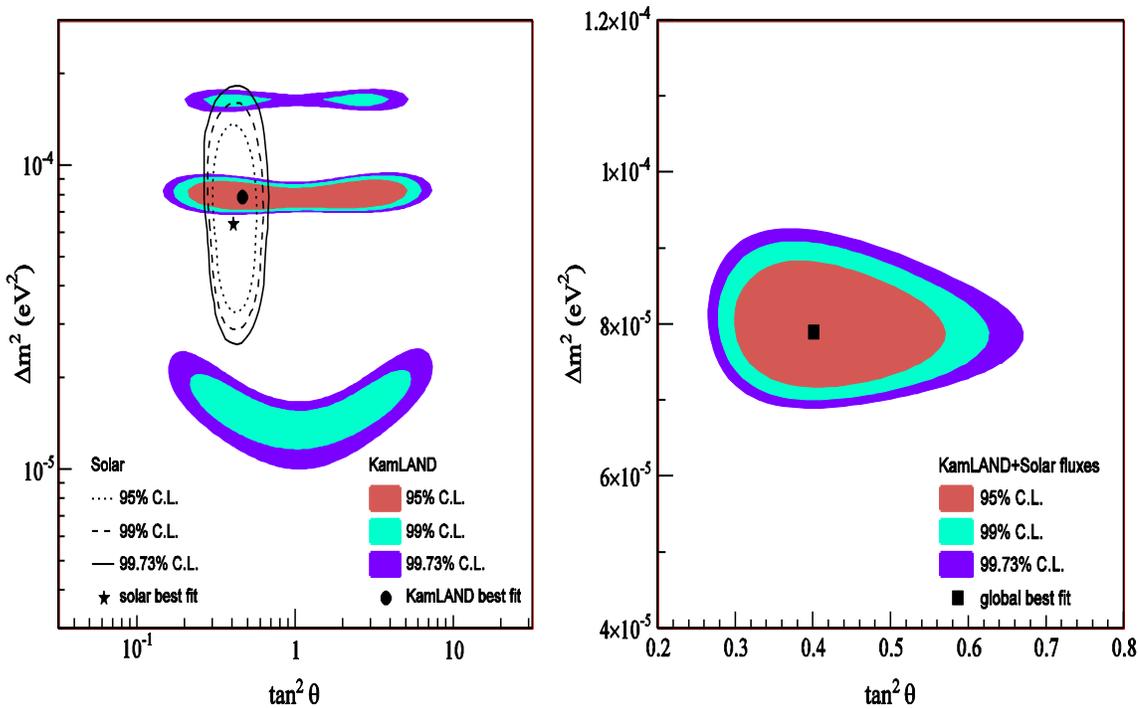

**Figure 28:** The best fit results for the oscillation mixing angle theta 12 from KamLAND (left, shaded contours) and solar neutrino data (left, line contours). The right plot shows the combined data[95].

### 8.1.2 Neutrino Mixing: $\theta_{13}$

$\theta_{23}$ has been measured to be very close to 45 degrees using a combination of cosmic ray and accelerator experiments[96,97,98].

$\theta_{13}$ is the only angle not yet measured. Since experiments that seek to measure the exact ordering of the neutrino masses, or that want to look for matter-antimatter asymmetry (CP violation) in neutrinos *must* first know the value of this parameter – there is intense scientific interest in making this measurement as soon as possible.

There are three different methods for approaching this problem:
1. a long-baseline (roughly 300-800 km) accelerator experiment with a neutrino beam with mean energy roughly 500-1500 MeV,
2. a precision reactor antineutrino experiment at a detector distance of about 1 km which looks for rate and spectral variations in the antineutrino signal, induced by oscillations,
3. a medium baseline experiment, 40-60 km from a reactor, that looks at antineutrino spectral distortions.

Currently, all three approaches are being pursued and are at various levels of maturity. The first approach is a major new thrust for the United States Deep Underground Science Laboratory Long-Baseline experiments (LBL-DUSEL)[99]. While the neutrino energy scales differ greatly from the MeV scale of interest for reactor monitoring, the LBL-DUSEL experiments are designing large water Cerenkov detectors which are of potential interest for nonproliferation applications.

Concerning reactor experiments of type (2), there are programs underway in China, Korea, and France. A typical detector (Double Chooz) is shown in figure 3. This detector is of moderate size – about 3.5 meters on a side. They will detect neutrinos at a rate of ~90 day at a distance of 1 km from a 8.5 GWt nuclear power station.

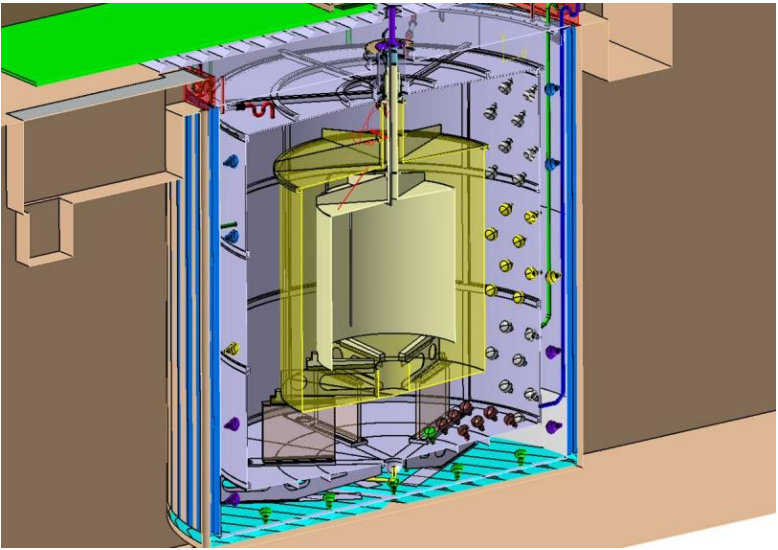

**Figure 29: One Detector of the Double Chooz experiment. This experiment will make a very precise measurement of the antineutrino spectrum starting in late 2008.**

The effect of $\theta_{13}$ on the measured reactor spectrum and rate is expected to be on the order of 5% or less. Therefore understanding the flux as a function of reactor power and fuel loading is critical. In addition, there are also basic physics parameters (mainly the amount of recoverable thermal energy per fission and the neutrino spectrum from different fuel components) that impact the final sensitivity. Table 1 shows the systematic and statistical uncertainties achieved by the Chooz experiment in the 1990's[100]. Also shown are the expectations for an improved Chooz detector design now under construction. While this will improve results somewhat, the positioning of an identical detector near the reactor cores (before neutrino oscillation has a chance to occur) will have the greatest improvement on sensitivity (last column).

| Uncertainty | Chooz | Improved Chooz | Double Chooz |
|---|---|---|---|
| Neutrino spectrum | 1.9% | 1.9% | <0.1% |
| Number of protons | 0.8% | 0.5% | 0.2% |
| Detector efficiency | 1.5% | 1.1% | 0.5% |
| Reactor power | 0.7% | 0.7% | <0.1% |
| Energy per fission | 0.6% | 0.6% | <0.1% |
| Statistical | 2.7% | 0.4% | 0.4% |
| **TOTAL** | **3.8%** | **2.5%** | **0.7%** |

Table 10: Systematic Uncertainties in the Chooz experiment, New Chooz detector, and Double Chooz (relative)

Thus this area has two potential impacts on the sensitivity of non-proliferation detectors: (1) the design, building, and calibration of advanced detectors with unprecedented stability and resolution, and (2) the measurement of the *absolute* flux of antineutrinos as a function of energy at a position nearby the core.

Note that the absolute flux measurement can be made because *identical* detectors are needed for the near and distant position, in order to cancel detector systematic uncertainties. This necessitates that a large detector (sized for 1 km distance) be built close to the reactor core (~200 meters). Thus these experiments will have as a by-product high statistics, well-calibrated, neutrino spectra complete with fuel loading dependence. This presents an excellent opportunity for the non-proliferation community to obtain the precise data that could be used for monitoring with reduced reliance on initial knowledge of the thermal power or the core design. Double Chooz collaborators from Brazil, France, Japan, Russia, and the U.S. are also interested in reactor monitoring applications.

Hanohano[101], with a proposed deployment location of 55 km from the SONGS reactor in California, is an example of a medium distance experiment of type (3). Its physics goals are to measure $\theta_{13}$ via a spectral analyses. It would demonstrate robust remote deployability of large (10,000 ton) detectors of interest for nonproliferation.

## 8.2 Supernovae Detection

Supernovae detection is another area of fundamental research that has considerable technology overlap with nonproliferation. Although supernovae emit some $10^{57}$ antineutrinos, the (fortunate) remoteness of the event implies detector sizes on the multikiloton to megaton scale, similar to the requirements established earlier for nonproliferation. Moreover, supernova antineutrinos have mean energies of about 15 MeV[102], close to the energy scale of reactor antineutrinos. Supernova neutrinos emerge from the core of the nascent neutron star rapidly—within 10 seconds or so, roughly consistent with the exponential pulse from nuclear explosion.

Detection of supernovae antineutrinos has already been accomplished with fairly large water Cerenkov detectors. In a remarkable breakthrough, the Kamiokande-II[103] and the 10,000 ton IMB[104] experiment both successfully recorded a few second burst of antineutrinos from supernovae 1987A. Next generation large water Cerenkov detector proposals, such as Hyperkamiokande[76], the proposed U.S. water detector at the Deep Underground Science and Engineering Laboratory (DUSEL) in South Dakota[77], and the European MEMPHYS detector[78], all include supernovae detection as an important element of their overall physics goals.

An additional feature of supernova neutrino detection, not shared by nuclear explosions, is the presence of all flavors of neutrinos and antineutrinos. It had been thought that these might provide geographic information about the nascent neutron star[105], but recent theoretical work[106,107] has suggested that flavor mixing and other effects will so thoroughly scramble the neutrino distributions that all flavors will emerge with roughly the same energy distribution. This wider range of neutrino flavors has motivated consideration of liquid scintillator based supernovae detectors. Liquid scintillator detectors would provide detection of both electron neutrinos and antineutrinos via charge current interactions, and of the mu and tau neutrinos and antineutrinos through neutral current interactions. Liquid scintillator detectors can detect lower energy antineutrinos than water detectors, somewhat improving detection efficiency for nuclear explosions and supernovae compared with water detectors. With such detectors, the energy spectra of the antineutrinos from an explosion could also provide information about the details of the actual weapon design.

A large segmented detector such as the 15 kiloton NOVA detector[108], with detector modules that are a few cm in cross section and 15 m long, might also provide additional background suppression and some directional sensitivity. The current NOVA detector design has only minimal shielding, relying on timing with a pulsed beam for background suppression. Substantial modification of the design, including 10-100 fold increases in size, and deep burial, would therefore be needed to enable explosion monitoring applications.

## 8.3  Geo-Antineutrinos

The 1000 ton KAMLand liquid scintillator detector described earlier, was the first to measure antineutrinos produced by radioactive decays of naturally occurring uranium and thorium isotopes from the Earth. The 10,000 ton Hanohano detector, described in Section 7 is also expected to acquire geoantineutrino signals at one or more locations in the ocean. The geoantineutrino signal is expected to offer rich insights into the origin and nature of the Earth's crust and mantle[109], and may ultimately be able to measure or rule out a geo-reactor postulated to exist in the Earth's core[110].

Geoantineutrino detectors are particularly relevant for nonproliferation since they are of necessity large - at minimum 1 kiloton -  and since the endpoint of the geoantineutrino energy spectrum, at roughly 3.3 MeV, is *lower* than that for nuclear reactors. The world's first geoantineutrino spectrum is shown in Figure 30, as measured by the KamLAND detector[111]. The large size and low energy threshold requirements for geoantineutrino detectors make them directly relevant for mid-field and far-field reactor monitoring applications.

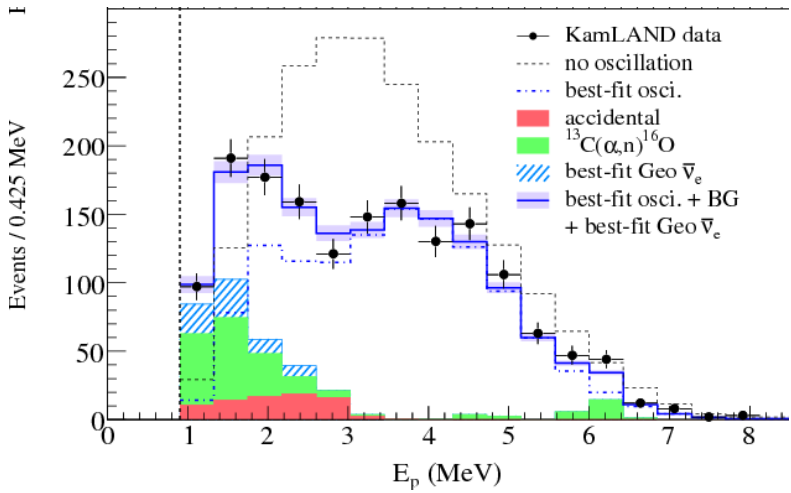

**Figure 30: A prompt energy spectrum from the KamLAND detector, reproduced from the Official KamLAND results website maintained at Lawrence Berkeley Laboratory[112]. The prompt energy (see 4.1.3) is closely related to the antineutrino energy spectrum. This spectrum shows the low energy geoantineutrino signal (prediction in hatched blue), superimposed onto the reactor antineutrino spectrum, and demonstrates that inverse beta detectors capable of detecting geoantineutrinos can also measure reactor antineutrinos.**

## 8.4 Coherent Elastic Neutrino Nucleus Scattering

There are several ongoing efforts to develop detectors to look for the predicted "coherent" interaction of neutrinos with nuclei[113,114,115]. This elastic scattering process, predicted by the SM, would give rise to a large neutrino interaction rate at low energies – even in small detectors. Detection at such low energies is difficult, however. To date no experiment has successfully confirmed the existence of this reaction. It is noteworthy, however, that most astrophysical theories of stellar collapse supernovae *require* this type of scattering in order to transfer energy from neutrinos to stellar mantles during the collapse process – else the supernova just "fizzles" and never explodes. Thus there is strong interest in the astrophysics community of verifying this reaction – and reactors are a promising venue for this type of measurement to take place.

If the wavelength of a low-energy neutrino is large enough it will react with an entire nucleus rather than with the individual nucleons. The cross-section for such a "coherent" process rises approximately as the *square* of the number of nucleons in the nucleus. Thus at low energies it can exceed the inverse beta interaction cross-section by 1-3 orders of magnitude, depending on the element. The increase in rate can be translated into smaller footprint detectors for reactor monitoring. For example, expressed in terms of the SONGS parameters described in Section 5.3.3, a 20 kg fiducial mass liquid argon detector could detect as many as 400 events per day, 25 meters from a typical commercial reactor core (3.4 GWt). After accounting for shielding requirements, this could lead to a factor of 5-10 smaller total footprint detector, when compared to detectors relying on the inverse beta interaction.

The main difficulty in making the measurement is that most of coherent nuclear recoils induced by fission neutrinos transfer very little energy, typically less than 1 keV. For an ionization detector, this meager energy transfer results in very few free electrons. Noise-free detection of a few electrons, even at relatively high rates, is a serious experimental challenge. Analogous difficulty confronts scintillation and calorimetry based approaches.

Figure 31 shows one such detector concept, now under development at LLNL[113]. It would make use of a dual-phase Ar detector, in which ionization electrons generated by coherent scatter interactions in liquid are drifted through a liquid-gas interface. Once in the gas, they are accelerated to create an electron cascade and an associated electroluminescence signal, which is detected by photomultiplier tubes. Detection using High Purity Germanium (HPGe) is also been pursued[114,115].

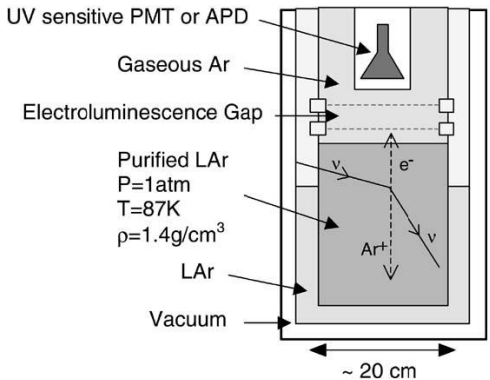

**Figure 31: A proposed dual-phase argon detector for measuring neutrino-nucleus coherent scattering.**

The small detector size and/or high rates that are potentially achievable through coherent scatter mechanism make such detectors of possible interest for reactor monitoring. However, because the coherent scatter interaction is identical for all flavors of neutrino and antineutrino, it is important to point out that they suffer from a probably irreducible limiting background, arising from solar neutrinos. At standoff distances beyond a few kilometers from a GWt scale power reactor, the solar neutrino signal, which is indistinguishable from the reactor antineutrino signal in these detectors, will dominate the measured signal in any coherent scatter detector. This means that the utility in a nonproliferation context is likely limited to near-field applications, out to a few kilometers.

## 8.5 Neutrino Magnetic Moment

In addition to neutrino mixing, there is also a program of using reactor antineutrinos to look for a non-zero neutrino magnetic moment. Neutrinos, like neutrons, are electrically neutral. Unlike neutrons, they behave as point particles and have no (as yet) measurable magnetic moment. In some theories, neutrinos *would* have a very small magnetic moment – which might be detected via a deviation of the "standard" weak interaction at low energies. This type of experiment typically requires detectors with extremely low backgrounds and very high neutrino rates. This is because they need to use very low energy reactor neutrinos to achieve the required sensitivity – below the threshold of the inverse beta decay reaction.[h] They therefore rely on antineutrino-electron scattering, which may be relevant for future non-proliferation detectors. The recoiling electron follows roughly the direction of the incident neutrino – allowing one to obtain the direction of the source.

---

[h] 1.8 MeV

Two recent experiments that have conducted a search for a non-zero magnetic moment are the TEXONO experiment at the Kuo-Sheng reactor[116] and the MUNU experiment at the Bugey reactor[117]. These experiments concentrate on neutrino electron scattering interactions at low energy, so rates are typically a few events/kg/day. TEXONO used a High Purity Germanium (HPGe) detector of mass ~1 kg surrounded by an extensive active and passive shield (figure 5). The analysis threshold was 12 keV recoil energy with a useable signal to about 120 keV.

The MUNU experiment used a one cubic meter $CF_4$ Time Projection Chamber (TPC) to image the tracks of recoil electrons (figure 6). Electron events associated with the reactor neutrino interactions were separated from background by making a cut on track direction. Although designed for 5 bars, this detector operated at 3 bars to improve tracking. Viable electron tracks were recorded to energies of ~ 1 keV when operating at 1 bar.

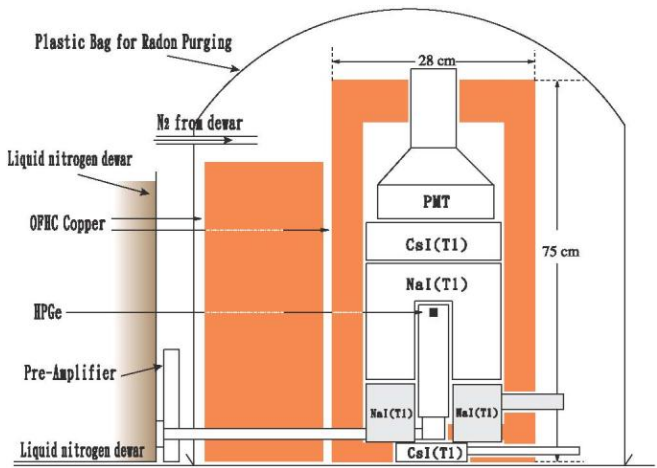

**Figure 32: The TEXONO detector at the Kuo-Sheng nuclear power station. The HPGe detector is used to detect recoil electrons down to an energy of 5 keV.[2]**

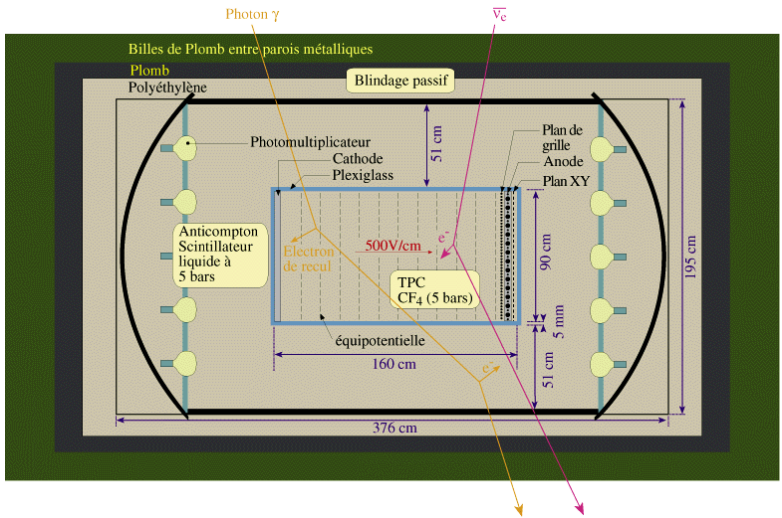

**Figure 33: The MUMU detector. A TPC installed at the Bugey reactor to look for evidence of a neutrino magnetic moment.**

These types of detectors may not be useful for non-proliferation purposes in their present form. However, the ability to deduce the direction of the parent neutrino is a feature of the TPC detectors and may warrant further study in the context of reactor antineutrino monitoring.

## 8.6 Other Physics

Beyond neutrino searches, large water Cerenkov and scintillator detectors may be used in pursuit of a range of other physics goals. Correlated signals with Gamma-ray bursts, searches for proton decay[118], monopoles[119], quark nuggets[120], and other phenomena have all been proposed using detectors similar in scale and design to those discussed in the preceding sections.

## 9 Conclusions

In this white paper, we have sought to demonstrate the breadth of ongoing activity in the area of antineutrino detection for nonproliferation, and the natural connection between this work and current and next generation detectors for particle astrophysics. The last decade has made near-field monitoring capability a reality. Albeit with considerable additional effort, the next decade may usher in reactor monitoring capabilities well beyond these cooperative near-field demonstrations. We hope this paper motivates the science and nonproliferation policy communities, as well as the global scientific community with an interest in neutrino and dark matter physics, to explore and where possible exploit the implications of this connection for both fields.

---

[1] A. Bernstein, et. al., J. Appl. Phys. 91 (2002)

[2] Y.V.Klimov, et. al., Atomnaya Energiya, 76 (1994)

[3] http://www.iaea.org/OurWork/SV/Safeguards/index.html